\documentclass[aps,prd,preprint,showpacs,preprintnumbers, amsmath,amssymb,floatfix, nofootinbib]{revtex4-2}
\usepackage{graphicx}
\usepackage{amsmath}
\usepackage{wasysym}
\usepackage{amssymb}
\usepackage{gensymb}
\usepackage{dcolumn}
\usepackage{bm}
\usepackage[normalem]{ulem}
\usepackage{slashed}
\usepackage{comment}
\usepackage{multirow}

\numberwithin{equation}{section}

\usepackage[dvipsnames]{xcolor}

\definecolor{persianblue}{rgb}{0.11, 0.22, 0.73}

\usepackage[
	colorlinks=true,
	citecolor=green!50!black,
        linkcolor=persianblue!70!black,
	urlcolor=green!50!black,
	hypertexnames=false]{hyperref}

\usepackage{subfigure}
\usepackage{float}

\usepackage{orcidlink}

\begin{document}

\preprint{FERMILAB-PUB-23-224-T, INT-PUB-23-001, N3AS-23-012}
\title{How Macroscopic Limits on Neutron Star Baryon Loss
  Yield Microscopic Limits on Non-Standard-Model Baryon Decay} 

\author{Jeffrey M. Berryman}
\email{jeffreyberryman0814@gmail.com}
\affiliation{Department of Physics, University of California, Berkeley, CA 94720, USA}
\affiliation{Institute for Nuclear Theory, University of Washington, Seattle, WA 98195, USA}
\affiliation{Center for Neutrino Physics, Physics Department, Virginia Tech, Blacksburg, VA 24061, USA}

\author{Susan~Gardner\orcidlink{0000-0002-6166-5546}}
\email{susan.gardner@uky.edu}
\affiliation{ Department of Physics and Astronomy, University of Kentucky, Lexington, KY 40506-0055, USA}

\author{Mohammadreza~Zakeri\orcidlink{0000-0002-6510-5343}}
\email{M.Zakeri@uky.edu}
\affiliation{Department of Physics and Astronomy, University of Kentucky, Lexington, 
KY~40506-0055, USA}

\date{{\today}}

\begin{abstract}
We investigate how our baryon-loss limits from anomalous binary-pulsar period lengthening can be interpreted microscopically to yield specific constraints on the particle physics of baryon number violation within a neutron star. We focus on the possibility of anomalous baryon disappearance via dark baryon processes and on scenarios in which the produced dark-sector particles do not survive to influence the response of the star to baryon-number-violating effects. We flesh out the conditions for which this may occur, as well as other key assumptions. We then turn to the analysis of particle processes in the dense nuclear medium found at the core of a neutron star, employing the techniques of relativistic mean-field theory. Using our study of in-medium effects and limits on macroscopic baryon number violation we extract limits on in-vacuum baryon-number-violating processes, and we determine them for various equations of state. We conclude by noting the implications of our results for models of dark-sector-enabled baryogenesis. 
\end{abstract}

\maketitle

\tableofcontents

\section{Introduction}
\label{sec:intro}

The cosmic excess of baryons over antibaryons is well established~\cite{Zyla:2020zbs}, but the theoretical mechanism by which it is produced is not. The essential theoretical ingredients are thought to be known: baryon number violation (BNV), along with C and CP violation, must all be present in a non-equilibrium environment~\cite{Sakharov:1967dj}. Thus BNV would seem to play an essential role, though in the Standard Model (SM) BNV is thought to occur appreciably only at extremely high temperature~\cite{Klinkhamer:1984di,Kuzmin:1985mm} --- and the existence of BNV at low energies has as yet to be established. In this paper we continue our scrutiny of such effects through observations of neutron stars, which contain enormous reservoirs of baryons. In earlier work we identified sensitive limits on BNV through the interpretation of precise observations of energy loss in isolated neutron stars and in neutron-star binary systems~\cite{Berryman:2022zic}. These studies limit the baryon-number-violating effects that occur across the entirety of a neutron star. In this sense they are {\it macroscopic} limits. In this paper we interpret these limits in a {\it microscopic} way, in that we develop a framework in which they can be translated to limits on the parameters of particular particle physics models that generate baryon-number-violating effects.

The particular models to which our studies are most sensitive are those in which baryons decay or otherwise transform to dark-sector fermions, of ${\cal O}(1\, \rm GeV)$ in mass, that carry baryon number. In such cases BNV becomes an apparent, rather than explicit, effect, because the dark-sector particles are unobserved, even if baryon number is not broken. Although the existence of dark matter is certainly established through astrometric observations, both its nature and origin continue to be open questions. It is possible that the origins of dark matter and of the cosmic baryon asymmetry are related, so that the loosely similar value of the cosmic baryon and dark matter energy densities today may follow from a single underlying model~\cite{Davoudiasl:2012uw}. The possibility of baryons that connect to hidden-sector baryons of comparable mass figure in many such explanations. In this paper we constrain this possibility through the study of neutron and hyperon transitions to final states with dark baryons in the neutron star. To our knowledge, an in-depth, quantitative study of non-SM processes within dense nuclear matter has not previously been realized~\footnote{Albeit studies of exotic light particle emission in dense matter, which possesses simplifying aspects, are of long standing~\cite{Raffelt:1996wa} and continue to be investigated~\cite{Dev:2021kje}.}, and its execution necessitates much care. The existence of neutron stars of about $2\, M_{\astrosun}$ in mass speaks to central densities in excess of three times nuclear matter saturation density, so that in this paper we employ relativistic mean field theory in baryonic degrees of freedom for our dense matter description, as its accuracy should improve with increasing density --- and thus it should work best at the core of the star. We note that a neutron star may become a hybrid star, i.e., one with a quark-based core predicated by a finite-density quark-hadron phase transition, if it is sufficiently heavy, and this possibility can also be constructed within this framework~\cite{Dexheimer:2020rlp}. Transitions to dark baryons could also occur within the quark-based core, though we will set aside this possibility in this paper --- and revisit it only in offering an assessment of our uncertainties in our concluding summary.

The broader possibility of dark decays of the neutron has been noted in explanation~\cite{Fornal2018PhRvL.120s1801F,Fornal:2020gto} of the long-standing neutron lifetime anomaly~\cite{Wietfeldt2011RvMP...83.1173W}, in which the lifetime inferred from counting surviving neutrons is significantly different from that inferred from counting the protons subsequent to ordinary neutron decay. Although the discrepancy may arise from experimental effects, the possibility that dark decays contribute to it in some measure is a continuing possibility~\cite{Berryman:2022zic}. In this paper we provide severe limits on the flavor structure of possible new-physics models with dark baryonic sectors, such as Refs.~\cite{Davoudiasl:2010am,Fornal2018PhRvL.120s1801F,Elor:2018twp,Fajfer:2020tqf,Alonso-Alvarez:2021oaj}, that arise from the interpretation of neutron-star energy loss constraints we developed in Ref.~\cite{Berryman:2022zic}. We also flesh out the general assumptions of that earlier analysis and note how the specific models we consider can satisfy them.

Let us conclude our introduction with a brief outline of the body of our paper. In Sec.~\ref{sec:dark_decay:models} we detail the models of baryon dark decays that we are able to constrain through our neutron star studies, and we note how they are distinct from models that we cannot. We also compute baryon dark decay rates in vacuum, for later reference, as well as dark baryon removal rates, because our analysis assumes that SM dynamics determine the response of the star in the presence of BNV. In Sec.~\ref{sec:macro_bnv} we consider macroscopic baryon number violation in neutron stars, revisiting our earlier work~\cite{Berryman:2022zic} and fleshing out constraints following from its assumptions in greater detail. In Sec.~\ref{sec:medium} we develop how to evaluate particle processes within dense matter, employing RMF, as developed in Refs.~\cite{Walecka:1974qa,Serot:1984ey,Serot:1997xg}, to describe the neutron star medium in $\beta$-equilibrium~\cite{Glendenning:1997wn,Dexheimer:2008ax}. In this context, uncertainties in our description of the dense medium are captured through variations in the equation of state (EoS). With these developments in hand, we evaluate particle processes within our framework for the dense nuclear medium of a neutron star in Sec.~\ref{sec:dark_decay:medium} and use our macroscopic limits on BNV from Sec.~\ref{sec:macro_bnv} to report limits on the parameters of the microscopic models we consider in Sec.~\ref{sec:dark_decay:med_limits_vacuum}. In Sec.~\ref{sec:dark_decay:baryogen_implications} we consider the implications of our results for models of dark-sector baryogenesis and dark matter, and we offer a summary and outlook in Sec.~\ref{sec:summary}. 

\section{Particle Physics Models of Baryon Dark Decays}
\label{sec:dark_decay:models}
The possibility of hadronic processes with dark-sector particles naturally emerges in models that explain both the origin of dark matter and the cosmic baryon asymmetry, particularly if the dark sector candidate carries a baryonic charge~\cite{Agashe:2004bm,Davoudiasl:2010am,Elor:2018twp}. Although it has long been thought that dark matter could also be described as a relic asymmetry~\cite{Petraki:2013wwa,Zurek:2013wia}, in these models, rather,  the two problems are solved simultaneously~\cite{Davoudiasl:2012uw}. More recently, highly testable scenarios~\cite{Barrow:2022gsu} have been developed~\cite{Allahverdi:2017edd,Alonso-Alvarez:2019fym,Elor:2020tkc,Alonso-Alvarez:2021qfd, Alonso-Alvarez:2021oaj,Elahi:2021jia}, and we probe their flavor structure through the studies of this  paper --- and in Sec.~\ref{sec:dark_decay:baryogen_implications} we consider the implications of the constraints that we find. Since the dark-sector particles are presumably SM gauge singlets, they could be light in mass, potentially with masses comparable to that of the known hadrons, and yet have escaped experimental detection thus far.

Our current discussion is loosely inspired by models connected to explanations of the neutron lifetime anomaly~\cite{Fornal2018PhRvL.120s1801F,Barducci:2018rlx,Alonso-Alvarez:2021oaj}, with neutrons decaying to a dark baryon with a photon or an $e^+e^-$ pair. Models with similar content have been considered for broader purposes~\cite{Arnold:2012sd,Heeck:2020nbq,McKeen:2020zni,Fajfer:2020tqf}, and alternative solutions have also been noted~\cite{Berezhiani:2018eds,Rajendran:2020tmw,Strumia:2021ybk}. The dark channels in the various models would impact the determined bottle lifetime, with a mirror neutron model~\cite{Berezhiani:2018eds,Berezhiani:2018udo} serving as a rare exception. There, neutron-to-mirror-neutron conversion occurs in a strong magnetic field, impacting the ability to detect protons in the beam-lifetime experiment. This last possibility has been excluded as a complete explanation of the anomaly by a direct experimental search~\cite{Broussard:2021eyr}, and we refer to further probes of mirror neutrons through existing and possible experiments~\cite{Berezhiani:2017azg,Kamyshkov:2021kzi,Ayres:2021zbh,Hostert:2022ntu}, through binary pulsar timing measurements~\cite{Goldman:2019dbq,Berezhiani:2020zck}, that we consider further here for a distinct
class of models, and through pulsar temperature observations~\cite{McKeen:2021jbh,Goldman:2022brt,Goldman:2022rth,Hostert:2022ntu}. The last set of 
constraints can be significantly weakened through the addition of visible-hidden-sector interactions~\cite{Goldman:2022rth}. We note that models that would explain the anomaly through neutron disappearance or decay to dark-sector final states can also be constrained by the close empirical agreement of the neutron lifetime with its measured $A$ decay correlation as interpreted in the SM~\cite{Czarnecki:2018okw,Dubbers:2018kgh,Berryman:2022zic}. This agreement limits the branching ratio on such exotic processes to~\cite{Czarnecki:2019mwq}:
\begin{equation}
    {\rm Br}(n\to \rm exotics) < 0.16\% \,\,(95\%\, \text{one-sided CL})\, ,
    \label{eq:nexoticCC}
\end{equation}
where we note that the neutron lifetime anomaly is roughly a 1\% effect~\cite{Fornal2018PhRvL.120s1801F}\footnote{The most precise measurement of the $A$ correlation coefficient yields the ratio of the axial-vector to vector coupling constants $|\lambda| = 1.27641(56)$~\cite{Markisch:2018ndu}, but recent measurements of the $a$ correlation do not completely fit this picture, yielding $|\lambda|=1.2677(28)$~\cite{Beck:2019xye} and $|\lambda|=1.2796(62)$~\cite{Hassan:2020hrj}.}. Direct experimental limits on $n\to \chi \gamma$~\cite{Tan:2019mrj} and $n\to \chi e^+e^-$~\cite{Sun:2019tid} decays also exist, removing ranges of parameter space as an explanation of the anomaly. We will be able to set much more severe limits through our studies, where we note the limit on $\Lambda\to \chi\gamma$ from SN1987 for reference~\cite{Alonso-Alvarez:2021oaj}. We regard the neutron lifetime anomaly as a motivation for further investigation of baryon dark decays, with new limits constraining the manner in which the co-genesis of dark matter and the cosmic baryon asymmetry could possibly occur. We now turn to the development of models of
dark baryon decays.

Following Ref.~\cite{Heeck:2020nbq}, we introduce a Dirac fermion $\chi$ with baryon number $B = +1$ which interacts with SM quarks via the generic form 
\begin{equation}
    {\cal L_\chi} = \bar\chi (i\slash{\!\!\!\partial} - m_{\chi})\chi 
    + \left( \frac{u_i d_j d_k \chi_L^c}{\Lambda_{ijk}^2}  
    + \frac{Q_i Q_j d_k \chi_L^c }{\tilde\Lambda_{ijk}^2} + {\rm h.c.} 
    \right) \,,
\label{eq:quark-chi-mix}
\end{equation}
where $i,j,k$ are generational indices, $Q$ and $q$ denote a left-handed quark doublet and a right-handed quark, respectively --- and color and Lorentz indices are left implicit. Such interactions can generate both decay and scattering processes involving dark final states, which we consider closely in this paper. First, though, we address their flavor structure. We could neglect this possibility altogether, dropping all subscript dependence, but simple, renormalizable models that produce Eq.~\eqref{eq:quark-chi-mix}, at energies below the mass scale of their new physics, show that strong flavor sensitivity can nevertheless exist. Turning to models with leptoquarks~\cite{Fornal2018PhRvL.120s1801F,Fajfer:2020tqf}, we consider colored scalars $S_1$ and ${\bar S}_1$ transforming as $(\bar{3}, 1, 1/3)$~\footnote{This variant was first considered in Ref.~\cite{Fornal2018PhRvL.120s1801F}.} and $(\bar{3}, 1, -2/3)$, respectively, under the SM gauge groups and SM invariant scalar-fermion interactions. Non-trivial flavor structure follows from the choice of leptoquark in that $S_1$ can mediate both $n\to \chi \gamma$ and $\Lambda \to \chi\gamma$ decay at tree level, whereas ${\bar S}_1$ can mediate $\Lambda \to \chi\gamma$ at tree level but mediating $n\to \chi \gamma$ would require a one-loop process with $W^\pm$ exchange as well~\cite{Fajfer:2020tqf}. Thus in this paper we strive to probe both $n\to \chi\gamma$ and $\Lambda \to \chi \gamma$ decay processes. These models also readily generate proton decay~\cite{Heeck:2020nbq,Fornal2018PhRvL.120s1801F,Fajfer:2020tqf}, noting $p\to \chi \pi^+$ or $p\to \chi K^+$ decay as examples, so that the possible range of $\chi$ masses is rather restricted as a result. We note that the stability of the $^9$Be nucleus~\cite{Fornal2018PhRvL.120s1801F}, particularly stability against $\rm ^{9}Be \to \chi \alpha\alpha$ decay~\cite{Pfutzner:2018ieu}, requires
\begin{equation}
    m_{\chi} 
    > 0.937993 \,\rm GeV \,,
    \label{eq:mchimin} 
\end{equation}
slightly in excess of the proton stability constraint $m_{\chi} > m_p - m_e$, and that atomic hydrogen is stable if $m_{\chi} > m_p + m_e = 0.93878\, \rm GeV$~\cite{McKeen:2020zni}. If either constraint were not satisfied, then the empirical limit on the pertinent lifetime would bound the parameters of the model. Within the SM both systems are absolutely stable, yet empirical tests of that, with a determined lifetime as an outcome, should be possible. We note H lifetime estimates, made finite through a model with a suitably light $\chi$, are made in Ref.~\cite{McKeen:2020zni}. Moreover, the radiative decay $\rm H \to \nu\chi\gamma$, which is subdominant relative to $\rm H \to \nu\chi$, can be probed through measurements at Borexino~\cite{McKeen:2020zni,McKeen:2020vpf}. Similar expectations follow from violating Eq.~(\ref{eq:mchimin}) --- and a concrete estimate of the $^9$Be lifetime can be found in Ref.~\cite{McKeen:2020oyr}.

In what follows we ignore the possible chiral structure of the quark-$\chi$ couplings and simply consider~\cite{Alonso-Alvarez:2021oaj}\footnote{In this reference $\chi$ has $B=-1$; here, rather, we define $\chi$ with $B=+1$ and
thus write $\chi \to \chi^c$, with $\chi^c\equiv C{\bar \chi}^T$~\cite{schwartz2014quantum}.}
\begin{equation}
{\cal L } \supset
\frac{    u^i d ^j d^k \chi^c}{\Lambda^2} + \rm h.c.\,
\label{eq:quarkchi_nochiral}
\end{equation}
Since the quarks carry electric charge, we have, at the energy scales for which baryonic degrees of freedom are pertinent, 
\begin{equation}
    {\cal L}_{\rm n} = 
    \bar{n} \left(i\slash{\!\!\!\partial} - m_n + \frac{g_n e}{8m_n} \sigma^{\alpha \beta} F_{\alpha \beta} \right)n
    + \bar{\chi} (i\slash{\!\!\!\partial} - m_{\chi} ) \chi
    + \varepsilon_{n\chi} ( \bar{n}\chi + \bar{\chi} n ) \,,
    \label{eq:nchimix}
    \end{equation}
noting $g_n=-3.826$ is the $g$-factor of the neutron~\cite{ParticleDataGroup:2022pth}. This form also holds for the $\Lambda$  upon the replacement $n \to \Lambda$, taking $g_\Lambda = -1.22$. After redefining the fields to remove the mixing term in Eq.~(\ref{eq:nchimix}), then if $\varepsilon \ll m_n - m_{\chi}$, with $m_{\chi} < m_n$, we have~\cite{Fornal2018PhRvL.120s1801F,Fajfer:2020tqf}
\begin{equation}
    {\cal L}_{n \to \chi\gamma} = \frac{g_n e}{8 m_n} \frac{ \varepsilon_{n\chi} }{m_n - m_{\chi}}
    \bar{\chi} \sigma^{\alpha \beta} F_{\alpha \beta} n 
    \,, 
    \label{eq:emdarkint}
\end{equation}
though potentially this operator could also stem from a distinct higher-energy source. Generally, the interaction of Eq.~(\ref{eq:quarkchi_nochiral}) can also generate transitions to dark baryon states with mesons, such as the decays $n \to \chi + {\rm meson}$ or $\Lambda \to \chi + \rm meson$. Ref.~\cite{Alonso-Alvarez:2021oaj} uses chiral effective theory~\cite{Claudson:1981gh} to relate the possibilities. We eschew this path because chiral effective theory ceases to be valid if the density of the neutron star medium much exceeds that of nuclear matter saturation density. Since our particular purpose is to set limits on microscopic models given BNV limits determined from observables associated with the entire neutron star, we set aside the study of final states containing both dark and hadronic degrees of freedom. They are distinct from the final states we do study, and cancellations cannot occur. We thus expect that including these additional decays with hadrons can only make our limits more severe, though the inclusion of hadronic channels would make our estimates less sure. 

\begin{figure}[htb]
    \centering
    \includegraphics[width=0.5\textwidth]{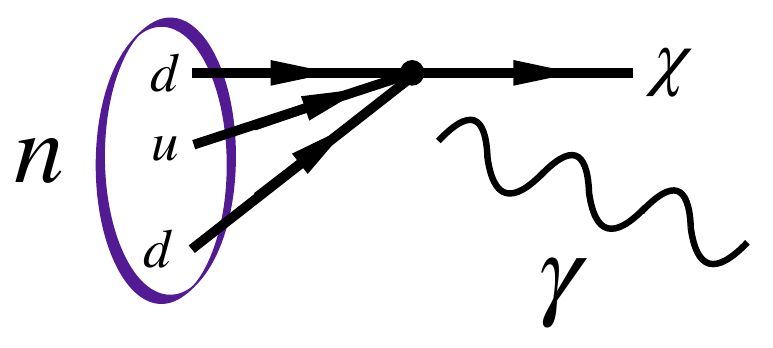}
    \caption{(Color Online) Illustration of 
    $n\to \chi\gamma$ decay in the degrees of 
    freedom of Eq.~(\ref{eq:quarkchi_nochiral}). 
    The decay $\Lambda\to \chi \gamma$ follows
    from the replacement of one $d$ quark with a
    $s$ quark. 
}
    \label{fig:nchigamma}
\end{figure}
At low energies, the magnetic interaction of Eq.~(\ref{eq:emdarkint}), employed in Refs.~\cite{Fornal2018PhRvL.120s1801F,Fajfer:2020tqf}, can be used to compute $n \to \chi\gamma$ or $\Lambda\to \chi \gamma$. A pertinent Feynman diagram is illustrated in Fig.~\ref{fig:nchigamma}. Denoting ${\cal B}$ as either $n$ or $\Lambda$, the total decay rate for ${\cal B}\to \chi\gamma$ is given by
\begin{align}
\label{eq:dark_decay:vacuum:vacdec}
    \Gamma \left({\cal B}\to \chi\gamma\right)
    = \frac{g_{{\cal B}}^2\, e^2\, \varepsilon_{{\cal B} \chi}^2}{128\pi} \frac{\left(m_{{\cal B}} + m_{\chi}\right)^2}{m_{{\cal B}}^5} \left( m_{{\cal B}}^2 - m_{\chi}^2\right),
\end{align}
in agreement with Ref.~\cite{Fajfer:2020tqf}. The mixing parameter $\varepsilon_{{\cal B} \chi}$ follows once the UV model is given, and it is what we constrain through the analysis of this paper.

To determine the impact of these microscopic processes on the neutron star requires further model building. Thus far, at low energies we have a dark baryon $\chi$, which we take to be a massive Dirac fermion. If it is a stable particle, then it can also be a dark matter candidate. If so, then it may already exist within the material that collapsed to form the protoneutron star, though likely only in small amounts, and through dark decays or adsorption on the star it may accumulate within the star. If it is able to give up its kinetic energy, then it may settle in the core of the star, ultimately impacting its properties and evolution. 

\begin{figure}[htb]
    \centering
    \includegraphics[width=0.5\textwidth]{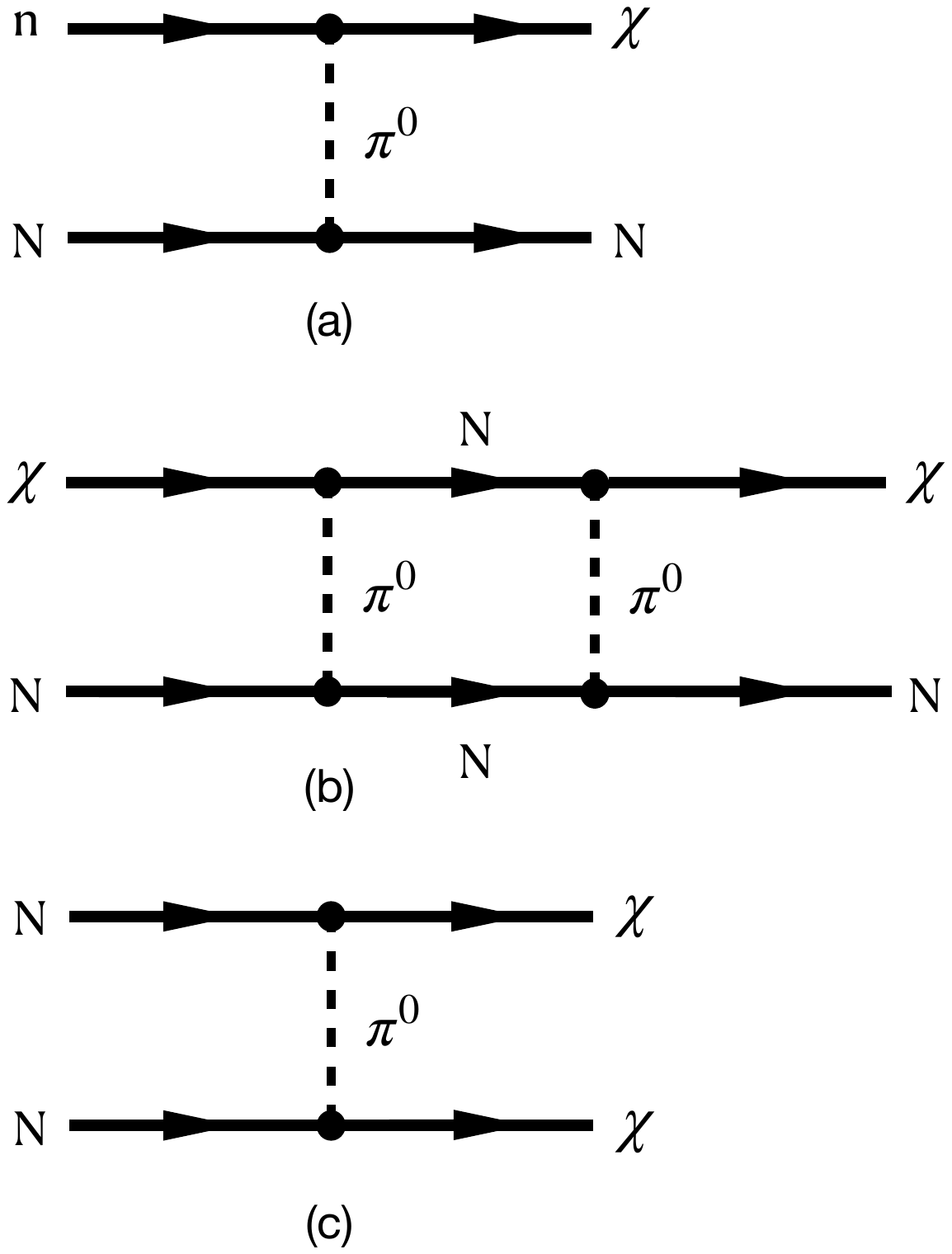}
    \caption{Illustration of various $\chi$-nucleon ($N$) processes at low energies, with the heavy
    black dot denoting the $n\chi\pi^0$ effective vertex
    noted in the text, namely, (a) $n-N$ scattering to produce $\chi-N$, (b) $\chi-N$ elastic 
    scattering, and (c)  $N-N$ annihilation 
    to produce $\chi-\chi$. 
    Processes with 
    $\gamma$ in place of $\pi^0$ are also 
    possible. The reverse of reactions (a) and
    (c) should be strongly suppressed by 
    Pauli-blocking effects in the interior of a neutron star. 
}
    \label{fig:sketches}
\end{figure}

\begin{figure}[htb]
    \centering
    \includegraphics[width=0.5\textwidth]{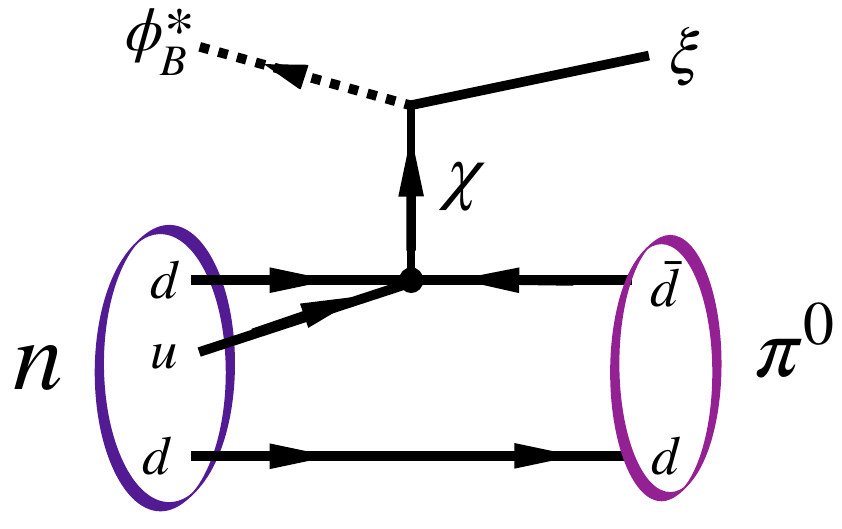}
    \caption{(Color Online) Feynman diagram 
    contributing to induced 
    neutron decay via a ${\bar\chi} \phi_B \xi$ 
    interaction, as per Ref.~\cite{Denner:1992me} 
    --- a $p$ decay channel follows from the
    replacement of $d\to u$ in the spectator quark. 
}
    \label{fig:darkinduced}
\end{figure}

\begin{figure}[htb]
    \centering
    \includegraphics[width=0.5\textwidth]{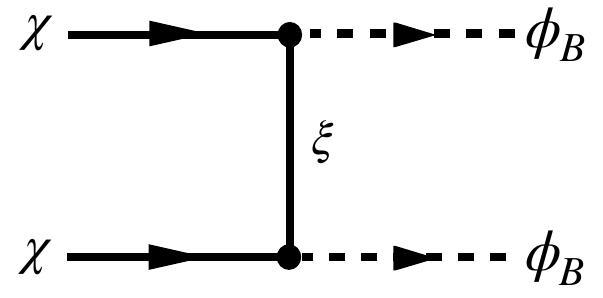}
    \caption{Feynman diagram contributing to 
        $\chi$-$\chi$ annihilation via $\xi$ exchange to yield $B$-carrying scalars, as per the conventions of Fig.~\ref{fig:darkinduced}. 
        Alternatively, $\chi\chi$ annihilation via 
    $\phi_B$ exchange in $t$-channel would yield a $\xi \xi$ final state, which could ultimately rematerialize as a $\bar\chi \bar\chi$ pair. 
}
    \label{fig:darkann}
\end{figure}

There are many processes in which $\chi$ could participate, though the interactions with baryons are severely limited by the cold, degenerate nature of the interior of the neutron star. In principle, given the $n\chi \gamma$ and $n\chi \pi^0$ effective interactions in the models we have noted, and using $N$ to denote either a neutron or a proton, $\chi$ could (i) be produced via $nN \to \chi N$ scattering, (ii) interact elastically with another nucleon via a $nN$ intermediate state, (iii) be formed via the annihilation $nn \to \chi \chi$ or (iv) it can decay via $\chi\to p + e^- + {\bar \nu}_e$ if it is heavy enough. The reverses of the reactions in (i) and (iii) could also occur. Pauli-blocking effects associated with the cold, dense neutron medium strongly suppress all of the reactions in which nucleons appear in the final state. Moreover, $\chi-N$ elastic scattering is further suppressed in that it occurs at ${\cal O}(\varepsilon_{n\chi}^2)$ at amplitude level. We note Fig.~\ref{fig:sketches} for an illustration. Given this, and our interest in limiting BNV  within the star in a model-independent way, implying that the response of the star to BNV ought be controlled by SM dynamics, we think that ensuring $\chi$ disappearance is important.  Thus we consider two different pathways to do just that. In the first, we add $\chi$-lepton interactions~\cite{Berryman:2022zic}, which intrinsically break baryon number and are intrinsically very poorly constrained. We would also want the rate for $\chi$ decay to be no less of that for $\chi$ production. This path, however, is potentially subject to severe constraints from proton decay experiments. For example, we could have $\chi \to e^+ e^- \nu$ or $\chi\to 3\nu$, and these channels could give rise to proton decay via an off-shell $\chi^*$ state as in 
\begin{equation}
    p \to \pi^+ \chi^* \to  \pi^+ e^+ e^- \nu \,.
\end{equation}
(Exotic proton decays of just this ilk also emerge in models with quark and lepton compositeness~\cite{Assi:2022jwg}.) Admittedly, this process, as well as the $p\to \pi^+ 3\nu $ channel, may evade severe constraints due to the particular nature of existing $|\Delta B|=1$ searches, both because of the final-states studied and the cuts on the final-state particle momenta needed to control backgrounds. Although this path could prove to be viable, we favor an alternate choice: we will allow $\chi$ to decay to other dark particles. A simple realization of this is given by~\cite{Elor:2018twp}
\begin{equation}
    {\cal L}_{\rm dark} \supset y_d {\bar \chi} \phi_B \xi + \rm h.c. \,,
    \label{eq:chidark}
\end{equation}
where $\phi_B$ is a complex scalar with $B=+1$ and $\xi$ is a Majorana fermion --- and both are dark-matter candidates. Introducing a $Z_2$ symmetry, so that $\chi$, $n$, and $p$ are all $Z_2$ even, but $\phi_B$ and $\xi$ are $Z_2$ odd, we see Eq.~\eqref{eq:chidark} is the only surviving hidden-sector interaction that traces to the visible sector, with $n-\xi$ oscillations, say, forbidden by the $Z_2$ symmetry. We note that such an interaction is needed for successful 
$B$-mesogenesis, and we also require 
$m_\xi + m_{\phi_B} > 0.937993$ to ensure nuclear stability~\cite{Alonso-Alvarez:2021oaj,Alonso-Alvarez:2021qfd}.
One interesting consequence of this new path is that dark decays can be induced in the scattering with either $\phi_B$ or $\xi$ in the initial state, as developed in Ref.~\cite{Berger:2023ccd} and illustrated in Fig.~\ref{fig:darkinduced}. A similar mechanism, considered in the context of the neutron lifetime anomaly, has been studied in Ref.~\cite{Rajendran:2020tmw}. 
The same process can destabilize the proton, with $|\Delta B|=1$ experimental studies constraining the model parameters~\cite{Berger:2023ccd}. We note that the ${\bar \chi} \phi_B \xi$ interaction can also induce $\chi\chi$ annihilation, as noted and illustrated in Fig.~\ref{fig:darkann}. The scope of possibilities can be limited through judicious choices of the parameters of the dark sector. For example, if $m_\xi + m_{\pi^0} > m_{\phi_B} +m_n$ then one would expect dark-matter-induced nucleon decay will not occur, and with $m_{\xi} > m_{\chi} > m_{\phi_B}$ Fig.~\ref{fig:darkann} depicts the only possible tree-level annihilation channel. This last effect acts to remove $\chi$ produced through neutron decay from the star, yet $\phi_B$ could potentially accumulate in its core --- and impact the survival of the neutron star~\cite{McDermott:2011jp}. If we suppose, rather, that $\phi_B$ is light enough to escape the star, then that outcome can be avoided. 
Finally, we note that $\phi_B$ and $\xi$ must each be stable to be dark-matter candidates, nor should they decay into each other~\cite{Alonso-Alvarez:2021qfd}. In our dark-sector scenario, $\xi \to \phi_B \chi$ could occur, with the subsequent decay $\chi \to {\bar p} \pi^-$ or $\chi \to {\bar n} \pi^0$ appearing if $\chi$ is too heavy, or if it has substantial coupling to either $u$ or $d$ quarks. This could act to dilute an asymmetry formed through $B$-mesogenesis. Thus successful baryogenesis within the dark sector scenario of Eq.~(\ref{eq:chidark}) can be realized if  $m_{\chi} \le 1.07784 \,\rm GeV$, though this is not the only possibility. We conclude by noting that this and nuclear stability considerations gives us the following window on $\chi$: 
\begin{equation}
0.937993 \,\rm GeV < m_{\chi} < 1.07784 \,\rm GeV \,,
\label{eq:chimasswindow}
\end{equation}
though in what follows we consider neutron-star constraints on $\chi$-baryon mixing over a  broader mass window, in anticipation of richer model-building solutions. 
We now turn to the explicit evaluation of processes that can remove $\chi$ from the neutron star. 

\subsection{Dark Baryon Removal Rates}
\label{sec:app:JMB:annihilation}

If the masses of $\xi$ and $\phi_B$ sum to less than the mass of $\chi$, then the decay $\chi \to \xi \phi_B$ is operative. Using Eq.~(\ref{eq:chidark}) and 
Refs.~\cite{Denner:1992me,Shtabovenko:2020gxv}, we calculate the width of this decay to be
\begin{equation}
    \Gamma_{\chi\to\xi\phi_B} = \frac{y_d^2}{16 \pi m_{\chi}^3} \left[ (m_{\chi} + m_\xi)^2 - m_{\phi_B}^2 \right]^{3/2} \left[(m_{\chi} - m_\xi)^2 - m_{\phi_B}^2 \right]^{1/2}.
\end{equation}
However, if this decay is operative and if $m_{\phi_B} + m_\xi < m_p - m_{\pi}$, then this allows for proton decay via $p^+ \to \pi^+ \xi \phi_B$. We avoid potentially running afoul of these constraints by insisting that this decay not be operative and thus require $m_\xi > m_{\chi}$.

Instead, we focus on possible annihilation processes of $\chi$, where we have assumed that only $\phi_B$ is lighter than $\chi$. Adopting the same tools to compute $\chi\chi \to \phi_B \phi_B$ we have:
\begin{align}
    \sigma_{\chi\chi\to\phi_B\phi_B}(s) = \frac{y_d^4 m_\xi^2}{64 \pi s} \Biggl[ & \frac{2\sqrt{(s-4m_{\chi}^2)(s-4m_{\phi_B}^2)}}{m_\xi^4 + m_\xi^2(s-2m_{\chi}^2-2m_{\phi_B}^2) + (m_{\chi}^2-m_{\phi_B}^2)^2} \\
    & + \frac{4}{s+2m_\xi^2-2m_{\chi}^2-2m_{\phi_B}^2} \nonumber \\
    & \quad \times \ln\left( \frac{s+2m_\xi^2-2m_{\chi}^2-2m_{\phi_B}^2 + \sqrt{(s-4m_{\chi}^2)(s-4m_{\phi_B}^2)}}{s+2m_\xi^2-2m_{\chi}^2-2m_{\phi_B}^2 - \sqrt{(s-4m_{\chi}^2)(s-4m_{\phi_B}^2)}} \right) \Biggr]. \nonumber
\end{align}
We note that this cross section goes to zero as $m_\xi \to 0$. This must occur, so that this outcome serves as a non-trivial check of our procedure. Our cross section result does not depends on whether the scalar is real or complex, but its interpretation does. If the scalar is real, it cannot carry baryon-number, and $\chi\chi$ annihilation to scalars would then break B by two units. This can only occur if $m_\xi$ has a nonzero baryon-number-violating mass. Thus its rate vanishes if $m_\xi$ does. 

We would like to understand how these annihilation processes operate within a neutron star. As we will see in Sec.~\ref{sec:macro_bnv:assumptions}, these cross sections would need to be averaged over the true distribution of $\chi$s produced in baryon decays within the star. Generically, $\chi$s need not be distributed thermally, and the process of thermalization would require self-interactions, which do not appear at tree level in our simple model. The problem of $\chi$ transport in the neutron star is beyond the scope of this paper, so that we assume that the thermally averaged cross section is a reasonable estimate of what the true averaged cross section would be.

We proceed by employing pertinent results from the seminal Ref.~\cite{Gondolo:1990dk}. The thermally averaged cross section $\langle \sigma v \rangle$ is given formally by
\begin{equation}
    \langle \sigma v \rangle = \frac{1}{8m_{\chi}^4 T_\chi K_2^2(m_{\chi}/T_\chi)} \int_{4m_{\chi}^2}^\infty ds \, \sigma(s) \times (s-4m_{\chi}^2) \sqrt{s} K_1(\sqrt{s}/T_\chi),
\end{equation}
where $T_\chi$ is the $\chi$ temperature (which is generically nonzero and may be different from the temperature of the rest of the neutron star) and $K_{1,2}$ are modified Bessel functions of the second kind. This expression assumes that it is appropriate to describe the $\chi$ fluid as abiding by a Maxwell-Boltzmann distribution; it would be inappropriate to apply this expression to a cold, degenerate population of $\chi$, but such a population does not occur in our framework. To perform the thermal averaging, we expand $\sigma (s) \times v$ in powers of $\epsilon \equiv s/(4m_{\chi}^2) - 1$:
\begin{equation}
    \sigma v = a^{(0)} + a^{(1)} \epsilon + \frac{1}{2} a^{(2)} \epsilon^2 + \ldots;
\end{equation}
this requires that $v = 2\sqrt{\epsilon(1+\epsilon)}/(1+2\epsilon)$. In the limit in which the $\chi$ fluid is nonrelativistic, the thermally averaged cross section can we written in terms of the coefficients $a^{(n)}$ as follows:
\begin{equation}
    \langle \sigma v \rangle = a^{(0)} + \frac{3}{2} a^{(1)} \left( \frac{T_\chi}{m_{\chi}} \right) + \frac{15}{8} a^{(2)} \left( \frac{T_\chi}{m_{\chi}} \right)^2 + \ldots
\end{equation}
This prescription is expected to be valid as long as $T_\chi \lesssim 3 m_{\chi}$ \cite{Gondolo:1990dk}. For $\chi\chi\to{\phi_B}{\phi_B}$, we find the leading-order contribution to the thermally averaged cross section in $T_\chi$ to be
\begin{equation}
    \langle \sigma v \rangle_{\chi\chi\to{\phi_B}{\phi_B}} = \frac{3}{2} \left( \frac{h^4 m_\xi^2 \sqrt{m_{\chi}^2 - m_{\phi_B}^2}}{8\pi m_{\chi} (m_\xi^2 + m_{\chi}^2 - m_{\phi_B}^2)^2} \right) \left( \frac{T_\chi}{m_{\chi}} \right) + \ldots
\end{equation}
Since the $a^{(0)}$ term vanishes, we conclude that the $s$-wave annihilation contribution vanishes, resulting in a suppression at low temperatures. We expect our $\chi$s to have a nonzero average kinetic energy from decays, so we do not expect to encounter a scenario in which these annihilations are completely quenched by the low energies of their parents, but it is an interesting feature to note. 

We conclude by noting some relevant qualitative features of this model. Since $\chi$ self-interactions do arise at the one-loop level as a result of interactions with ${\phi_B}$ and $\xi$, we can expect the $\chi$ population would thermalize, but that timescale is likely slow relative to that of their annihilation to scalars. There are many more interesting phenomenological consequences of this model that one could explore, but for our purposes, it is enough to assume that the masses and coupling conspire such that $\chi$ can be removed from neutron stars quickly enough that our formalism is valid.

\section{Macroscopic Baryon Number Violation in Neutron Stars}
\label{sec:macro_bnv}
We set out this section by elaborating the main assumptions for our analysis, followed by a description of the resulting formalism, which we flesh out in greater detail than in Ref.~\cite{Berryman:2022zic}. We then discuss the observable effects associated with our framework, along with methods of interpreting pulsar observations to yield limits on BNV in such systems. 
We use the limits derived at the end of this section to constrain specific baryon dark decay rates in Sec.~\ref{sec:dark_decay:med_limits_vacuum}, though we develop our description of dense matter, as well as of particle processes within it, in intervening sections before doing so. 
\subsection{Assumptions}
\label{sec:macro_bnv:assumptions}
The structure of a neutron star can be approximated by a static and spherically symmetric metric ($g_{\mu\nu}$) with a line element given by~\cite{tolman1987relativity}
\begin{equation}
\label{eq:macro_bnv:assum:g:def}
    d\tau^2 = g_{\mu\nu} dx^{\mu} dx^{\nu} = e^{2\nu(r)} \, dt^2 - e^{2\lambda(r)}\, dr^2 - r^2\, d\theta^2 - r^2\, \sin^2\theta\, d\phi^2,
\end{equation}
in which $\nu(r), \lambda(r)$ are solutions to the Einstein field equations~\cite{Einstein:1916vd}, $G^{\mu\nu} = -8 \pi G T^{\mu\nu}$, in which $G^{\mu\nu}$ is Einstein's tensor, $G$ is the gravitational constant, and $T^{\mu\nu}$ is the stress-energy tensor. The rotation effects on the neutron star structure, which are $\mathcal{O}\left(\Omega^2 / \left(G\, M / R^3\right)\right)$~\cite{Hartle:1967he}, amount to less than $3\%$ for the fastest spinning pulsar (J1614$-$2230) that we consider in this work. Furthermore, the inclusion of quasi-static BNV processes, which are sourced by the matter in the star, would keep the spherical symmetry intact and changes to the metric ($g_{\mu\nu}$) very slow in time, such that the use of Eq.~\eqref{eq:macro_bnv:assum:g:def} is warranted. 

We also assume that the medium in the neutron star can be described by a perfect fluid with 
\begin{equation}
    \label{eq:macro_bnv:assum:tensor:perfect}
    T_{0}^{\,0} = {\cal E}, \quad T_{i}^{\,i} = -P \qquad (i = 1,2,3),
\end{equation}
as the only nonzero components of the stress-energy tensor. We note $P$ and ${\cal E}$ are the local pressure and energy density of the fluid respectively which in general depend on the local baryon number density ($n$) and temperature ($T$) via the EoS. In the standard picture, neutron stars cool down to internal temperatures $T \lesssim 10^{11}\, {\rm K} \ll E_{\rm F} \lesssim {\rm GeV}$ within a minute after formation~\cite{Yakovlev:2004iq}, such that the thermal contribution to the pressure and energy density can be neglected. The neutron star fluid can then be described as a cold degenerate Fermi gas at $\beta$--equilibrium. The existing terrestrial constraints on neutron dark decay, Eq.~(\ref{eq:nexoticCC}), along with the BNV limits we find in Table~\ref{tab:macro_bnv:interp:psrbinary}, show that BNV rates should be slow with respect to other dynamical processes in the neutron star. We have also devised a model in which $\chi$, the dark baryon-like particle, can be removed efficiently from the star. Thus we expect the deviations from a degenerate state at $\beta$--equilibrium due to BNV should be negligibly small, and we leave a more detailed study of possible thermal effects on neutron stars from BNV to future work.

In order to be able to apply our model-independent formalism~\cite{Berryman:2022zic}, we are going to focus on a subset of models in which the dark contributions to the EoS are negligible relative to the energy density and pressure of the visible sector. In other words, we demand that the following (local) conditions 
\begin{equation}
    \frac{{\cal E}_{\chi}(r)}{{\cal E}(r)} \ll 1, \qquad {\rm and} \qquad \frac{P_{\chi}(r)}{P(r)} \ll 1, \label{eq:macro_bnv:assum:local:noacc}
\end{equation}
hold throughout the neutron star at all times, which can be equivalently written as a condition on the local number density of $\chi$: $n_{\chi}(r) \ll n(r)$. This means that $\chi$ has to decay or annihilate either back to the visible sector or to some other dark particles that can escape the neutron star. We assume that $\chi$ participates in self-annihilation to lighter dark particles that can escape the neutron star (see Sec.~\ref{sec:app:JMB:annihilation} for more details). 

We can express the condition $n_{\chi}(r)\ll n(r)$ in terms of the BNV rate, $\Gamma_{\rm BNV}$, and the annihilation cross section that is averaged over $\chi$ distribution, which we denote by $\langle \sigma v\rangle$. We note that the exact distribution of $\chi$ in the neutron star can in principle be found by solving the Boltzmann transport equation in the star, but this is not practical for our estimation purposes. We instead consider two scenarios for $\chi$: one in which the annihilation rate is much faster than the self-interactions which help establish a thermal equilibrium, and another in which self-interactions of $\chi$ are much faster than its annihilation rate, and we present them in App.~\ref{app:dark_removal}.

With the assumptions set forth in this section, we only need to specify the EoS of hadronic matter to find the neutron star structure. Once the EoS is specified, the Tolman-Oppenheimer-Volkoff (TOV)~\cite{Tolman:1939jz,Oppenheimer:1939ne} equations can be integrated with the initial conditions $M(0) = 0$ and ${\cal E} (0) = {\cal E}_c$ up to the surface of the star, corresponding to $P(r) = 0$. In other words, to study BNV effects on neutron stars generated by a fixed EoS, we focus on the unique family of stars, each parameterized by its central energy density (${\cal E}_c$), known as the \textit{single-parameter sequence}~\cite{Glendenning:1997wn} of stars.

The baryon decay rate (per baryon) in a small volume ($V$) in the nuclear matter (n.m.) rest frame ($\Gamma_{\rm nm}$) is defined by $d(n V)/d\tau = - \Gamma_{\rm nm} \, n \, V$, in which $\tau$ is the fluid's proper time, and $n$ is the proper baryon number density. We can define a baryon number-flux vector by $j^{\mu} = u^{\mu} n$~\cite{Misner:1973prb}, in which $u^{\mu}$ is the four-velocity of the fluid ($u^{\mu}u_{\mu} = 1$) and use the definition of $\Gamma_{\rm nm}$ to write $j^{\mu}_{;\mu} = - n\, \Gamma_{\rm nm}$, in which `;' denotes the covariant derivative. We then use the relationship $\sqrt{-g}\, j^{\mu}_{;\mu} = \left(\sqrt{-g}\, j^{\mu}\right)_{,\mu}$~\cite{Glendenning:1997wn}, in which `,' denotes ordinary partial derivative and $g \equiv {\rm det}|g_{\mu\nu}|$, to arrive at
\begin{equation}
    \frac{\partial}{\partial t} \int \sqrt{-g} \, n\, u^0\, d^3x + \int \left(\sqrt{-g} \, n\, u^i\right)_{, i}\, d^3x = -\int d^3x \sqrt{-g} \, n \Gamma_{\rm nm}, \label{eq:macro_bnv:assum:B:cons:general}
\end{equation}
in which $i=1,2,3$. For a static ($u^i=0$), spherically symmetric neutron star with radius $R$, we can use the metric in Eq.~\eqref{eq:macro_bnv:assum:g:def} to simplify Eq.~\eqref{eq:macro_bnv:assum:B:cons:general} as
\begin{equation}
    \frac{\dot B}{4\pi} \equiv \frac{\partial}{\partial t}\left[\int_0^R \left[1 - \frac{2M(r)}{r}\right]^{-\frac{1}{2}}\, n(r)\, r^2\, dr \right] = -\int e^{\nu(r)}\, \left[1 - \frac{2M(r)}{r}\right]^{-\frac{1}{2}}\, \Gamma_{\rm nm}(r)\, n(r)\, r^2\, dr, \label{eq:macro_bnv:assum:B:cons:static}
\end{equation}
where $B$ is the total baryon number of the neutron star. We have used $\sqrt{-g} = \exp(\nu(r)+\lambda(r))\, r^2\, \sin\theta$, with $\exp(2\lambda(r)) = (1 - 2M(r)/r)^{-1}$, and $M(r)$ is the total mass included within radius $r$:
\begin{equation}
\label{eq:macro_bnv:assum:m:def}
M(r') = 4\pi \int_0^{r'} {\cal E}(r) r^2 dr.
\end{equation}
Given a particle physics model for BNV we can evaluate $\Gamma_{\rm nm}(r)$ and use Eq.~\eqref{eq:macro_bnv:assum:B:cons:static} to find the resulting $\dot{B}$.
 
 \subsection{Framework}
 \label{sec:macro_bnv:framework}
\begin{figure}[!t]
    \centering
    \subfigure[]{\includegraphics[width=0.495\linewidth]{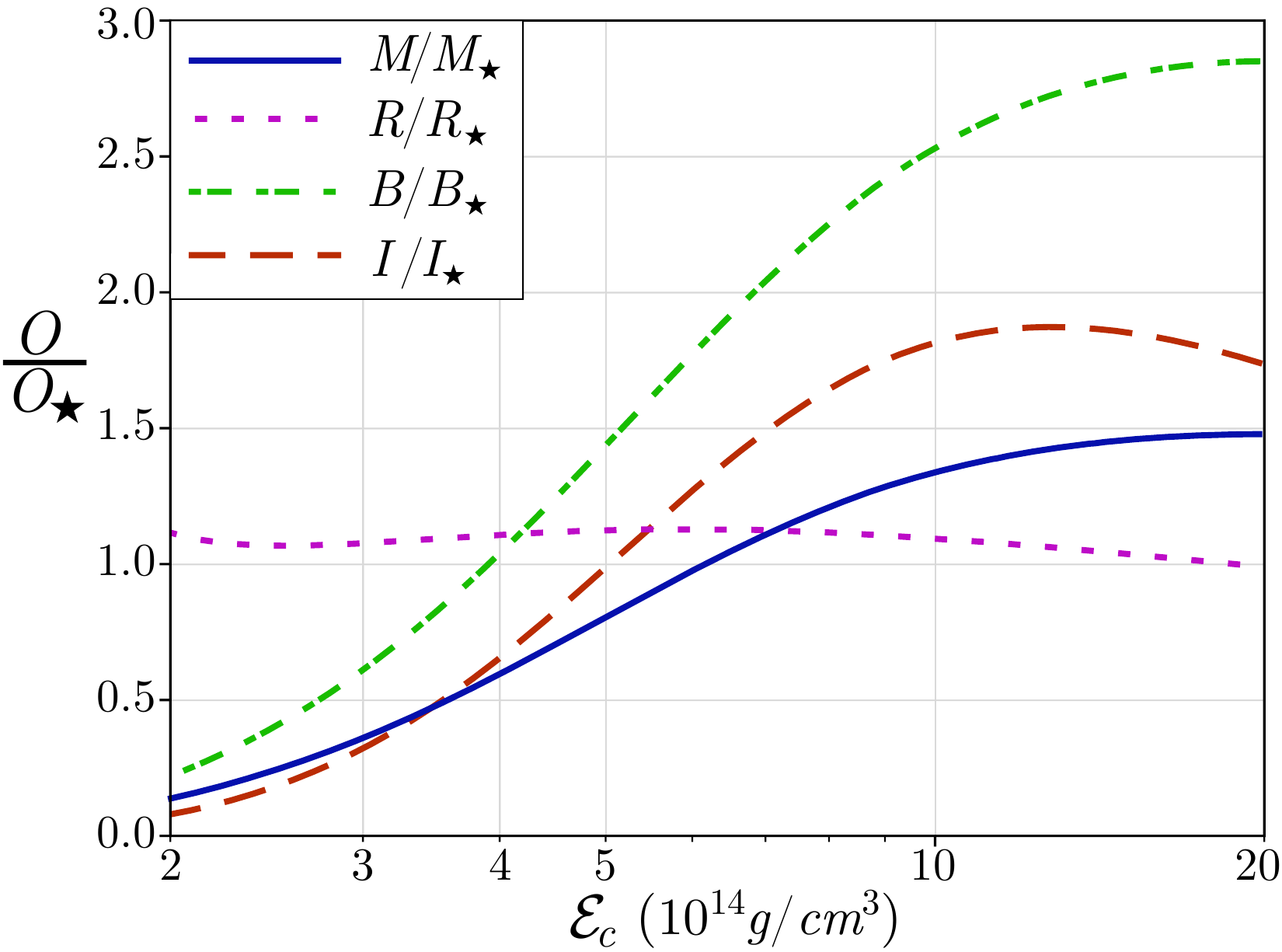}}
    \subfigure[]{\includegraphics[width=0.495\linewidth]{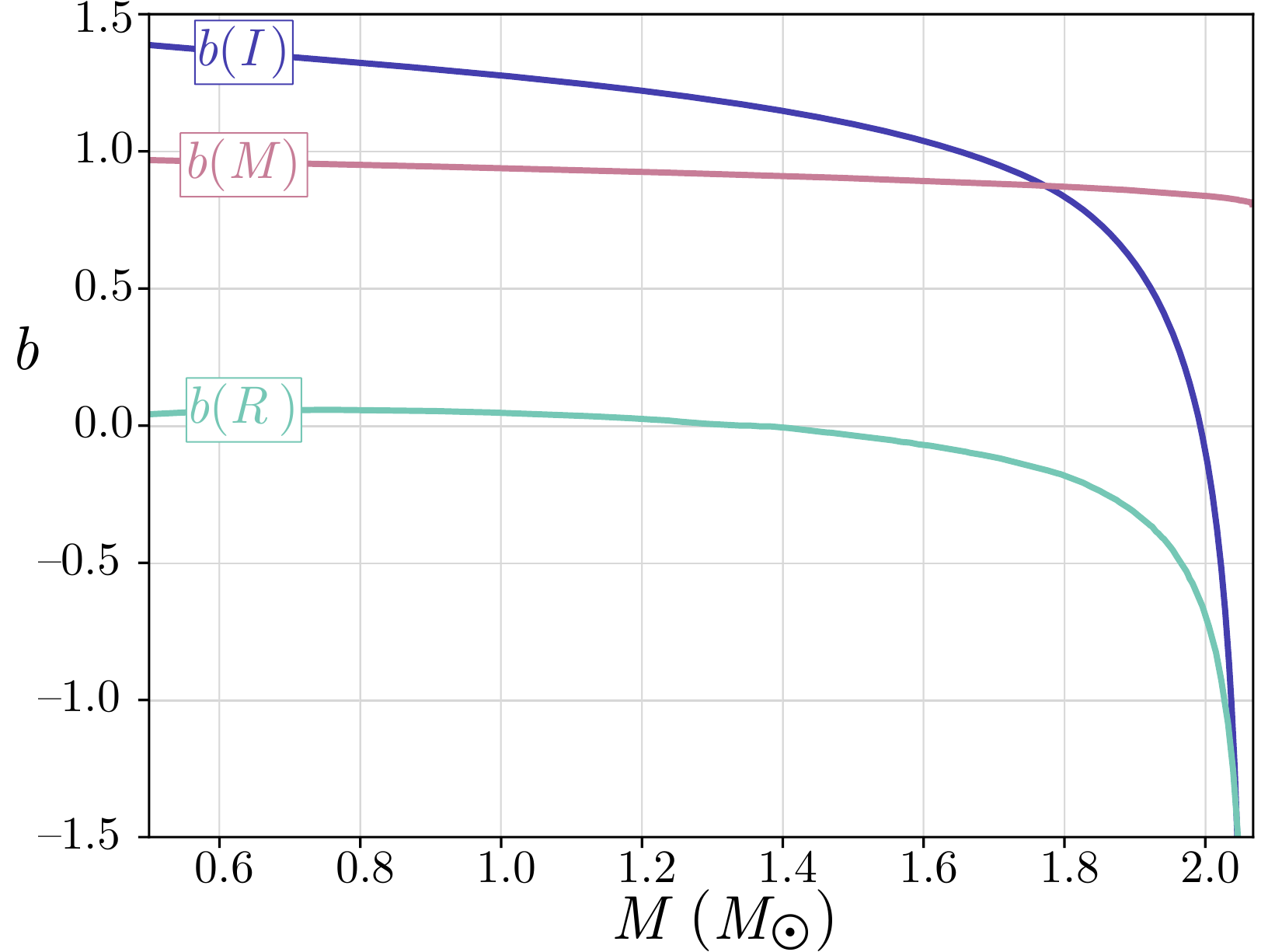}}
    \caption{(Color online) (a) The set of observable quantities ($O$): mass $(M)$, radius $(R)$, baryon number $(B)$, and moment of inertia $(I)$ for a sequence of neutron stars as a function of the central energy density (${\cal E}_c$) relative to their canonical values ($O_*$): $M_{\star} = 1.4\, M_{\astrosun}$, $R_{\star} = 12\, {\rm km}$, $B_{\star} = 10^{57}$, $I_{\star} = 70\, ({\rm M_{\astrosun}\, {km}^2})$ assuming the DS(CMF)-1 EoS. (b) The $b(\mathcal{O})$ factors for three parameters ($\mathcal{O} = M, R, I$) as a function of neutron star masses for the DS(CMF)-1 EoS. See Sec.~\ref{sec:medium:MFT} for more details on our choice of EoS.  }
    \label{fig:macro_bnv:framework:b_factors}
\end{figure}

It was shown in Ref.~\cite{Berryman:2022zic} that the conditions in Sec.~\ref{sec:macro_bnv:assumptions} are necessary for a model-independent analysis of BNV effects on neutron star. These conditions can be summarised as: (1) BNV is slower than chemical and dynamical responses in the neutron star, and (2) the contributions to the EoS from any new particles (e.g., $\chi$) are negligible (see Eq.~\eqref{eq:macro_bnv:assum:local:noacc}). The overall effect of BNV within this framework is to relocate the neutron star along its single-parameter sequence prescribed by the chosen baryon-number-conserving EoS. The rate of change in any neutron star observable $\mathcal{O}$ as a result of this quasi-equilibrium evolution can be written as
\begin{equation}
    \dot{\mathcal{O}} \equiv \left(\frac{d {\cal E}_c}{dt}\right) \left(\frac{\partial \mathcal{O}}{\partial {\cal E}_c} \right).\label{eq:macro_bnv:framework:Odot:general}
\end{equation}
Here, we ignore any possible dependence of $\mathcal{O}$ on the angular velocity $\Omega$, i.e., we assume $\mathcal{O}$ evolves along a one-dimensional trajectory with $\Omega=0$ on the general two-dimensional space parameterized by ${\cal E}_c$ and $\Omega$. We can solve for $\dot{\cal E}_c$ in terms of the rate of baryon loss, $\dot{B}$, such that
\begin{equation}
    \frac{\dot{\mathcal{O}}}{\mathcal{O}} = \left(\frac{B}{\mathcal{O}} \times \frac{\partial_{{\cal E}_c} \mathcal{O}}{\partial_{{\cal E}_c} B} \right) \frac{\dot{B}}{B} \equiv - b (\mathcal{O}) \times \Gamma_{\rm BNV}, \label{eq:macro_bnv:framework:Odot:general:2}
\end{equation}
in which we defined the effective BNV rate $\Gamma_{\rm BNV} \equiv -\dot{B}/B$ and the dimensionless parameter $b(\mathcal{O})$ encodes the relative rate of change in $\mathcal{O}$ with respect to $\Gamma_{\rm BNV}$. We pick hadronic versions of the DS(CMF) EoS~\cite{Dexheimer:2008ax} that includes a crust~\cite{Gulminelli:2015csa} from the CompOSE database ~\cite{CompOSECoreTeam:2022ddl}. The details of these EoS including their Lagrangians and particle contents are given in Sec.~\ref{sec:medium:MFT}. In order to evaluate the $b(\mathcal{O})$ factors, we 
generate a sequence of neutron stars on a grid of ${\cal E}_c$ values, and then find the derivative of $\mathcal{O}$ using the central finite difference method. The resulting $b (\mathcal{O})$ is plotted in Fig.~\ref{fig:macro_bnv:framework:b_factors} for various observable quantities as a function of neutron star masses for the DS(CMF)-1 EoS.
In Sec.~\ref{sec:macro_bnv:observables}, we use this formalism to show how changes in neutron star parameters due to BNV would affect pulsar-binary orbital decay rates.

 \subsection{Observables}
\label{sec:macro_bnv:observables}
Baryon loss in pulsars may lead to observable effects on their individual spin-down rate ($\dot{P}_s$), and their orbital period lengthening ($\dot{P}_b$) if they belong to a binary system~\cite{Berryman:2022zic, Goldman:2009th, Goldman:2019dbq,Berezhiani:2018udo}. The BNV modifications to $\dot{P}_s$ are caused by the quasi-equilibrium changes in the moment of inertia ($I$), and angular momentum loss due to light particles (e.g., $\phi_B$) escaping the pulsar. While the first contribution can be expressed in a model-independent manner, the latter depends on the specific BNV model and the masses of particles involved. Therefore, we focus our attention on BNV modifications to $\dot{P}_b$, which can still be formulated in a model independent way.

The energy loss due to BNV can modify the orbital period decay rate in a binary system, assuming it is active in one or both of the components. This energy loss can be written as~\cite{Berryman:2022zic}
\begin{equation}
\label{eq:macro_bnv:observables:mdot:eff}
    \dot{M}^{\rm eff} \equiv \frac{d}{dt} \left( M + \frac{1}{2} I \Omega^2 \right) = \underbrace{\, b(M)\left(\frac{\dot{B}}{B}\right) M + b(I)\left(\frac{\dot{B}}{B}\right) \left(\frac{2 \pi^2 I}{P_s^2}\right) \, }_\textrm{BNV} - \frac{4\pi^2 I \dot{P_s}}{P_s^3},
\end{equation}
in which $b(M)$ and $b(I)$ are defined in Eq.~\eqref{eq:macro_bnv:framework:Odot:general:2}, $P_s$ and $\dot{P_s}$ are the observed pulsar spin period and its observed rate of change respectively. Note that the rates of change in $I$ due to spin-down, $(dI/d\Omega) \dot{\Omega}$, are negligible in the pulsars that we consider. The relative rate of change in a binary period due to energy loss in its components is given by~\cite{1963ApJ...138..471H,10.1093/mnras/85.1.2, 10.1093/mnras/85.9.912}
 \begin{equation}
 \label{eq:macro_bnv:observables:jeans:mloss}
    \left(\frac{\dot{P}_b}{P_b}\right)^{\dot{E}} = - 2 \left(\frac{\dot{M}_1^{\rm eff} + \dot{M}_2^{\rm eff}}{M_1 + M_2}\right),
 \end{equation}
 in which $1$ and $2$ refer to the components of the binary system. After plugging Eq.~\eqref{eq:macro_bnv:observables:mdot:eff} into \eqref{eq:macro_bnv:observables:jeans:mloss}, we get the following BNV and spin-down contributions to the energy-loss term
\begin{align}
  \left(\frac{\dot{P}_b}{P_b}\right)^{\rm BNV}  =& \frac{- 2}{M_1 + M_2}\sum_{i=1,2}  \left(\frac{\dot{B}_i}{B_i}\right) \left[ b_i(M) M_i + b_i(I) \left(\frac{2 \pi^2 I_i}{P_{s,i}^2}\right) \right],\label{eq:macro_bnv:observables:pbdot:bnv}\\
  \left(\frac{\dot{P}_b}{P_b}\right)^{\dot{\Omega}}  =& \frac{8\pi^2}{M_1 + M_2} \left(\frac{ I_1 \dot{P}_{s,1}}{P_{s,1}^3} + \frac{ I_2 \dot{P}_{s,2}}{P_{s,2}^3}\right).\label{eq:macro_bnv:observables:pbdot:spindown}
\end{align}
We should note that the second term in Eq.~\eqref{eq:macro_bnv:observables:pbdot:bnv}, which is due to changes in the moment of inertia, is $\mathcal{O}(10^{-3}) \times b(I)\, M_{\astrosun}$ for J1614$-$2230 and even smaller for the other two systems considered in this work. Given that $b(M) \approx 1$ and $|b(I)| \sim \mathcal{O}(1)$, and the spin-down contributions from Eq.~\eqref{eq:macro_bnv:observables:pbdot:spindown} are usually subdominant, we conclude that our limits would be mainly controlled by the first term in Eq.~\eqref{eq:macro_bnv:observables:pbdot:bnv}. For this reason, our inferred limits on $\dot{B}/B$ are not sensitive to the specific choices of EoS. We can use the observed pulsar binary period decay rate to limit the contributions from Eq.~\eqref{eq:macro_bnv:observables:pbdot:bnv}, but first we need to identify other sources of binary orbital decay.

\subsection{Interpretation}
\label{sec:macro_bnv:interp}
The dominant contributions to the observed relative rate of orbital period decay can be written as~\cite{1991ApJ...366..501D}:
\begin{equation}
    \left(\frac{\dot{P}_b}{P_b}\right)^{\rm obs}  =  \underbrace{\, \left(\frac{\dot{P}_b}{P_b}\right)^{\rm GR} + \left(\frac{\dot{P}_b}{P_b}\right)^{\dot{E}}\, }_\textrm{intrinsic} +
    \left(\frac{\dot{P}_b}{P_b}\right)^{\rm ext}, \label{eq:Pbdot}
\end{equation}
in which the first term is due to gravitational radiation~\cite{PhysRev.136.B1224}, and the third term includes extrinsic effects, e.g., due to the relative motion of a binary pulsar with respect to the solar system barycenter. The numerical values for each of these contributions and the limits on $\dot{P}_b^{\dot{E}}$, which is found by subtracting the general relativity (GR) contribution, $\dot{P}_b^{\rm GR}$, from the intrinsic orbital-period decay rate, $\dot{P}_b^{\rm int} \equiv \dot{P}_b^{\rm obs} - \dot{P}_b^{\rm ext}$, are given in Table~\ref{tab:macro_bnv:interp:psrbinary} for three binary systems. 

Tens of binary pulsar systems with measured $P_b$ and ${\dot P}_b^{\rm obs}$ are known~\cite{Manchester:2004bp}, and from those we choose ones for which there is no mass transfer between the components, because that would modify $\dot{P}_b^{\rm obs}$~\cite{1986Natur.319..383R} in a complicated way. We also choose systems for which the individual masses are known. 
With those in place, our subsequent choices 
are motivated by wanting 
better sampling of ($m_{\chi}, 
\varepsilon_{\cal B})$ space 
(see Fig.~\ref{fig:combining_limits}). The precision of the orbital parameters in the binary system under consideration 
determines the minimum value of the excluded mixing parameter. Therefore, we include the double pulsar J0737$-$3039A/B in our analysis due to its high precision. (Although the Hulse-Taylor pulsar system~\cite{Hulse:1974eb} has also been precisely measured~\cite{Weisberg:2016jye}, it gives redundant information in this case.)
Furthermore, binaries with a heavier pulsar would yield a better kinematical (horizontal) reach, i.e., the maximum value of excluded $\chi$'s mass for which the decays are kinematically open. Selecting binary systems with heavy pulsars is also crucial for setting limits on the $\Lambda$ hyperon dark decays. This is because in EoSs that contain hyperon degrees of freedom, hyperons start populating the medium only at very high densities~\cite{1960SvA.....4..187A}, i.e., in the cores of heavy neutron stars (see Fig.~\ref{fig:dark_decay:medium:rest-e:J0348}). Among the heaviest known pulsar candidates we exclude PSR J0952--0607 since it is a ``black widow" pulsar with relatively large errors in its mass $M_p = \left(2.35 \pm 0.17\right)\, M_{\astrosun}$~\cite{Romani:2022jhd}, and a notable ($\sim 11\%$) rotation effect on the static structure of a neutron star given by the metric in Eq.~\eqref{eq:macro_bnv:assum:g:def}. The next candidates are PSR J0740+6620 with $M_p = \left(2.08 \pm 0.07\right)\, M_{\astrosun}$~\cite{Fonseca:2021wxt}, PSR J0348$+$0432 with $M_p = \left(2.01 \pm 0.04\right)\, M_{\astrosun}$~\cite{Antoniadis:2013pzd} and PSR J1614--2230 with $M_p = \left(1.908 \pm 0.016\right)\, M_{\astrosun}$~\cite{Arzoumanian:2017puf}. For our study we pick the latter two pulsars which have smaller errors in $M_p$ and $\dot{P}_b^{\rm obs}$. 

\begin{enumerate}
    \item \textbf{PSR J0348$+$0432:} A pulsar–white dwarf binary discovered in 2007 with the Robert C. Byrd Green Bank Telescope~\cite{lynch2013green} with an orbital period of about $2.4$ hr.  We use the results from the analysis in Ref.~\cite{Antoniadis:2013pzd}, in which it was shown that the kinematic, spin-down (Eq.~\eqref{eq:macro_bnv:observables:pbdot:spindown}), and tidal ($\dot{P}_b^T \lessapprox 10^{-16}$) contributions to $\dot{P}_b$ are negligible and the observed $\dot{P}_b$ should be mainly caused by the gravitational wave (GW) emission. We use the value from Ref.~\cite{Antoniadis:2013pzd} for the intrinsic period decay rate, $\dot{P}_b^{\rm int} = -0.275(45) \times 10^{-12}$. 
    
    \item \textbf{PSR J1614$-$2230:} A pulsar–white dwarf binary discovered in 2006 with the Parkes radio telescope~\cite{Hessels:2004ps}. We use the Shapiro delay mass estimates from Ref.~\cite{Arzoumanian:2017puf}, and the binary parameters from the NANOGrav 12.5 yr data set~\cite{NANOGrav:2020gpb} at $56323$ MJD. The observed value of $\dot{P}_b^{\rm obs} = 1.57(13) \times 10^{-12}$ is dominated by the Doppler shift due to the pulsar motion which is itself mainly caused by the Shklovskii effect~\cite{1970SvA....13..562S}:
    \begin{equation}
        \dot{P}_b^{\rm Shk} = \frac{\mu^2\,d}{c} P_b = 1.24(9) \times 10^{-12},
    \end{equation}
    in which we input the value for proper motion $\mu = 32.4(5)\, {\rm mas\, yr^{-1}}$, and used the parallax distance $d = 0.65 \pm 0.04\, {\rm kpc}$~\cite{NANOGrav:2020qll}. We use Eq.~(16) from Ref.~\cite{10.1111/j.1365-2966.2009.15481.x} to estimate the contribution due to the Galactic potential, namely, 
    \begin{equation}
        \left( \frac{\dot{P}_b}{P_b}\right)^{\rm Gal} = - \frac{K_z |\sin(b)|}{c} - \frac{\Omega_{\astrosun}^2 R_{\astrosun}}{c} \left( \cos(l) + \frac{\beta}{\beta^2 + \sin^2 (l)}\right) \cos(b),
    \end{equation}
    in which $\beta \equiv (d / R_{\astrosun}) \cos(b) - \cos(l)$, $R_{\astrosun} = 8.0(4)$ is the Sun’s Galactocentric distance, $\Omega_{\astrosun} = 27.2(9)\, {\rm km}\, {\rm s}^{-1}\, {\rm kpc}^{-1}$ is its Galactic angular velocity, $K_z$ is the vertical component of Galactic acceleration approximated by
    \begin{equation}
        K_z \left(10^{-9}\, {\rm cm}\, {\rm s}^{-2}\right) \approx 2.27 z_{\rm kpc} + 3.68 (1 - \exp(-4.31\,z_{\rm kpc}))
    \end{equation}
    for Galactic heights $z \equiv |d\, \sin(b)| \leq 1.5\, {\rm kpc}$. We use the pulsar's coordinates $(l, b) = (352.64\degree, 20.19\degree)$ to find $z = 0.223(14)\, {\rm kpc}$, and $\dot{P}_b^{\rm Gal} = 1(5) \times 10^{-14}$. These extrinsic effects  combine to yield $\dot{P}_b^{\rm ext} = \dot{P}_b^{\rm Gal} + \dot{P}_b^{\rm Shk} = 1.25(10) \times 10^{-12}$. Our resulting estimate for the period derivative, $\dot{P}_{b}^{\rm int} = 0.32(16)\times 10^{-12}$, is positive at $2\sigma$ significance, pointing to a possible underestimation of extrinsic effects and their errors. However, we note that if, for example, we instead assume a negligible value for $\dot{P}_{b}^{\rm int}\approx 0$ and double our error estimates, then we would still obtain the same limits. We also evaluate the relatively small GW contribution which for circular orbits is given by~\cite{PhysRev.136.B1224}
    \begin{equation}
    \begin{split}
        \dot{P}_b^{\rm GW} &= -\frac{192 \pi}{5} \left(\frac{2 \pi T_{\astrosun}}{P_b}\right)^{5/3}\frac{M_p M_c}{(M_p + M_c)^{(1/3)}} = -4.17(4) \times 10^{-16},
    \end{split}
    \end{equation}
    in which we used the pulsar and white dwarf masses from Ref.~\cite{Arzoumanian:2017puf}, $T_{\astrosun}=  4.92549094\times 10^{-6}\, s$, and we neglected the small eccentricity of the orbit $e=1.333(8)\times 10^{-6}$~\cite{Fonseca:2016tux}. In estimating $\dot{P}_b^{\dot{\Omega}}$ using Eq.~\eqref{eq:macro_bnv:observables:pbdot:spindown} we assumed the canonical value $I = 10^{45}\, {\rm g\, cm^2}$ for the pulsar's moment of inertia.
    
    \item \textbf{PSR J0737$-$3039A/B:} A double pulsar discovered in 2003~\cite{Burgay:2003jj}, 
    comprised of two radio pulsars (A and B) with pulse periods of $22.7$ ms and $2.8$ ms, respectively. We use the data from Ref.~\cite{PhysRevX.11.041050} and the inferred limits on BNV contributions from Ref.~\cite{Berryman:2022zic}. 
\end{enumerate} 
\begin{table}[t!]
    \centering
    \begin{tabular}{|c|c|c|c|}
    \hline 
    Name & J0348$+$0432  & J1614$-$2230 & J0737$-$3039A/B\\
        \hline\hline                                          
        $M_p \, (M_{\astrosun})$ & $2.01(4)$ & $1.908(16)$ & $1.338\, 185(+12, -14)$ [A] \\\hline
        $M_c \, (M_{\astrosun})$ & $0.172(3)$ & $0.493(3)$ & $1.248\, 868(+13, -11)$ [B] \\\hline
        $P_s\, ({\rm ms})$ & $39.122\, 656\, 901\, 780\, 6(5)$ & $3.150\, 807\, 655\, 690\, 7$ & $22.699\, 378\, 986\, 472\, 78(9)$ [A] \\\hline
        $\dot{P}_s^{\rm obs} \, (10^{-18})$ & $0.240\, 73(4)$ & $9.624 \times 10^{-3}$ & $1.760\, 034\, 9(6)$ [A]\\\hline
        $P_b \, ({\rm days})$ & $0.102\, 424\, 062\, 722(7)$ & 
        $8.686\, 619\, 422\, 56(5)$
        & $0.102\, 251\, 559\, 297\, 3(10)$\\\hline
        $\dot{P}_b^{\rm obs} \, (10^{-12})$ & $-0.273(45)$ & 
        $1.57(13)$
        & $-1.247\, 920(78)$ \\\hline
        $\dot{P}_b^{\rm ext} \, (10^{-12})$ & $1.6(3) \times 10^{-3}$ & 
        $1.25(10)$  
        & $-1.68(+11, -10) \times 10^{-4}$\\\hline
        $\dot{P}_b^{\rm int} \, (10^{-12})$ & $-0.275(45)$ & 
        $0.32(16)$ 
        & $-1.247\, 752(79)$\\\hline
        $\dot{P}_b^{\rm GR} \, (10^{-12})$ & $-0.258(+8,-11)$ & $-4.17(4) \times 10^{-4}$ & $-1.247\, 827(+6,-7)$\\\hline\hline
         $(\frac{\dot{P}_b}{P_b})^{\dot{E}}_{2\sigma} \, ({\rm yr^{-1}})$ & $2.7\times 10^{-10}$  &
         $2.7 \times 10^{-11}$ 
         & $8.3\times 10^{-13}$ \\\hline
         $(\frac{\dot{P}_b}{P_b})^{\dot{\Omega}}\, ({\rm yr^{-1}})$ & $<1.4\times 10^{-13}$ & $\approx 4.2 \times 10^{-15}$ & $1.04(7)\times 10^{-13}$\\\hline
        $(\frac{\dot{P}_b}{P_b})^{\rm BNV}_{2\sigma} \, ({\rm yr^{-1}})$ & $2.7\times 10^{-10}$ & 
        $2.7 \times 10^{-11}$ 
        & $7.3\times 10^{-13}$\\\hline\hline
        $|\frac{\dot{B}}{B}|^{\rm BNV}_{2\sigma} \, ({\rm yr^{-1}})$ & $1.8\times 10^{-10}$ & 
        $2.0 \times 10^{-11}$ 
        & $4.0\times 10^{-13}$\\\hline
    \end{tabular}
    \caption{The relevant binary parameters for J0348$+$0432~\cite{Antoniadis:2013pzd}, J1614$-$2230~\cite{Arzoumanian:2017puf, NANOGrav:2020gpb}, and J0737$-$3039A/B~\cite{PhysRevX.11.041050}. See the discussion in Sec.~\ref{sec:macro_bnv:interp} for more details.}
    \label{tab:macro_bnv:interp:psrbinary}
\end{table}

We can now translate the bounds on $(\dot{P}_b/P_b)^{\rm BNV}$ from Table~\ref{tab:macro_bnv:interp:psrbinary} to limits on $(\dot{B}/B)$ using Eq.~\eqref{eq:macro_bnv:observables:pbdot:bnv}, which are presented in the last row of Table~\ref{tab:macro_bnv:interp:psrbinary}. In deriving these limits, we assumed that BNV is only active in the pulsars. We also note that we can only infer a model-independent limit on a linear combination of BNV in pulsars A and B of the double pulsar system (J0737$-$3039A/B). However, we expect that the rates of BNV (per baryon) would be about the same in both pulsars, i.e., $(\dot{B}_1/B_1) \approx (\dot{B}_2/B_2)$, since their masses are very close and the composition of light neutron stars ought not change much over $0.1\, M_{\astrosun}$.  In Sec.~\ref{sec:dark_decay:med_limits_vacuum}, in which we adopt a specific BNV model ($\cal B \to \chi \gamma$), our inferred limits on the mixing parameter ($\varepsilon_{{\cal B} \chi}$) are found by evaluating the individual BNV rates in each of the two pulsars J0737$-$3039A and J0737$-$3039B, which 
we then sum to compare to the observational limit on BNV in this system. We also observe that changing between the DS(CMF) EoSs (see Table~\ref{tab:medium:MFT:CMF:choices}) induces variation in, at most, the last significant digit in our limits (see the discussion below Eq.~\eqref{eq:macro_bnv:observables:pbdot:spindown}). 

\section{Dense Matter Considerations for Particle Processes} 
\label{sec:medium}
Different lines of evidence reveal that dense matter environments can be discriminating probes of non-SM processes. For example, limits on $\Lambda \to \chi \gamma$, as well as other decay channels with dark particles, follow by noting that the duration of the observed neutrino pulse in SN 1987A should not be significantly impacted by dark sector emission~\cite{Alonso-Alvarez:2021oaj}. We, too, have found severe limits on BNV from binary pulsar period lengthening, as shown in Table \ref{tab:macro_bnv:interp:psrbinary}. In this context we note the limit of Ref.~\cite{Goldman:2009th} as a constraint on dark-sector emission.  Here we sharpen BNV studies 
by computing particle processes within a theoretical framework suitable to the description of the dense matter in the interior of a neutron star.

To compute particle processes in dense matter we might first turn to chiral effective theory to describe the low-energy interactions of such hadrons~\cite{Bedaque:2002mn,Epelbaum:2008ga}. At the simplest level, these studies exploit the symmetries of QCD to systematize the interactions of mesons and baryons in a momentum expansion in powers of $(Q/\Lambda_{\chi})$, in which $Q$ is the momentum or pion mass and $\Lambda_{\chi}$ is the chiral-symmetry breaking scale ($\Lambda_{\chi} \approx 1\, {\rm GeV}$), with experiments fixing the value of the unknown low-energy constants (LECs) that appear. This framework can also be extended to the determination of the EoS of neutron stars~\cite{Hebeler:2013nza,Drischler:2021kxf}. The empirical nature of the LEC determinations limit the applicability of chiral effective theory to densities no more than $2 n_{\rm sat}$~\cite{Huth:2021bsp}. Moreover, in neutron stars, the central densities can easily exceed that of saturation density by a factor of a few, making the nucleons {\it relativistic}. As a result, we turn to relativistic mean-field (RMF) theory in hadronic degrees of freedom to describe the dense matter at the core of a neutron star. In what follows we first describe how a RMF treatment emerges from a simple, covariant quantum field theory description of hadronic interactions before describing the specific, state-of-the-art chiral mean-field (CMF) EoS that we employ for generating our numerical results, showing how this specific choice maps onto the RMF treatment of the simpler model. We then show how particle decays can be computed within that framework. 
\subsection{Modelling Dense Matter}
\label{sec:medium:MFT}
\begin{figure}[!t]
    \centering
    \includegraphics[width=0.81\linewidth]{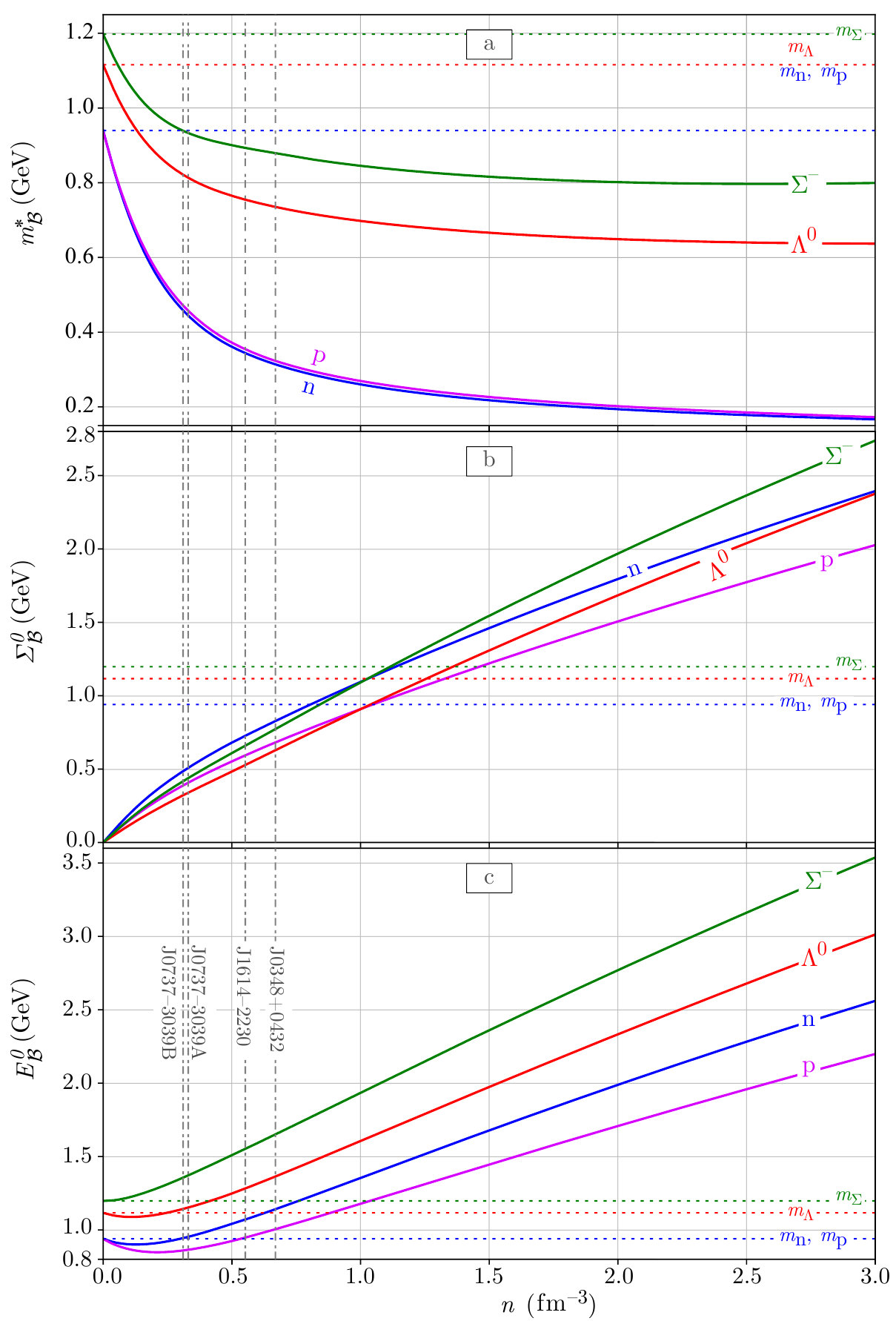}
    \caption{(Color Online) The effective masses (a), vector self-energies (b), and the energy of baryons at rest in the n.m.~frame (c) as a function of density in the DS (CMF)-1 EoS. The horizontal lines correspond to the vacuum masses of baryons, and the vertical lines indicate the central number density ($n_c$) of the pulsars we consider 
    in this work.}
    \label{fig:medium:MFT:meff_sigma}
\end{figure}

A prototypical choice is the Walecka model~\cite{Walecka:1974qa,Serot:1984ey,Serot:1997xg}, namely, 
\begin{eqnarray}
{\cal L}_{\varphi/V} &=&{\bar \psi} [(i \slashed{\partial} - g_V \slashed{V}) - (m_N - g_s \varphi) ] \psi + \frac{1}{2} (\partial_\mu \varphi \partial^\mu \varphi - m_s^2 \varphi^2) \nonumber \\
&& -\frac{1}{4} F^{\mu\nu} F_{\mu \nu} + \frac{1}{2} m_v V_\mu V^\mu + \delta {\cal L} \,, 
\end{eqnarray}
where $F_{\mu\nu} = \partial_\mu V_\nu - \partial_\nu V_\mu$ and $\delta {\cal L}$ is a counterterm, as the model is renormalizable. It is similar to massive QED with a scalar extension and a conserved current (baryon number). Both a neutral scalar meson ($\varphi$) and a neutral vector meson ($V^{\mu}$), describing the attractive and repulsive features, respectively, of the nucleon-nucleon force appear. The equations of motions (EoMs) take the form 
\begin{eqnarray}
&& \left(\slashed{\partial}^2  + m_s^2 \right) \varphi(x) = 
 g_s {\bar \psi } \psi \label{eq:medium:MFT:Sscr}
 \\
 && \partial_\nu F^{\nu \mu} + m_v^2 V^\mu = g_v 
 {\bar\psi} \gamma^\mu \psi
\label{eq:medium:MFT:Vscr}
 \\
&&    \left\{ \left[i\slashed{\partial} - g_v \slashed{V}(x)\right] - \left[m_N - g_s \varphi(x)\right]\right\} \psi(x) = 0. \label{eq:medium:MFT:dirac:general}
\end{eqnarray}
The EoMs are nonlinear and thus complicated. Working in the mean field limit is grossly simplifying, however. 
That is, 
at high baryon number densities, the sources for $\varphi(x)$ and $V^{\mu}(x)$ fields become large, and these field operators can be replaced by their vacuum expectation values (VEV) in the n.m.~frame: $\varphi(x) \to \langle \varphi(x) \rangle \equiv \overline{\varphi}
$, and $V_{\mu}(x) \to \langle V_{\mu}(x) \rangle \equiv \delta_{\mu 0} \overline{V}_0$. In doing this, we assume rotational invariance and note that in static uniform matter, as in a neutron star, $\overline{\varphi}$ and $\overline{V}_0$ become constants that only depend on density. The solutions to Eq.~\eqref{eq:medium:MFT:dirac:general} would then take the form of that of the free Dirac equation if suitable replacements of the baryon (canonical) four-momentum $k^\mu$ (in a uniform medium) and mass $m$ are made.
Thus the medium effects in the RMF limit are captured by a shift in the baryon momenta and masses, namely, by mapping 
$k_{\mu} \to k^{*}_{\mu} \equiv k_{\mu} - g_v V_\mu$ with 
$V_\mu \to \delta_{\mu 0} \overline{V}_0$ 
and $m \to m^* \equiv m - g_s {\overline \varphi}$. 
In generalizing this result for broader use, we note that the Lagrangian of interactions for a more realistic hadronic model would have more ingredients (e.g., mesons). However, we would still be able to add up the scalar meson VEVs that modify the baryon's mass in a similar manner and denote the effective baryon mass by $m^*$, independent of the specific scalar mesons in our model. Similarly, we can combine all the contributions to the baryon's momentum from vector mesons and denote them by $\Sigma^{\mu}$, such that in going from the vacuum to the in-medium formalism we would replace $k_{\mu} \to 
k^{*}_{\mu} \equiv k_{\mu} - \Sigma_{\mu}$ in the mean-field limit. 
Equipped with this result, we can write the wave-function for a baryon 
in a uniform medium as 
\begin{equation}
    \psi (x) = e^{-ik\cdot x} u(k^*, \lambda),
    \label{eq:medium:MFT:modspinor}
\end{equation}
in which $k^{*\mu} \equiv k^{\mu} - \Sigma^{\mu} = \left\{E^*(k^*), \vec{k} - \vec{\Sigma}\right\}$ is defined to be the 
{\it kinetic} four-momentum and the vector self-energy ($\Sigma^{\mu}$) is generated by the vector meson VEVs, with $\vec{\Sigma} = 0$ in the n.m.~frame. The time-component of $k^{*\mu}$ is defined by $E^*(k^*) \equiv \sqrt{{m^*}^2 + |\vec{k}^*|^2}$, in which $m^*$ is generated by the scalar meson VEVs. The baryon spinor $u(k^*, \lambda)$ satisfies the Dirac equation
\begin{equation}
    \left(\slashed{k}^* - m^*\right) u(k^*, \lambda) = 0, \label{eq:medium:MFT:dirac}
\end{equation}
which has the following solution in Dirac-Pauli representation
\begin{equation}
    u(k^*, \lambda) = \sqrt{E^*(k^*) + m^*}  
    \begin{pmatrix}
    1 \\
    \frac{\vec{\sigma}\cdot \vec{k}^{\,*}}{E^*(k^*) + m^*}
   \end{pmatrix} 
\chi_{\lambda},\label{eq:medium:MFT:spinor}
\end{equation}
in which $\vec{\sigma}$ contains the Pauli matrices, and $\chi_{\lambda}$ is the Pauli spinor with $\chi_{\uparrow} = (1, 0)^T$ and $\chi_{\downarrow} = (0, 1)^T$. Note that $u$ has a Lorentz-invariant normalization given by $\overline{u}(k^*, \lambda) u(k^*, \lambda) = 2 m^*$. The wave-function for antibaryons can be similarly constructed. The energy spectrum of baryons ($k^0$) is given by 
\begin{equation}
    E(k) = \sqrt{{m^*}^2 + |\vec{k} - \vec{\Sigma}|^2} + \Sigma^0; \label{eq:medium:MFT:dirac:energy}
\end{equation}
in the mean-field approximation, $\Sigma^{\mu}$ and $m^*$ do not depend on $k^{\mu}$ but they do vary with density. The values for $m^*$ and $\Sigma^0$ (in the n.m.~frame) decrease and increase respectively (see Fig.~\ref{fig:medium:MFT:meff_sigma}) in such a way that the total energy of baryons in Eq.~\eqref{eq:medium:MFT:dirac:energy} increases at higher densities. As we will see shortly, this brings about in-medium baryon decays to particles that are heavier than the baryon's vacuum mass since $E(0) > m_{\cal B}$ at high densities. In general, the increase in the repulsion between baryons in a RMF framework can be understood by comparing the time-like component of vector (repulsive) interactions, which are proportional to $u^{\dagger}u$, with scalar (attractive) interactions, which are parameterized by $\overline{u}u = (m^*/ E^*)u^{\dagger}u$. As the density increases, $m^*$ decreases and the strength of the attractive forces relative to the repulsive ones diminishes~\cite{HOROWITZ1987613}. However, we should note that having a highly repulsive nuclear interaction at extremely high densities (compared to $n_{\rm sat}$) is a reasonable expectation, regardless of the specific dense matter formalism.
Having explained the formalism utilized in this work, we now describe the specific EoS that we use for generating our numerical results.

We choose an EoS based on a non-linear hadronic SU($3$) CMF model~\cite{Papazoglou:1998vr}, in which the baryonic degrees of freedom include nucleons ($n$, $p$), hyperons ($\Lambda$, $\Sigma$, $\Xi$) and the spin-3/2 resonances ($\Delta$, $\Sigma^*$, $\Xi^*$, $\Omega$). These baryons interact via exchange of scalar ($\sigma$, ${\bm \delta}$, $\zeta$, $\chi$) and vector mesons (${\bm \rho}^{\mu}$, $\omega^{\mu}$, $\phi^{\mu}$), in which ${\bm \rho}^{\mu}$ and ${\bm \delta}$ are both isovectors.  In the RMF limit, the mesons become classical fields, and in the n.m.~frame only the zeroth components of vector mesons develop VEV. We employ
the CMF model of Ref.~\cite{Dexheimer:2008ax}, using 
the CMF EoS variants taken from Ref.~\cite{compose_CMF1,compose_CMF8}. 
We refer to them as DS(CMF)-1 through DS(CMF)-8 as detailed in App.~\ref{app:CMF} and note that Appendix for further discussion of this class of models. 
The numerical values for $m^*$ and $\Sigma^0$  in the DS(CMF-1) EoS are plotted in Fig.~\ref{fig:medium:MFT:meff_sigma}. We note that the reduction of the effective baryon masses at high densities as shown in Fig.~\ref{fig:medium:MFT:meff_sigma} is due to chiral symmetry restoration at high densities. 

The conventional approach to determining the coupling constants in RMF models, which we also use here, relies on an extrapolation from symmetric finite nuclei to infinite neutron matter. We would like to contrast this with an alternative that we may wish to employ in the future, which is based on fitting uniform pure neutron matter properties determined through the use of chiral effective field theory~\cite{Alford:2022bpp}. The latter procedure involves fitting the RMF couplings with the synthetic neutron matter data generated using Quantum Monte Carlo (QMC) many-body methods~\cite{Carlson:2014vla}, in addition to reproducing $n_{\rm sat}$, $B/A$, and $K$.

\subsection{Medium Effects: Effective Masses and Beyond}
\label{sec:medium:modific}
In this section, we discuss some of the notable features that emerge in studying processes in the medium, and make comparisons with the vacuum formalism. We start with the quantization of baryon fields in the medium followed by the rate and cross section calculation formalism. We then discuss the electromagnetic form factors of the baryons that are needed for our calculations in Sec.~\ref{sec:dark_decay:medium:rates}. 

The presence of the baryon Fermi sea modifies the quantization procedure of the baryon fields, $\psi(x)$, in the medium~\cite{Serot:1984ey} compared with the usual procedure in the vacuum~\cite{bjorken2013relativistic}. 
Once the coefficients behind the Fourier modes of $\psi(x)$ are promoted to baryon creation ($a^{\dagger}(k)$) and annihilation ($a(k)$) operators (likewise $b^{\dagger}(k)$ and $b(k)$ for antibaryons), we conclude that the action of these operators on the medium ground state $|\Omega\rangle$, which contains baryon levels filled to a Fermi momentum ($k_F$), should be given by
\begin{equation}
    \begin{split}
        b(k)|\Omega\rangle =& 0 \quad \forall \vec{k},\\
        a^{\dagger}(k)|\Omega\rangle =& 0 \quad |\vec{k}| < k_F,\\
        a(k)|\Omega\rangle =& 0 \quad |\vec{k}| > k_F.
    \end{split}
\end{equation}
This leads to a different form (compared to in the vacuum) for the baryon propagator which is given by~\cite{Serot:1984ey}
\begin{equation}
\label{eq:medium:modific:baryon-propagator}
    G(p) \equiv \left( \slashed{p}^* + m^*\right) \Bigg\{ \frac{1}{p^{*2} - m^{*2} + i\varepsilon } + 2\pi i \delta(p^{*2} - m^{*2}) \theta(p^*_0)\, \theta\Big[k_F^2 + p^{*2} - \frac{\big(p^*_{\mu} B^{\mu}\big)^2}{B_{\mu}B^{\mu}}\Big]\Bigg\},
\end{equation}
in which $\theta$ is the Heaviside step function, $B^{\mu}$ is the baryon current density, which in the n.m.~frame is given by $B^{\mu}_{nm}  = \delta^{\mu 0} n_B$, and the second term in Eq.~\eqref{eq:medium:modific:baryon-propagator} allows for the propagation of holes in the Fermi sea. Using this modified propagator and the spinors in Eq.~\eqref{eq:medium:MFT:spinor}, one can derive Feynman rules~\cite{Serot:1984ey} for calculating the amplitudes for various processes (see Sec.~\ref{sec:dark_decay:medium:rates}). However, in calculating rates via phase space integrals, we should first observe that an on-shell ($p^{*2} = m^{*2}$) and positive energy ($p^0 >0$) Lorentz-invariant integral over the four-momentum is given by
\begin{equation}
\begin{split}
    \int d^4p\,  \delta \left( \left(p^{\mu} - \Sigma^{\mu}\right)^2 - m^{*2} \right) \theta (p^0) f(p^{\mu}) =& \int d^3p\, \frac{f\left(\sqrt{|\vec{p} - \vec{\Sigma}|^2 + m^{*2}} + \Sigma^0, \vec{p} \right)}{2 \sqrt{|\vec{p} - \vec{\Sigma}|^2 + m^{*2}} }.
    \label{eq:medium:modific:phase_space_int}
\end{split}
\end{equation}
Therefore, we identify the Lorentz-invariant (on-shell) volume element in the medium as $d^3p/2E^*(p)$. This means that the normalization factors in the in-medium phase space integrals should contain $\left(2 E^*\right)^{-1}$ in place of the usual vacuum expression. 

We also note that the velocity of a baryon is defined in terms of the kinetic momentum as opposed to the canonical one, i.e., $v_{\mu} \equiv k^{*}_{\mu} / E^*$. This velocity should be used for calculating the cross section of 
two-body scattering involving a baryon (see App.~\ref{app:compton}). We can explicitly show this by performing an integration over the longitudinal ($\hat{z}$) components of the incident beams' momenta ($\bar{k}^z_{\mathcal{A}}$ and $\bar{k}^z_{\mathcal{B}}$). Let us assume for the moment that only one baryon ($\mathcal{B}$) is involved, in which case we have (see Eq.~(4.77) of Ref.~\cite{peskin2018introduction})
\begin{equation}
    \begin{split}
    \label{eq:medium:modific:incident_beam_velocity}
        &\int d\bar{k}^z_{\mathcal{A}} d\bar{k}^z_{\mathcal{B}} \, \delta \Big( \bar{k}^z_{\mathcal{A}} + \bar{k}^z_{\mathcal{B}} - \sum_f p_f^z\Big) \delta\Big( \bar{E}_{\mathcal{A}} + \bar{E}_{\mathcal{B}} - \sum_f E_f\Big) \\
        &= \int d\bar{k}^z_{\mathcal{A}} \, \delta\Big( \sqrt{\bar{k}^2_{\mathcal{A}} + m_{\mathcal{A}}^2} + \sqrt{\big(\bar{k}_{\mathcal{B}} - \vec{\Sigma}_{\mathcal{B}} \big)^2 + m_{\mathcal{B}}^2}  + \Sigma^0_{\mathcal{B}} - \sum_f E_f\Big)\bigg|_{ \bar{k}^z_{\mathcal{B}}=  \sum p_f^z - \bar{k}^z_{\mathcal{A}} } \\
        &= {\left| \frac{\bar{k}^z_{\mathcal{A}}}{\bar{E}_{\mathcal{A}}} - \frac{ \bar{k}^z_{\mathcal{B}} - \Sigma^z_{\mathcal{B}} }{\bar{E}^*_{\mathcal{B}}} \right|}^{-1} =  {\left| \frac{\bar{k}^z_{\mathcal{A}}}{\bar{E}_{\mathcal{A}}} - \frac{ \bar{k}^{*z}_{\mathcal{B}}}{\bar{E}^*_{\mathcal{B}}} \right|}^{-1} \equiv {\left| v_{\mathcal{A}} - v_{\mathcal{B}}\right|}^{-1},
    \end{split}
\end{equation}
in which in the last line we are assuming $\bar{k}^z_{\mathcal{B}}=  \sum p_f^z - \bar{k}^z_{\mathcal{A}}$ and have identified the baryon velocity using the kinetic momentum, such that $\left| v_{\mathcal{A}} - v_{\mathcal{B}}\right|$ is the relative velocity of the beams as viewed from the laboratory frame. The generalization to the case with two baryons is straightforward. The fact that the velocity of a baryon is zero when $\vec{k}^{\,*}=0$ could have also been deduced by inspecting the kinetic energy component in Eq.~\eqref{eq:medium:MFT:dirac:energy}. For this reason, the frame in which $\vec{k}^{\,*}=0$ holds is called the center of velocity (c.v.) frame which is distinct from the center of mass (c.m.) frame defined by $\vec{k}=0$. Therefore, the decay rate of a baryon in an arbitrary frame ($\Gamma$) is found by boosting ($\gamma$) the rate evaluated in the c.v.~frame using $\Gamma = \gamma^{-1} \, \Gamma_{\rm c.v.}$.

Since we study processes that involve electromagnetic (EM) interactions with baryons, the generalization of the usual form factors from the vacuum to within the medium should be checked. The in-medium spinors in Eq.~\eqref{eq:medium:MFT:spinor} are different from their vacuum counterparts. Therefore, certain commonly used properties (e.g., the Gordon decomposition) in vacuum need to be reestablished. However, we note that the general form of these interactions is determined by the structure of Dirac algebra. While important for formulating our analyses, this is slightly tangential to the broader narrative of this work; we thus relegate the details to the Appendices, but we encourage the reader to study them nonetheless. In App.~\ref{app:form_fac}, we explicitly show that the vacuum EM vertex form can be generalized to its in-medium form if one replaces $m\to m^*$, $p \to p^*$, and identifies the electric charge and magnetic moment of a baryon from the scattering amplitudes in the c.v.~frame. Our numerical results in Sec.~\ref{sec:dark_decay:medium} assume the vacuum values for the in-medium form factors $F_{1,2}^*$ of neutron and $\Lambda$. We also derive the non-relativistic limit of baryon's EM interactions and their elastic scattering formalism in App.~\ref{app:non-rel_scatt}. We present the calculations for in-medium Compton scattering in App.~\ref{app:compton}, as a demonstration of the RMF formalism utilized in this work.

\section{Baryon Dark Decay Rates in Dense Matter}
\label{sec:dark_decay:medium}

In this section, we develop the procedures for evaluating particle physics processes, such as neutron decays and neutron-neutron scattering, in the neutron-star medium. Our particular interest is in radiative decays such as ${\cal B} \to \chi\gamma$ in the core of the star. In the absence of a matter environment, a common procedure, adopted in many contexts, is to assume the mixing is weak and to redefine the fields, here $B^i$ and $\chi$~\cite{McKeen:2018xwc}, so that they no longer mix, and then to analyze $B^i \to \chi$ transitions in that new basis. In Sec.~\ref{sec:dark_decay:medium:method}, we show why and how this procedure can fail in strongly interacting dense matter, and we argue for a Feynman diagram analysis in its place. Subsequently, starting in Sec.~\ref{sec:dark_decay:medium:rates}, we show how the transition rates can be evaluated explicitly and consider their implications.

\subsection{General Considerations}
\label{sec:dark_decay:medium:method}
To illuminate the essential points, we consider 
the possibility of $n$-$\chi$ mixing in a background
field $\Sigma^\mu$, the vector self-energy of a neutron in the neutron-star medium, which 
interacts with the neutron field $\psi_n$, 
but not the $\chi$ field $\psi_\chi$. Thus we adopt 
the following simple model: 
 \begin{equation}
     \mathcal{L} = \overline{\psi}_n \left( i 
     \slashed{\partial}
     - \slashed{\Sigma} - m_n^* \right) \psi_n + \overline{\psi}_\chi \left(  i 
     \slashed{\partial}
     - m_{\chi} \right) \psi_{\chi} - \varepsilon \left( \overline{\psi}_n \psi_{\chi} + \overline{\psi}_{\chi} \psi_n\right) \,. \label{eq:dark_decay:medium:method:toy_mix_L}
 \end{equation}
Under a field redefinition, $\psi \to \psi'$, prescribed by
\begin{equation}
\begin{pmatrix}
    \psi_n' \\
    \psi_{\chi}'
\end{pmatrix} = \begin{pmatrix}
    \cos \theta & \sin\theta\\
    -\sin\theta & \cos \theta
\end{pmatrix}
\begin{pmatrix}
    \psi_n \\
    \psi_{\chi}
\end{pmatrix},
\end{equation}
Eq.~\eqref{eq:dark_decay:medium:method:toy_mix_L} becomes 
 \begin{equation}
 \begin{split}
     \mathcal{L}' =& \overline{\psi}_n' \left( i \slashed{\partial}
     -  \slashed{\Sigma} \cos^2 \!\theta - m_n^* \cos^2 \!\theta - m_{\chi} \sin^2 \!\theta \right) \psi_n'  \\
     &+ \overline{\psi}_\chi' \left(  i 
     \slashed{\partial}
     -  \slashed{\Sigma} \sin^2 \!\theta - m_{\chi}\cos^2 \!\theta - m_{n} \sin^2 \!\theta \right) \psi_{\chi}' \\
     &+ \overline{\psi}_n' \left[ \frac{\sin (2\theta) }{2} \left(m_n^* - m_{\chi} + \slashed{\Sigma}\right)  - \varepsilon \cos(2\theta)  \right] \psi_{\chi}' \,.
     \label{eq:dark_decay:medium:method:toy_mix_L_rotated}
\end{split}
 \end{equation}
If $\Sigma^{\mu}$ were absent, and with $\varepsilon$ real, then for $\tan \left(2 \theta\right) = {2 \varepsilon}/({m_{n}^* - m_{\chi}})$, $\cal L'$ describes two decoupled fields with a modified energy spectrum. These fields can then map to the asymptotic  (``in" and ``out") states 
needed to define the $S$-matrix~\cite{schwartz2014quantum}. To do this, any interactions with these fields should vanish as $t \to \pm \infty$. For the neutron (and other SM baryons), we note that the effect of the vector self-energy can be absorbed into the definition of a modified single-particle spinor, as discussed in Sec.~\ref{sec:medium:MFT}, and thus suitable ``in" and ``out" states can still be constructed. In the current case, $\Sigma^\mu$ mediates an interaction between the rotated $n$ and $\chi$ fields, putting the utility of our field redefinition procedure into question. After all, even in the mean-field limit, $\Sigma^0$ can greatly exceed the $n$ and $\chi$ masses at the high densities reached within a neutron star, and it cannot vanish as $t \to \pm \infty$, since we work within a medium of infinite extent. Since $\Sigma^\mu$ is not a Lorentz scalar, we cannot extend our field redefinition approach to include it. Therefore, there would seem to be no advantage to 
following a field redefinition approach in neutron matter. Moreover, in the small mixing limit ($\varepsilon \ll |m_n^* -  m_{\chi}|$), the mass ($n'$, $\chi'$) and interaction ($n$, $\chi$) eigenstates are nearly the same. Working with Eq.~(\ref{eq:dark_decay:medium:method:toy_mix_L}), we can treat $\varepsilon \overline{\psi}_n \psi_{\chi}$ as a tiny interaction that mediates $n \leftrightarrow \chi$ transitions within perturbation theory. This Feynman diagram analysis, through the in-medium baryon propagator, Eq.~(\ref{eq:medium:modific:baryon-propagator}), naturally includes the impact of momentum dependence and of the neutron self-energy on $n$-$\chi$ mixing. We emphasize that both effects are absent in the field redefinition procedure. As a result, too, we do not have large enhancements in our predictions should the in-medium neutron and $\chi$ states become degenerate in energy --- the imaginary part of the neutron self-energy effectively eliminates that possibility. Nevertheless, $n$-$\chi$ mixing within the neutron-star medium could potentially lead to effects not possible in terrestrial experiments, and we consider those possibilities more carefully in Sec.~\ref{sec:dark_decay:medium:med_enabled}.


\subsection{Dark Decay Rate Estimates}
\label{sec:dark_decay:medium:rates}
We now turn to the explicit evaluation of rates of particle processes within the neutron-star medium, with a particular focus on dark decay rates. As long known, the background field associated with matter leads to a spontaneously breaking of Lorentz symmetry, but as a consequence of our Lorentz covariant description, discussed in Sec.~\ref{sec:medium:MFT}, our expressions always have definite Lorentz transformation properties. In what follows, we exploit our freedom to choose a frame to simplify our analysis. 

Generally, processes of the form ${\cal B} + \{X\} \to \chi + \{Y\}$ lead to the following rate of change of the local baryon density $n_B$ (with respect to the proper time, $\tau$, referenced to that spacetime point):
\begin{align}
    \label{eq:master_rate}
    \dfrac{dn_B}{d\tau} & = -\int d\Pi_{\cal B} \biggl( \prod_{\{X\}} d\Pi_{X} \biggr) d\Pi_\chi \biggl( \prod_{\{Y\}} d\Pi_{Y} \biggr)  \\
    & \quad \qquad \times f_{\cal B}(\vec{p}_{\cal B}) \biggl( \prod_{\{X\}} f_X(\vec{p}_X) \biggr) \Bigl(1-f_\mathcal{\chi}(\vec{k}_\chi)\Bigr) \biggl( \prod_{\{Y\}} \Bigr[1\pm f_Y(\vec{k}_Y) \Bigr] \biggr) & \nonumber \\
    & \quad \qquad \times |\mathcal{M}|^2 \times (2\pi)^4 \delta^{(4)}\biggl(p_{\cal B} + \sum_{\{X\}} p_X - k_\chi - \sum_{\{y\}} k_Y \biggr), \nonumber
\end{align}
where $\{X\}$ ($\{Y\}$) is some set of other states in the initial (final) state --- which may be empty.
Moreover, $d\Pi_i = d^3 \vec{p}_i/[(2\pi)^3 (2E^*_i)]$ is the Lorentz-invariant phase space  measure, $f(\vec{p})$ are the species-dependent occupation numbers\footnote{The occupation factor for $Y$ depends on whether or not $Y$ is a boson ($+$) or fermion ($-$).}, and $|\mathcal{M}|^2$ is the \emph{spin-summed} (as opposed to \emph{spin-averaged}) squared matrix element. We denote final-state momenta with $k_i$ instead of $p_i$. Consistent with our assumption that there is no appreciable background of $\chi$, we set its occupation factor $f_\mathcal{\chi}(\vec{k}_\chi)$ to zero. 
All baryonic species abide by zero-temperature Fermi distributions characterized by distinct Fermi momenta $p_{F,{\cal B}}$.

We briefly discuss important qualitative features of the evaluation of Eq.~\eqref{eq:master_rate} for the decay process ${\cal B} \to \chi \gamma$ and present the corresponding results. We relegate details of the calculation to App.~\ref{sec:app:JMB:decay}. We work in the interaction basis, so that the decay proceeds via the Feynman diagram containing the $n-\chi$ interaction and the baryon magnetic dipole moment operator, which we write as 
\begin{equation}
	\mathcal{O}_{{\cal B}\gamma}  = \frac{g_{\cal B} e}{8m_{\cal B}^*} \overline{{\cal B}} \sigma^{\mu\nu} {\cal B} F_{\mu\nu},
\end{equation}
noting $g_n = 3.826$ and $g_\Lambda = -1.226$ \cite{ParticleDataGroup:2022pth}. This computation is made in a background mean-field of neutron matter, and the associated decay amplitude, as developed in Sec.~\ref{sec:medium:MFT}, is determined by replacing the canonical momenta of the in-vacuum computation with kinetic momenta as per Eq.~(\ref{eq:medium:MFT:modspinor}). Labeling canonical momenta as ${\cal B}(p_{\cal B}) \to \chi (k_\chi) + \gamma(k_\gamma)$, the corresponding spin-summed squared matrix element is
\begin{equation}
    |\mathcal{M}|^2 = \frac{\varepsilon_{{\cal B}\chi}^2 g_{\cal B}^2 e^2}{2(m_{\cal B}^*)^2} \left[ (p_{\cal B}^* \cdot k_\chi) + m_{\cal B}^* m_{\chi} \right] \,,
\end{equation}
noting that both the kinetic momentum $p^*_{\cal B}$ and the in-medium mass $m_{\cal B}^*$ appear. It then remains to insert this into Eq.~\eqref{eq:master_rate} and integrate. The  integral takes the form
\begin{equation}
    \label{eq:master_decay_rate}
    \frac{dn_{\cal B}}{d\tau} = -\int \frac{d^3 \vec{p}_{\cal B}}{(2\pi)^3 (2E^*_{\cal B})} \frac{d^3 \vec{k}_\chi}{(2\pi)^3 (2E_\chi)} \frac{d^3 \vec{k}_\gamma}{(2\pi)^3 (2E_\gamma)} f_{\cal B}(\vec{p}_{\cal B}) \times |\mathcal{M}|^2 \times (2\pi)^4 \delta^{(4)}\left(p_{\cal B} - k_\chi - k_\gamma \right) \,,
\end{equation}
and it can be computed in different ways. The Lorentz-invariance of each measure $d\Pi_i$ affords the opportunity of performing different parts of the integration in different frames. We note that the integration over $d\Pi_{\cal B}$ is simplest in the n.m.~frame: there, we have $f_{\cal B}(\vec{p}_{\cal B}^{\,\rm(n.m.)}) = \Theta(p_{F,{\cal B}} - |\vec{p}_{\cal B}^{\,\rm(n.m.)}|)$, and the integrand is isotropic. Contrariwise, it is simplest to evaluate the integration over $d\Pi_\chi ~ d\Pi_{\gamma}$ in the c.v.~frame. Moreover, as discussed in Sec.~\ref{sec:medium:modific}, the width of an individual baryon is most simply interpreted in its respective c.v.~frame, since 
the baryon is not moving. The baryon width in the c.v.~frame takes the form 
\begin{equation}
    \label{eq:CV_decay_width}
   \Gamma_{\rm c.v.}(|\vec{p}_{\cal B}^{\,\rm(n.m.)}|) = \frac{1}{2m_{\cal B}^*} 
\int \frac{d^3 \vec{k}_\chi}{(2\pi)^3 (2E_\chi)} \frac{d^3 \vec{k}_\gamma}{(2\pi)^3 (2E_\gamma)} \times \frac{1}{2} |\mathcal{M}|^2 \times (2\pi)^4 \delta^{(4)}\left(p_{\cal B} - k_\chi - k_\gamma \right)\,,\,
\end{equation}
where the argument of $\Gamma_{\rm c.v.}$ follows from our earlier frame choice.\footnote{The additional factor of $\frac{1}{2}$ in Eq.~\eqref{eq:CV_decay_width} arises because $|\mathcal{M}|^2$ has been spin-summed and not spin-averaged.} Henceforth we abbreviate $p_{\cal B} \equiv |\vec{p}_{\cal B}^{\,\rm(n.m.)}|$. We have $p_{\cal B}^{*,\rm{(c.v.)}} = (m_{\cal B}^*, 0)$ and $p_{\cal B}^{\rm (c.v.)} = (m_n^* + \Sigma_{\cal B}^{\rm{(c.v.)},0}, \vec{\Sigma}_{\cal B}^{\rm (c.v.)})$, with $\Sigma_{\cal B}^{\rm(c.v.)}$ the baryon vector self-energy in the c.v.~frame. The results in different frames are connected by Lorentz boosts, yielding
\begin{equation}
    \Gamma_{\rm c.m.} (p_{B})= \left( \frac{m_{\cal B}^*}{E_{\cal B}^{*, \rm{(c.m.)}}} \right) \Gamma_{\rm c.v.}(p_{B}) = \left( \frac{E_{\cal B}^{*, \rm{(n.m.)}}}{E_{\cal B}^{*, \rm{(c.m.)}}} \right) \Gamma_{\rm n.m.}(p_{B})\,. \label{eq:medium:decay:rate_boosts}
\end{equation}

The total rate of baryon loss in the n.m.~frame is then given by integrating over all baryons in the local fluid, accounting for the contraction of their individual widths by a factor $\gamma^{-1} = m_{\cal B}^*/E_{\cal B}^*$, with $E_{\cal B}^* = \sqrt{p_{\cal B}^2 + (m_{\cal B}^*)^2}$ the n.m.-frame energy associated with
the kinetic momentum: 
\begin{equation}
    \label{eq:NM_decay_rate}
    \dfrac{dn_{\cal B}}{d\tau} = - 2 \times \int_0^{p_{F,{\cal B}}} \frac{p_{\cal B}^2 dp_{\cal B}}{2\pi^2} \gamma^{-1} \Gamma_{\rm c.v.}(p_{\cal B})\,.
\end{equation}
The prefactor of 2 comes from the baryon's two spin degrees of freedom, and the factor of $2\pi^2$ in the denominator ensures that if the function $\gamma^{-1} \Gamma_{\rm c.v.}(p_{\cal B})$ were a constant, then the result would be $(\gamma^{-1} \Gamma_{\rm c.v.}) \times n_B$. Combining Eqs.~\eqref{eq:CV_decay_width} and \eqref{eq:NM_decay_rate}, we find that this procedure leads to the same result as the direct evaluation of Eq.~\eqref{eq:master_decay_rate} shown in App.~\ref{sec:app:JMB:decay}. In either case, we arrive at the following result:
\begin{align}
    \dfrac{dn_{\cal B}}{d\tau} = - \frac{\varepsilon_{B\chi}^2 g_{\cal B}^2 e^2}{128\pi^3} (m_{\cal B}^*)^2 \int_{1}^{x_F} & dx \, \sqrt{x^2-1} \times \frac{1+2x\sigma+\sigma^2-\mu^2}{(1+2x\sigma+\sigma^2)^2} \nonumber \\
    & \times \left[ (1+2x\sigma+\sigma^2)(1+x\sigma+2\mu) + \mu^2(1+x\sigma) \right]\,, \label{eq:app:JMB:decay:tot_rate}
\end{align}
in which
\begin{align}
    x\equiv \frac{E_{\cal B}^{*, \rm{(n.m.)}}}{m_{\cal B}^*}, \quad \quad
    x_F\equiv  \frac{E_{F,{\cal B}}^{*, \rm{(n.m.)}}}{m_{\cal B}^*}, \quad\quad 
    \sigma \equiv \frac{\Sigma_{\cal B}^{\rm{(n.m.)}, 0}}{m_{\cal B}^*}, \quad\quad
    \mu \equiv \frac{m_{\chi}}{m_{\cal B}^*}
   \label{eq:dark_decay:medium:dim_less_def} 
    \,.
\end{align}
The corresponding c.m.-frame single baryon decay rate $\Gamma_{\rm c.m.} (p_{B})$ is given in Eq.~\eqref{eq:app:decay:cm_rate}.

\begin{figure}[!t]
    \centering
    \includegraphics[width = \textwidth]{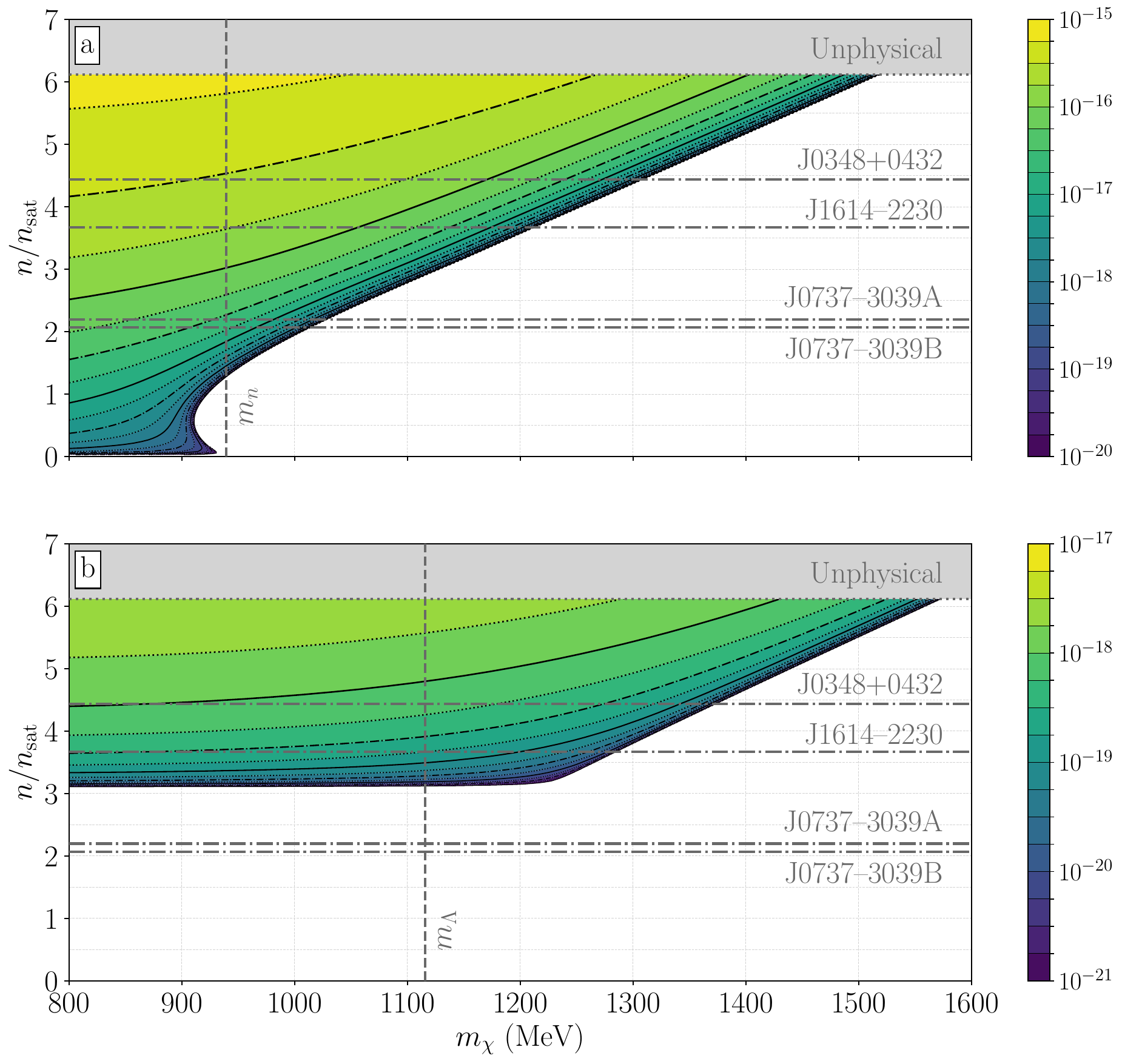}
    \caption{(Color Online) The proper rates for ${\cal B} \to \chi\gamma$ decays, $(-dn_{\cal B}/d\tau)$ (in fm$^{-3}$ s$^{-1}$), for neutrons (a) and $\Lambda$s (b) assuming the DS(CMF)-1 EoS. In either panel, we fix the corresponding $\varepsilon_{B\chi}$ to be $10^{-16}$ MeV. Note that the color scales are different between the two panels. See text for additional details.}
    \label{fig:decay:local_rates}
\end{figure}

\begin{figure}[!t]
    \centering
    \includegraphics[width=0.7\linewidth]{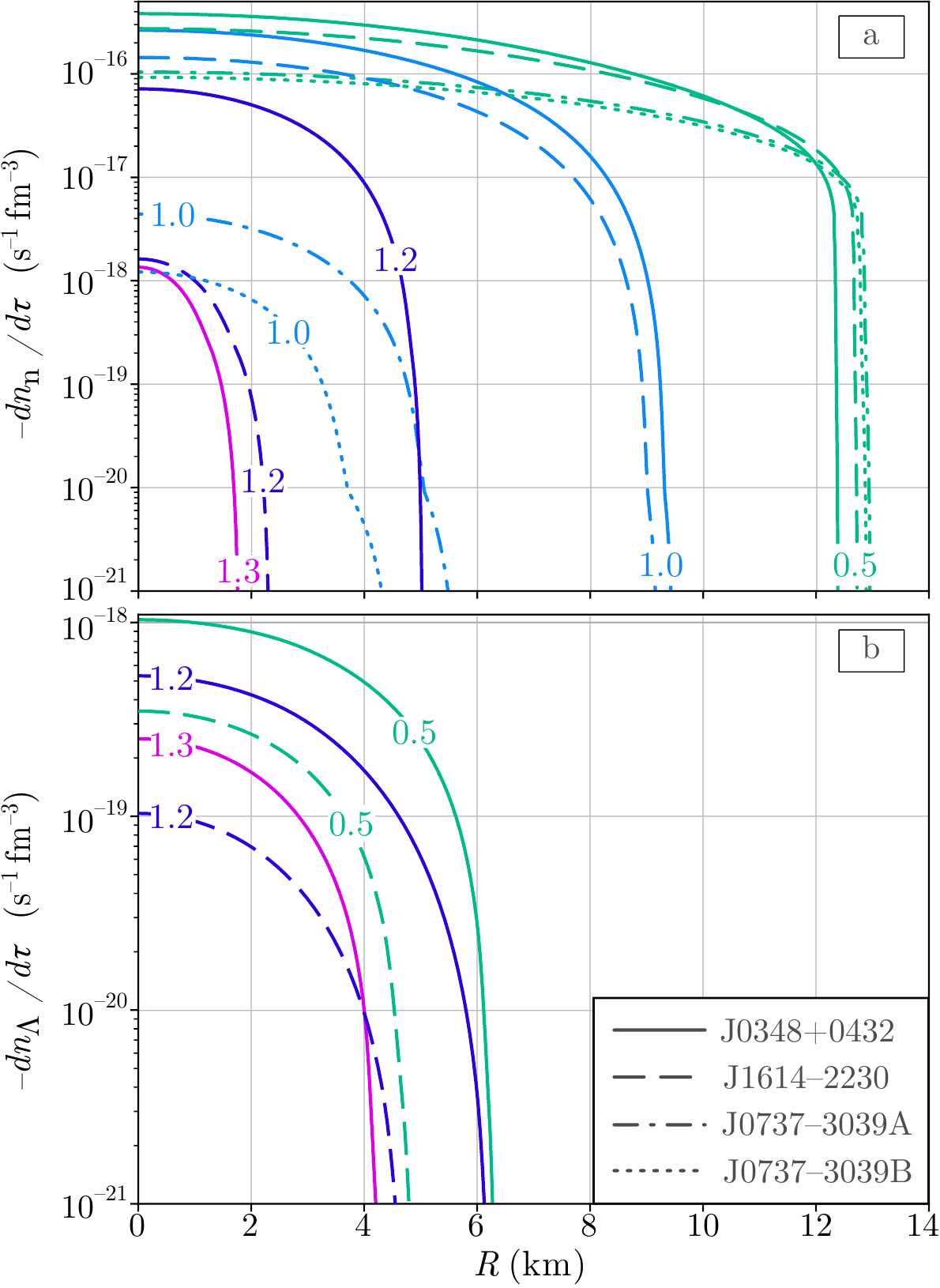}
    \caption{(Color Online) The baryon proper decay (${\cal B} \to \chi \gamma$) rate $(-dn_{\cal B}/d\tau)$ (per unit volume) assuming $\varepsilon_{{\cal B}\chi} = 10^{-16}$ MeV for (a) neutrons and (b) $\Lambda$s as a function of radius for four pulsars using the DS(CMF)-1 EoS. The numbers next to curves (colors) indicate different values of $m_{\chi} = \{0.5, 1.0, 1.2, 1.3\}\, {\rm GeV}$. }
    \label{fig:chi_phot:rate_v_r}
\end{figure}

We illustrate the rates of ${\cal B} \to \chi \gamma$ as a function of $m_{\chi}$ and $n$ in Fig.~\ref{fig:decay:local_rates}, for both neutrons (a) and $\Lambda$s (b). (In contrast, in Fig. 8, we show how these proper decay rates change across the different pulsars of interest for various values of the $\chi$ mass.) These calculations are for the DS(CMF)-1 EoS, but the results are qualitatively similar for the other EoS in this family.\footnote{Of course, the EoS that do not contain hyperons will not lead to $\Lambda \to \chi\gamma$ decays within neutron stars.} The vertical axes have been normalized to the value of nuclear saturation density in this EoS, $n_{\rm sat} = 0.15$ fm$^{-3}$. The respective color scales are shown at right, assuming $\varepsilon_{{\cal B}\chi} = 10^{-16}$ MeV; the units are fm$^{-3}$ s$^{-1}$ and we emphasize that these rates scale as $\varepsilon_{{\cal B}\chi}^2$. To guide the eye, we have also added black contours every quarter order of magnitude. Solid contours correspond to integer numbers; dot-dashed contours correspond to half-integer numbers; and dotted lines correspond to quarter-integer numbers. In either panel, the dashed vertical line indicates the vacuum mass of the corresponding baryon. The dotted horizontal line corresponds to the central density of the heaviest stable neutron star within this EoS, corresponding to $M_{\rm TOV} \approx 2.07 \, M_\odot$; the region above this line has been grayed out because these densities do not occur in a stable neutron star. Similarly, the dot-dashed horizontal lines correspond to the central densities of neutron stars with the masses of J0348+0432, J1614--2230 and J0737--3039A/B. We observe that when $n\to\chi\gamma$ is operative, it is almost always numerically larger than the rate of $\Lambda\to\chi\gamma$ (for $\varepsilon_{n\chi} = \varepsilon_{\Lambda\chi}$). This is a simple consequence of larger neutron number fractions at these densities, and the two rates often differ by several orders of magnitude. However, $\Lambda$s have a further reach in $m_{\chi}$ when they are present than neutrons do, owing to the larger total energy of $\Lambda$s in neutron matter.

\subsection{Medium-Enabled Dark Decay Processes}
\label{sec:dark_decay:medium:med_enabled}
\begin{figure}[!t]
    \centering
    \includegraphics[width=0.7\textwidth]{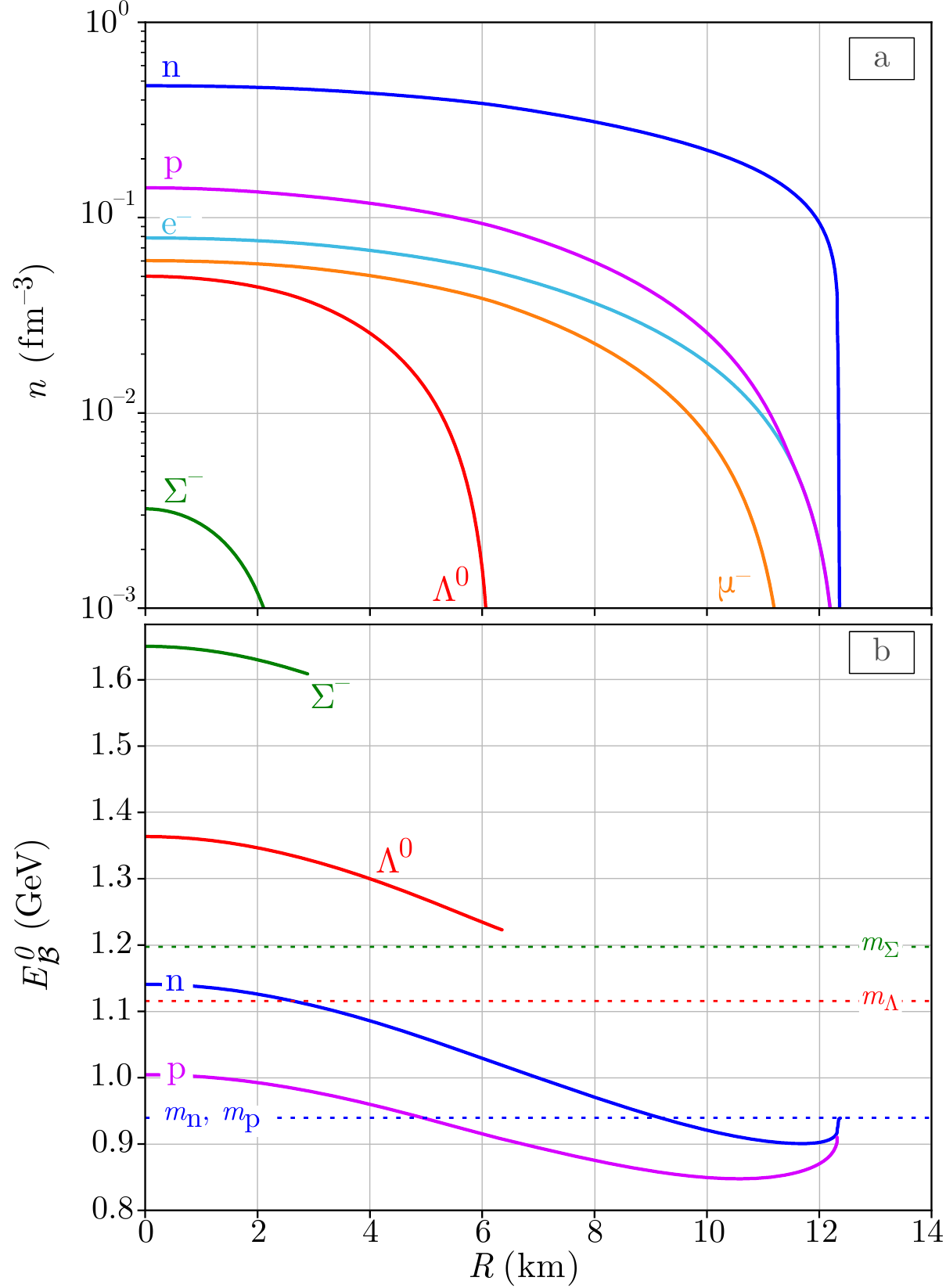}
    \caption{(Color Online) The particle composition (a), and the energy of baryons at rest in the n.m. frame (b) as a function of radius in PSR J0348$+$0432 assuming the DS (CMF)-1 EoS. The horizontal lines correspond to the vacuum masses of baryons. We choose models for which we expect the steady-state admixture of dark states to be completely negligible.}
    \label{fig:dark_decay:medium:rest-e:J0348}
\end{figure}

It was shown in Sec.~\ref{sec:medium:MFT} that baryons in neutron stars have a lower effective mass ($m_{\cal B}^*$) and a higher self-energy ($\Sigma^0_{\cal B}$) at higher densities (see Fig.~\ref{fig:medium:MFT:meff_sigma}), but their overall energy can be much higher than their vacuum rest mass ($m_{\cal B}$). In order to illustrate this for a heavy neutron star, we plot the baryon rest-energies ($E_{\cal B}^0 \equiv E_{\cal B}(p=0)$ in the n.m.~frame) for PSR J0348$+$0432 as a function of radius in Fig.~\ref{fig:dark_decay:medium:rest-e:J0348}. We can see that baryon decays containing a final state $\chi$ with $m_{\chi} > m_{{\cal B}}$, which would be forbidden in vacuum, can occur at the core of heavy neutron stars. This enables a novel way of analyzing models with $m_{\chi}$ values for which nuclear and vacuum decays are kinematically forbidden. Furthermore, constraints derived from heavy neutron stars can still be applicable in the vicinity of $m_{\chi} \approx m_{\cal B}$ and beyond that. This should be contrasted with limits derived from processes in vacuum and within nuclei, which diminish at $m_{\chi} \approx m_{\cal B}$ or even at much lower values of $m_{\chi}$ due to the binding energy and possible energy cuts on the final states. For example, when inferring limits on $n \to\chi \gamma$ via the non-detection of $\gamma$ there is an energy cut $E_{\gamma}^{\rm min}$~\cite{Tang:2018eln}, which means $m_{\chi}$ values larger than $m_n - E_{\gamma}^{\rm min}$ cannot be constrained. 

\subsubsection*{1. Spontaneous ${\cal B}\to\chi$ Conversion}

The existence of $\chi$ raises the possibility that the baryons to which they couple might undergo spontaneous conversion to $\chi$ in the neutron-star medium as they propagate. Such an effect could prove loosely analogous to empirically observed matter-enhanced neutrino oscillations~\cite{SNO:2002tuh} or to the possibility of neutron-antineutron oscillations~\cite{Kuzmin:1970nx,Glashow:1979kx,Mohapatra:1980qe}, breaking baryon number by two units. In the latter case the presence of external interactions from matter or magnetic fields modify the energy of the $n$ and ${\bar n}$ differently, severely reducing the spontaneous oscillation probability for a fixed source of new physics~\cite{Mohapatra:1980de}, and the cross-section for scattering-mediated $n$-${\bar n}$ conversion is also very small~\cite{Gardner:2017szu}. In this section, we note the distinct features of ${\cal B}$-$\chi$ conversion.

The essential physics is thus: ${\cal B}$ and $\chi$ constitute a two-level quantum system. As we have noted in Sec.~\ref{sec:dark_decay:medium:method}, if the coupling $\varepsilon_{{\cal B}\chi}$ is nonzero, then ${\cal B}$ and $\chi$ constitute the interaction basis, whereas the eigenstates of this Hamiltonian, which we term $f_1$ and $f_2$ for this discussion, constitute the mass basis. Formally, the strong interactions that operate in neutron matter only ever produce $n$ --- this is what it means for ${\cal B}$ to be an interaction eigenstate. This ${\cal B}$ is, however, a coherent superposition of $f_1$ and $f_2$ at the moment of its creation. The subsequent evolution of this coherent wavepacket depends on the details of the ${\cal B}-\chi$ system. These details are discussed in depth in  App.~\ref{app:mixing}; we pick out the most relevant results as they pertain to this discussion.

The Hamiltonian that describes our two-state system depends on the local environment: the total energy of the baryon depends on the density through $m_{\cal B}^*$ and $\Sigma_{\cal B}$, and baryons with different n.m.-frame momenta will mix differently with $\chi$ because Lorentz invariance is spontaneously broken by the background. There exists a \emph{resonance} in this system wherever the condition, which follows from energy-momentum conservation of the canonical momenta, 
\begin{equation}
    \label{eq:resonance}
    \sqrt{m_{\chi}^2 + |\vec{p}^{\,\rm (n.m.)}|^2} \approx \sqrt{(m_{\cal B}^*)^2 + |\vec{p}^{\,\rm (n.m.)}|^2 } + \Sigma_{\cal B}^{0, {\rm (n.m.)}}
\end{equation}
is satisfied. We expect that this condition will occur for at most one value of the (magnitude of the) baryon momentum for a given density. Moreover, Eq.~(\ref{eq:resonance}) cannot be satisfied if $\Sigma_{\cal B}^{0, {\rm (c.v.)}}$ is \emph{complex}. In what follows, we set this latter possibility aside, because, as we will see, other effects act to suppress the likelihood of ${\cal B}$-$\chi$ conversion. At resonance, the offset between the interaction and mass bases is maximal, corresponding to a mixing angle of $\theta = 45^\circ$; however, if Eq.~\eqref{eq:resonance} is violated by more than a few times $\varepsilon_{{\cal B}\chi}$, then the mixing angle is parametrically small: $\theta^2 \sim \varepsilon_{{\cal B}\chi}^2/(\omega^{(+)}_{\cal B} - \omega^{(+)}_\chi)^2 \equiv (\varepsilon_{{\cal B}\chi}/\delta\omega^{(+)})^2$, where $\omega^{(+)}_\chi$ and $\omega^{(+)}_{\cal B}$ are the left- and right-hand sides of Eq.~\eqref{eq:resonance}, respectively (cf.~Eqs.~\eqref{eq:app:mixing:fmix:zeroth_chi} and \eqref{eq:app:mixing:fmix:zeroth_n}).

We first consider what happens when the system is not close to resonance. In this case, the eigenvalues of the system are given by Eqs.~\eqref{eq:app:mixing:fmix:normal:chi_pos} and \eqref{eq:app:mixing:fmix:normal:n_pos}, which are very nearly given by $\omega^{(+)}_\chi$ and $\omega^{(+)}_{\cal B}$ up to $\mathcal{O}(\varepsilon_{{\cal B}\chi}^2)$ corrections. If the system is far from resonance, then these eigenvalues are well separated. As a result, the ${\cal B}$ states produced in scattering processes will essentially immediately decohere into their component $f_1$ and $f_2$ with, respectively,  probabilities of $\cos^2 \theta$ and $\sin^2 \theta$. As such, the state that emerges from the scattering process manifests as either $f_1$ with probability $\cos^2\theta \sim 1$ or $f_2$ with probability $\sin^2 \theta \sim (\varepsilon_{{\cal B}\chi}/\delta \omega)^2$, and the latter may be vanishingly small  --- and thus so would be any yield in $\chi$. 

The situation is richer if the state is close to resonance. In this case, if the canonical momentum of the baryon is fixed by Eq.~(\ref{eq:app:mixing:fmix:transition:mom}), then the eigenvalues of the Hamiltonian are better given by Eq.~\eqref{eq:app:mixing:fmix:transition:all_orders}. They are nearly identical to each other, but they are split by a small factor, 
\begin{equation}
    \Delta \omega^* = 2\varepsilon_{{\cal B} \chi} \Sigma_{\cal B}^{0, ({\rm n.m.})} \sqrt{\frac{\left(m_{\cal B}^*+m_{\chi}\right)^2-(\Sigma_{\cal B}^{0, ({\rm n.m.})})^2}{\left[(m_{\cal B}^*)^2 - m_{\chi}^2\right]^2-(\Sigma_{\cal B}^{0, ({\rm n.m.})})^4}} + \mathcal{O}(\varepsilon_{{\cal B}\chi}^3) \,.
\end{equation}
This means that when ${\cal B}$ is produced in some strong interaction, the wavepacket containing $f_1$ and $f_2$ may remain coherent over relatively long timescales. This is analogous to how neutrino mass eigenstates remain coherent as they propagate in terrestrial oscillation experiments, despite being formed in an interaction eigenstate.\footnote{We point the interested reader to Ref.~\cite{Akhmedov:2007fk} for a discussion about the role of (de)coherence in understanding neutrino oscillations specifically, as well as Ref.~\cite{Smirnov:2016xzf} for a comparative analysis of neutrino oscillations with adiabatic conversion.} As in the case of neutrino oscillations, the $f_1$ and $f_2$ components of the ${\cal B}$ state generically evolve with different phases; over time, this leads to nonzero overlap between the evolved state and either $\cal{B}$ or $\chi$. The state is then measured, in a sense, at its next interaction some time $t$ later, either by its environment or by some experimental apparatus. It is appropriate, in this case, to invoke the concept of an oscillation probability; this is estimated by 
\begin{equation}
    \label{eq:osc_prob2}
    P_{{\cal B} \to \chi}(t) = \sin^2 2\theta \times \sin^2\left[ \frac{(\Delta \omega^*) t}{2} \right]\,.
\end{equation}
When the state is observed, however, it collapses to the combination of $f_1$ and $f_2$ appropriate to either ${\cal B}$ or $\chi$ with probabilities given by Eq.~\eqref{eq:osc_prob2}, and the process repeats for further interactions. While the oscillations have a large amplitude ($\sin^2 2\theta \sim \mathcal{O}(1)$) in this regime, the probability to convert will remain small if the time between successive measurements $\delta t_{\rm meas}$ is small, in the sense $(\Delta \omega^*) (\delta t_{\rm meas}) \ll 1$. This is precisely the quantum Zeno effect \cite{Misra:1976by, Itano:1990zz}.

It remains to determine the timescale of the interactions in the nuclear medium in order to estimate the rate of ${\cal B} \to \chi$ conversions. We estimate this to be the light time of the mean interparticle separation around nuclear saturation density: $\delta t_{\rm strong} \sim n^{-1/3}_{\rm sat} c^{-1} \sim \mathcal{O}(10^{-23})$ s. For a benchmark value $\varepsilon_{{\cal B}\chi} = 10^{-16}$ MeV, the argument of the latter sine function in Eq.~\eqref{eq:osc_prob2} is $\sim \mathcal{O}(10^{-39})$ MeV s $\sim \mathcal{O}(10^{-18})$; this is safely approximated as small, and we see that the quantum Zeno effect is indeed operative under these conditions. Therefore, even if the mixing angle is large, we estimate the probability to be
\begin{equation}
    P_{{\cal B} \to \chi}(\delta t_{\rm strong}) \sim \mathcal{O}(10^{-36}) \times \left( \frac{\varepsilon_{{\cal B}\chi}}{10^{-16} {\rm \, MeV}} \right)^2 \times \left( \frac{\delta t_{\rm strong}}{10^{-23} {\rm \, s}} \right)^2.
\end{equation}
This implies an approximate per-baryon conversion rate of
\begin{equation}
    R_{{\cal B} \to \chi}(\delta t_{\rm strong}) \sim \mathcal{O}(10^{-13}) {\rm \, s^{-1}} \times \left( \frac{\varepsilon_{{\cal B}\chi}}{10^{-16} {\rm \, MeV}} \right)^2 \times \left( \frac{\delta t_{\rm strong}}{10^{-23} {\rm \, s}} \right).
\end{equation}
One might expect that this would multiply the large density of baryons to yield a macroscopically relevant rate. However, the near-resonance region occupies a thin shell (parametrically of width $\sim\varepsilon_{{\cal B}\chi}$) within the baryon Fermi sphere; the fraction of baryons relevant for this phenomenon is fantastically small, even in the best case scenario. Thus we summarize by emphasizing that we do not expect ${\cal B}-\chi$ conversion to be a phenomenologically relevant mechanism for the production of $\chi$.

\begin{figure}[!t]
    \centering
    \includegraphics[width = \textwidth]{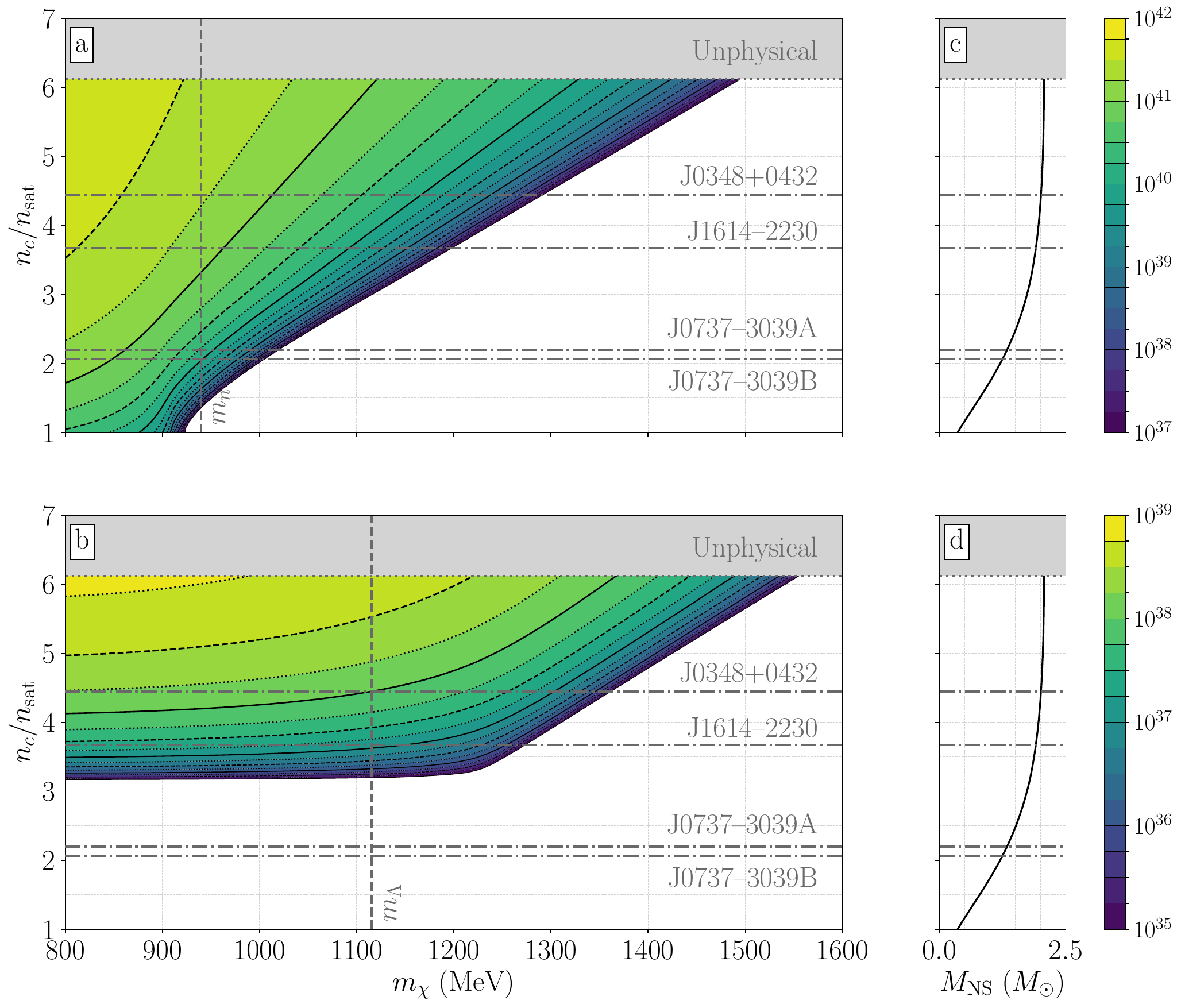}
    \caption{(Color Online) The volume-integrated rates for ${\cal B} \to \chi\gamma$ decays, $(-d{\cal B}/dt)$ (in s$^{-1}$) for neutrons (a) and $\Lambda$s (b) assuming the DS(CMF)-1 EoS. As in Fig.~\ref{fig:decay:local_rates}, we have fixed $\varepsilon_{{\cal B}\chi} = 10^{-16}$ MeV. The right panels (c \& d) show the relationship between the neutron star mass and its central density. }
    \label{fig:decay:integrated_rates}
\end{figure}

\begin{figure}[!t]
    \centering
    \includegraphics[width=0.75\textwidth]{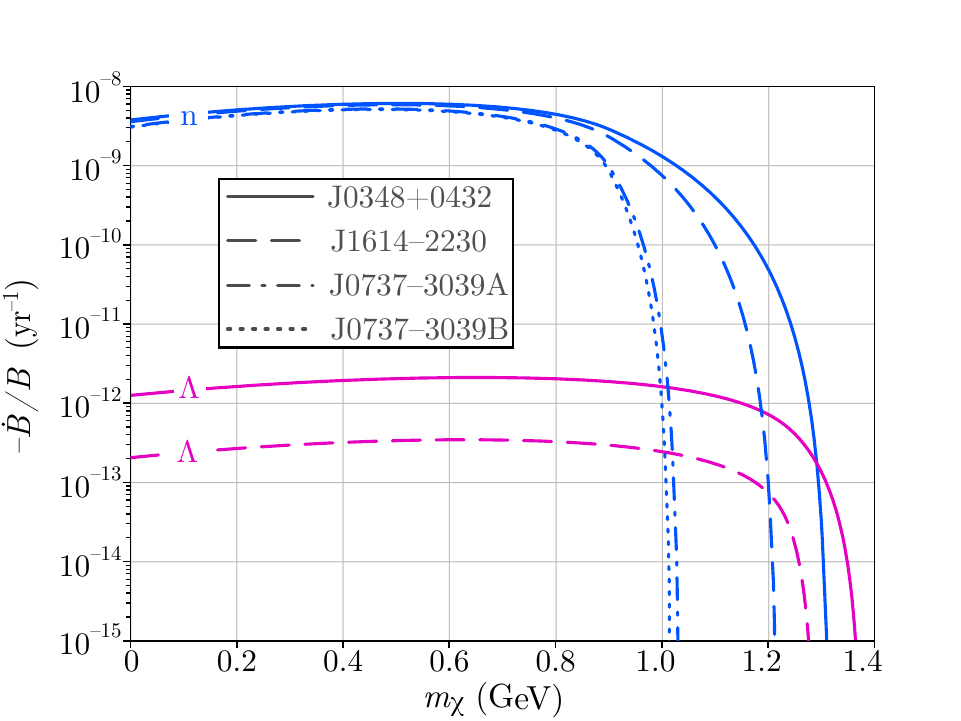}
    \caption{(Color Online) The total baryon loss rate per baryon $(-\dot{{\cal B}}/{\cal B})$ due to ${\cal B} \to \chi \gamma$ decays assuming $\varepsilon_{{\cal B}\chi} = 10^{-16}$ MeV for neutron and $\Lambda$ as a function of $m_{\chi}$ for four pulsars using the DS(CMF)-1 EoS. }
    \label{fig:chi_phot:rate_v_mchi}
\end{figure}

\subsection{Total Rates}
\label{sec:dark_decay:medium:total_rates}
In this section, we report the total baryon decay rates that emerge after integrating our earlier results over the structure of a neutron star with a given central density, $n_{\rm c}$. For example, in Fig.~\ref{fig:decay:integrated_rates}, we show the rates that result from integrating the local BNV rates in Fig.~\ref{fig:decay:local_rates} over the neutron star volume using Eq.~\eqref{eq:macro_bnv:assum:B:cons:static} and report these results as a function of $m_{\chi}$ and $n_c$. Panel (a) is for neutron decays, while panel (b) is for $\Lambda$ decays; similar to Fig.~\ref{fig:decay:local_rates}, the contours correspond to constant (base-ten log of the) integrated rate of $\cal{B} \to \chi\gamma$. We have again fixed $\varepsilon_{{\cal B}\chi} = 10^{-16}$ MeV, and note that the results are in s$^{-1}$. We have coded the black contours in the same way as in Fig.~\ref{fig:decay:local_rates}, and we have again indicated the central densities of J0348+0432, J1614--2230, and J0737--3039A/B within this EoS. The right panels, (c) and (d), contextualize these results by showing the neutron star mass, $M_{\rm NS}$, on the horizontal axis as a function of the central density on the vertical axis. Note that Figs.~\ref{fig:decay:local_rates} and \ref{fig:decay:integrated_rates} together imply that J0737--3039A/B are both too light to contain hyperons.

\section{Inferred Limits on Baryon Dark Decays}
\label{sec:dark_decay:med_limits_vacuum}
\begin{figure}[!t]
    \centering
    \includegraphics[width = \textwidth]{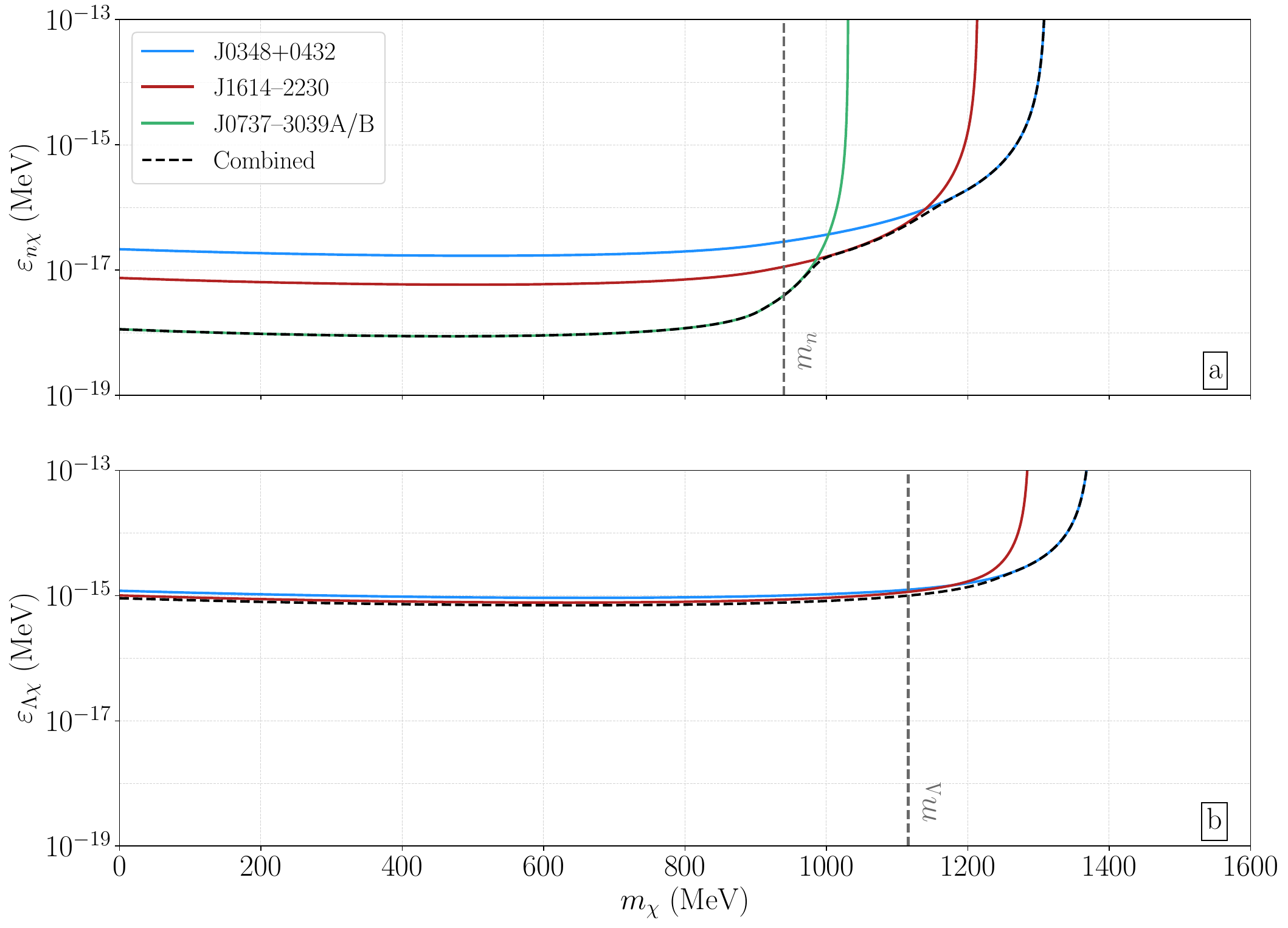}
    \caption{(Color Online) Exclusion constraints at 2$\sigma$ on the $n-\chi$ (upper panel) and $\Lambda-\chi$ (lower panel) mixing parameters as functions of $m_{\chi}$ for the DS(CMF)-1 EoS. The blue, red and green curves correspond to the $2\sigma$ limits derived from J0348+0432, J1614--2230, and J0737--3039A/B, respectively. The dashed black curves correspond to the combined limits, realized as per the discussion in the text. The vertical dashed lines indicate the in-vacuum baryon mass in each case.}
    \label{fig:combining_limits}
\end{figure}

We now turn to the task of assessing the limits on the ${\cal B}-\chi$ mixing parameters that emerge from our numerical assessment of the stellar-volume-integrated  baryon dark decay rates, as shown in Fig.~\ref{fig:chi_phot:rate_v_mchi}, and the macroscopic baryon number loss limits we have determined from astrophysical observations and their analysis. The latter, namely, are limits on anomalous binary-pulsar period lengthening, to which we refer as ``binary spin-down,'' and they are given in Table \ref{tab:macro_bnv:interp:psrbinary}. We show the limits we find for each astrophysical system as well as that associated with a final combined limit. To make our presentation more compact, we first discuss how the individual limits on $\varepsilon_{{\cal B}\chi}$ can be combined before showing all of these results. Note, too, that since our constraint depends on the square of $\varepsilon_{{\cal B}\chi}$ that its sign is left unconstrained --- we choose $\varepsilon_{{\cal B}\gamma} > 0$ in reporting our limits.

\subsection*{Combining Individual Limits}

Here we briefly describe our statistical procedure for combining limits on $\varepsilon_{{\cal B}\chi}$ derived from different pulsar binary 
systems. The limits we show have implicitly been determined as contours of constant $\chi^2(m_{\chi}, \, \varepsilon_{{\cal B}\chi})$. Our assumed-true hypothesis is that rate of BNV-induced binary spin-down vanishes in these systems, so we have $\chi^2 = 0$ for $\varepsilon_{{\cal B}\chi} = 0$. As such, each $\chi^2$ function is generically of the form
\begin{equation}
    \chi^2(m_{\chi}, \varepsilon_{{\cal B}\chi}) = \left(\frac{\dot{P}_b}{P_b}(m_{\chi}, \, \varepsilon_0) \right)^2 \frac{(\varepsilon_{{\cal B}\chi}/\varepsilon_0)^4}{\sigma^2} \equiv F(m_{\chi}) \times \varepsilon_{{\cal B}\chi}^4 \,.
\end{equation}
The first equality follows from the fact that $\dot{P}_b/P_b \propto \varepsilon_{{\cal B}\chi}^2$, noting Eq.~(\ref{eq:macro_bnv:observables:pbdot:bnv}), and we emphasize that $F$ is a function of $m_{\chi}$ only. The limits we have shown correspond to $\chi^2 = c$;\footnote{For two degrees of freedom, the $2\sigma$ exclusion curves we have shown correspond to $c=6.18$.} we call the resulting curve $\widetilde{\varepsilon}(m_{\chi})$. From this, we determine
\begin{equation}
    F(m_{\chi}) = \frac{c}{\widetilde{\varepsilon}(m_{\chi})^4};
\end{equation}
this allows to determine the $\chi^2$ function over the entire parameter space.

The combined limit, then, corresponds to the contour along which the sums of the individual $\chi^2$ functions also equals $c$. Using the definitions above, we determine the combined limit $\widetilde{\varepsilon}_{\rm comb}(m_{\chi})$ as follows:
\begin{eqnarray}
    & \chi^2_{\rm comb}(m_{\chi}, \varepsilon_{{\cal B}\chi}) = \sum_i F_i(m_{\chi}) \varepsilon_{{\cal B}\chi}^4 =\sum_i c \left(\dfrac{\varepsilon_{{\cal B}\chi}}{\widetilde{\varepsilon}_i(m_{\chi})}\right)^4 & \\
    & \chi^2_{\rm comb} \bigl(m_{\chi}, \widetilde{\varepsilon}_{\rm comb}(m_{\chi})\bigr) = c \implies \widetilde{\varepsilon}_{\rm comb}(m_{\chi}) = \Bigl( \sum_i \widetilde{\varepsilon}_i(m_{\chi})^{-4} \Bigr)^{-1/4} &
\end{eqnarray}
This discussion has assumed that all $\widetilde{\varepsilon}_i$ are defined at the same level $c$, and that the desired combined limit is also at $c$. This result can be generalized for distinct individual significances $c_i$ and combined significance $C$:
\begin{equation}
    \widetilde{\varepsilon}_{\rm comb} = \Bigl( \sum_i c_i/C \times \widetilde{\varepsilon}_i(m_{\chi})^{-4} \Bigr)^{-1/4}.
\end{equation}
We show our individual pulsar limits as well as our combined limits, realized via our described procedure, for the DS(CMF)-1 EoS in  Fig.~\ref{fig:combining_limits}.
\begin{figure}[!t]
    \centering
    \includegraphics[width = \textwidth]{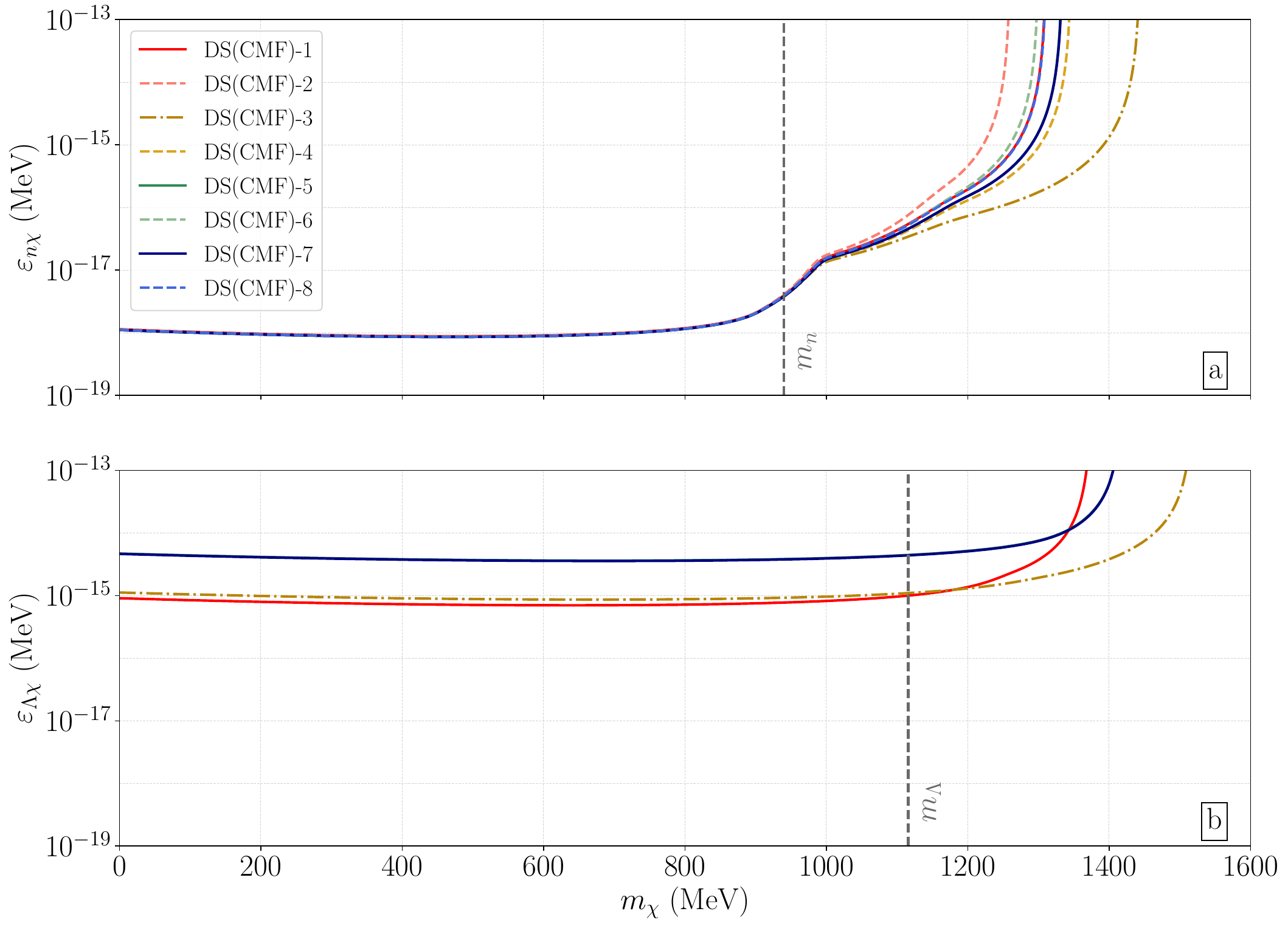}
    \caption{(Color Online) Combined exclusion constraints at 2$\sigma$ on the $n-\chi$ (upper panel) and $\Lambda-\chi$ (lower panel) mixing parameters as functions of $m_{\chi}$ for the eight equations of state in the DS(CMF) family.}
    \label{fig:DS(CMF)_eps_results}
\end{figure}

\bigskip

Figure \ref{fig:DS(CMF)_eps_results} depicts our results for the constraints on $\varepsilon_{n\chi}$ (upper panel) and $\varepsilon_{\Lambda\chi}$ (lower panel) as functions of $m_{\chi}$, calculated for each of the eight EoS in the DS(CMF) family. Equations of state that do not include hyperons are indicated with dashed curves in the upper panel. We also note that the DS(CMF)-3 EoS formally cannot support a neutron star with a mass of 2.01$M_\odot$ --- its maximum TOV mass is 2.00$M_\odot$ (note Table \ref{tab:medium:MFT:CMF:choices} in App.~\ref{app:CMF}). However, this is within 1$\sigma$ of the observed mass of J0348+0432; we therefore elect to include it in this figure, but instead show the constraint derived for this maximal neutron star. This constraint has been shown in dot-dashing to indicate that it is qualitatively different from the others.

We underscore that we have fixed the masses of these neutron stars to their best-fit values to construct these limits. A more statistically complete analysis would propagate the uncertainty in the inferred masses of the observed pulsars into the determinations of their central densities (within the context of a given EoS), and thus into the predicted baryon loss and binary spin-down rates. The mass uncertainties on J0348+0432 and J1614--2230 are $\mathcal{O}(10^{-2}) M_\odot$; we anticipate that there would be $\mathcal{O}(1)$ corrections to the limits whenever these are the only operative constraints, though the orders of magnitude are expected to be correct. That said, the uncertainties on the masses of J0737--3039A/B are $\mathcal{O}(10^{-5}) M_\odot$~\cite{PhysRevX.11.041050}, so that we expect the limits on $\varepsilon_{n\chi}$ for $m_{\chi} \lesssim 1000$ MeV, noting Fig.~\ref{fig:combining_limits}, to be quite robust.

\begin{figure}[!t]
    \centering
    \includegraphics[width = \linewidth]{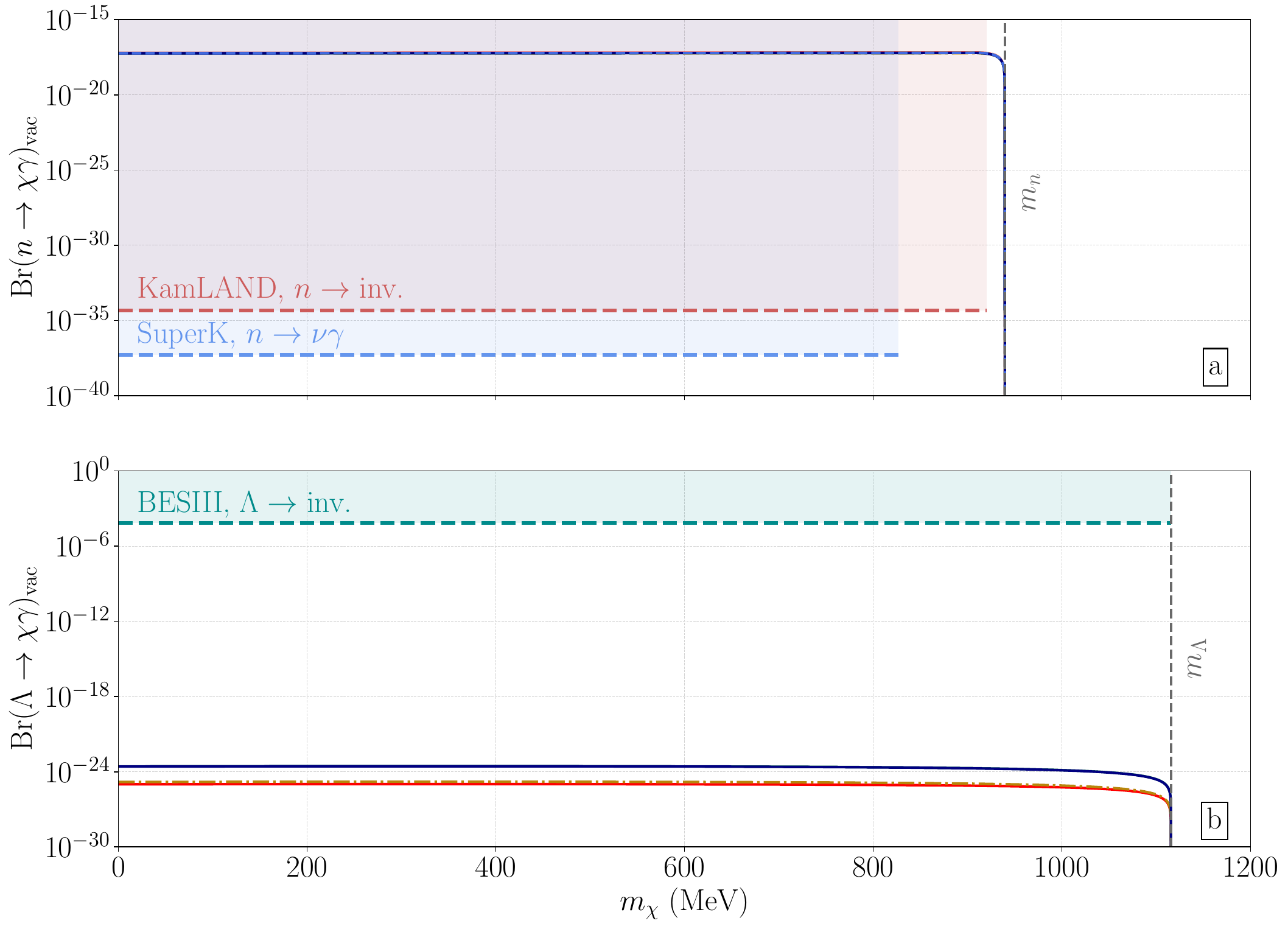}
    \caption{(Color Online) Exclusion limits at 2$\sigma$ on the vacuum branching fraction for ${\cal B}\to\chi\gamma$ for neutrons (upper panel) and $\Lambda$s (lower panel). The results for each EoS are color-coded as in Fig.~\ref{fig:DS(CMF)_eps_results}. We also show constraints from KamLAND~\cite{KamLAND:2005pen}, SuperKamiokande~\cite{Super-Kamiokande:2015pys}, and BESIII~\cite{BESIII:2021slv}.} 
    \label{fig:BR_limits_wide}
\end{figure}

In Fig.~\ref{fig:BR_limits_wide}, we reinterpret our constraints on $\varepsilon_{{\cal B}\chi}$ as constraints on the branching fractions for ${\cal B} \to \chi \gamma$ in vacuum and contrast them against laboratory constraints, with neutrons ($\Lambda$s) in the upper (lower) panel. For neutrons, we also show the KamLAND constraint on invisible neutron decay~\cite{KamLAND:2005pen} in red and the SuperKamiokande constraint on $n\to\nu\gamma$~\cite{Super-Kamiokande:2015pys} in light blue. 
(We note that the KamLAND constraint~\cite{KamLAND:2005pen} is also pertinent to our study of $n\to \chi\gamma$ decay because 
the prompt photon would not pass the correlation cuts and would 
remain undetected.)
These are as much as twenty orders of magnitude stronger than the constraints we have derived, but we note that these are only operative up to $m_{\chi}$ = 920 and 827 MeV, respectively, as a result of experimental cuts.
We emphasize, in particular, that these experiments cannot probe the region $m_{\chi} > m_n$; while they are more powerful when they are operative, they are fundamentally constrained in ways that astrophysical probes of new physics are not. For $\Lambda$s, we show the constraint on invisible decays from BESIII~\cite{BESIII:2021slv} in dark cyan. In this case, we find the opposite result: pulsar binaries are able to probe this branching ratio as much as twenty orders of magnitude more severely than laboratory constraints! The caveat is that this requires hyperons to appear in neutron stars, which is still a matter of debate, simply because EoSs without hyperons exist that confront current observational data successfully. However, if hyperons appear in an appreciable amount in these objects, then one can expect vast improvements on laboratory searches.

\begin{figure}[!t]
    \centering
    \includegraphics[width=\linewidth]{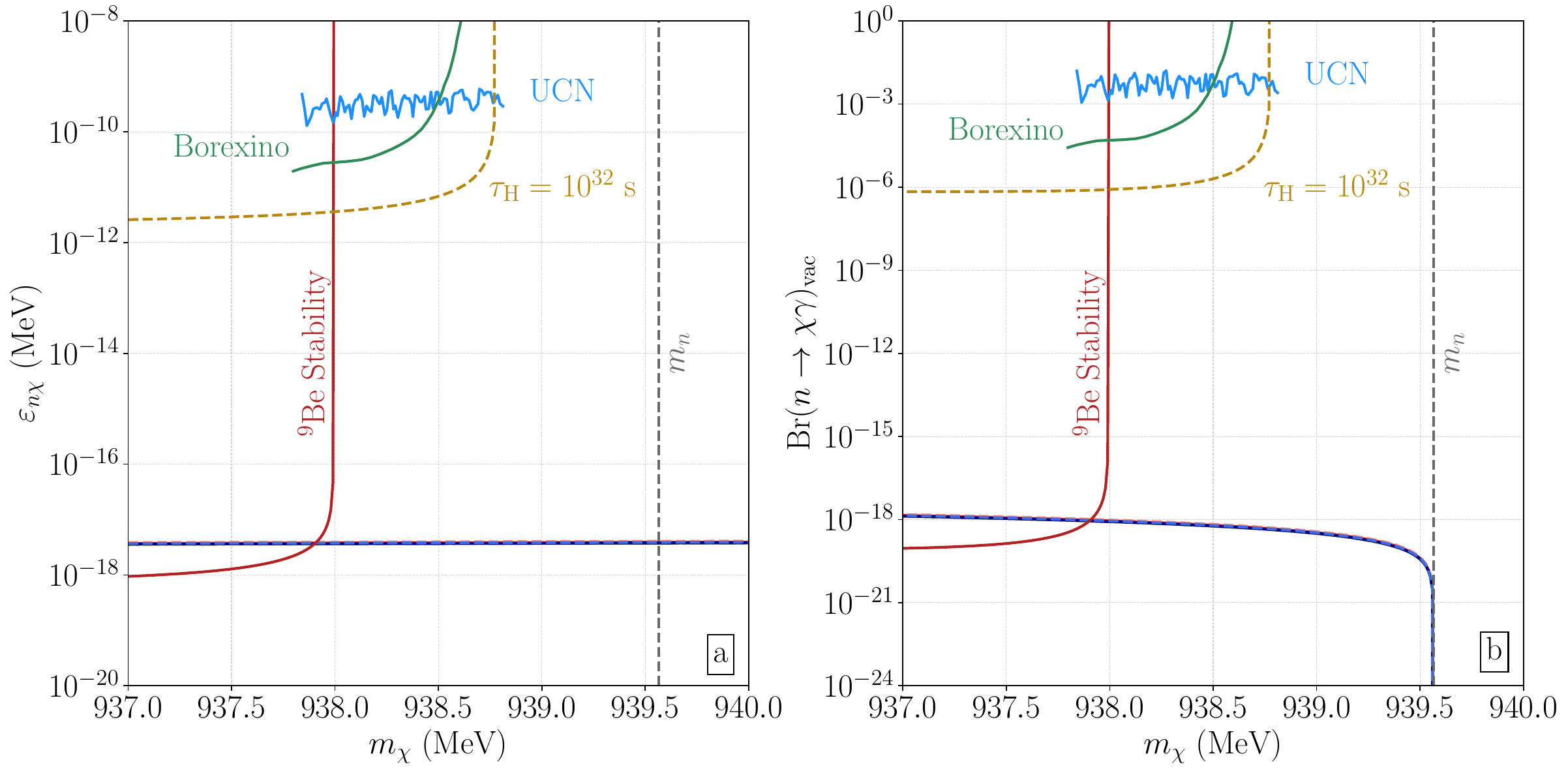}
    \caption{(Color Online) Exclusion limits at 2$\sigma$ on 
    (a) $\varepsilon_{n\chi}$ and (b) the vacuum branching fraction for $n \to \chi \gamma$ as per  Fig.~\ref{fig:BR_limits_wide}, in the particular $\chi$ mass region pertinent to an explanation of the neutron lifetime anomaly. Additional constraints and expected limits have been included as detailed in the text, after Ref.~\cite{McKeen:2020zni}.}
    \label{fig:zoomed_in_limits}
\end{figure}

The upper panel of Fig.~\ref{fig:BR_limits_wide} is incomplete in that there are additional constraints around $m_{\chi} \approx m_n$, a region that has become of interest in recent years as a result of tests of new-physics explanations~\cite{Fornal2018PhRvL.120s1801F} of the neutron lifetime anomaly~\cite{Wietfeldt2011RvMP...83.1173W}. We examine this region more closely in Fig.~\ref{fig:zoomed_in_limits}; panel (a) casts these searches in terms of constraints on $\varepsilon_{n\chi}$, while panel (b) casts them in terms of constraints on Br$(n\to\chi\gamma)$. We show in blue the estimated constraint from a direct search for $n \to\chi\gamma$ using ultra-cold neutrons (UCN)~\cite{Tang:2018eln}, and in green we show a constraint from Borexino from searches for hydrogen decay, both from Ref.~\cite{McKeen:2020zni}. We also show the curve along which the free hydrogen lifetime is supposed to be $\tau_H = 10^{32}$ s in dashed gold, also from Ref.~\cite{McKeen:2020zni}. (The constraints from Ref.~\cite{McKeen:2020zni} are reported at 90\% CL, though the differences between those and limits at $2\sigma$ should be very small given the ranges shown in the figure.) Clearly, neutron stars are more sensitive to these decays than these (would-be) laboratory constraints by many orders of magnitude.

It was noted in Ref.~\cite{Fornal2018PhRvL.120s1801F} that the existence of $\chi$ can destabilize nuclear matter, including $^9$Be. This constraint was calculated more precisely in Ref.~\cite{McKeen:2020oyr}, assuming that the lifetime of $^9$Be is longer than $3\times10^9$ years to account for the presence of $^9$Be in old, metal-poor stars \cite{Rich:2009gj}. This constraint is shown in red in Fig.~\ref{fig:zoomed_in_limits} and is competitive with (if not dominant to) our neutron star constraints in the region of its operation, $m_{\chi} < 937.993$ MeV. We note that other probes of dark decays of nuclei with low neutron separation energies have been discussed in, e.g., Ref.~\cite{Pfutzner:2018ieu}. Particular attention has been paid to decays of $^{11}$Be, with experimental efforts underway at CERN-ISOLDE \cite{Lopez-Saavedra:2022vxh} and ISAC-TRIUMF \cite{Ayyad:2019kna}, though we are unaware of any efforts to interpret these experimental results as constraints on new physics. As a side note, it is curious that there are no laboratory constraints, as far as we can tell, on the lifetime of $^9$Be. We find the arguments about the presence of $^9$Be in old stars compelling and agree that this is a valid constraint, but we are surprised, frankly, that the lifetime is only constrained at the billion-year scale. While experimentalists of yore would have had little reason to interrogate the stability of $^9$Be -- or indeed, any species thought to be stable in the SM -- we regard the observation that the stability of these systems has not been tested in a detailed way in the laboratory as a potentially promising avenue for constraining new physics.

We conclude by noting that Ref.~\cite{McKeen:2020oyr} has also presented constraints on $n \to \chi\gamma$ from cosmology and from neutron star cooling. The former is a combination of constraints coming from modifications to Big Bang Nucleosynthesis (BBN) and the Cosmic Microwave Background (CMB); this treatment includes the reverse decay $\chi\to n \gamma$ when $m_{\chi} > m_n$, and so constrains the region shown. However, in their calculations, $\chi$ is assumed to constitute (at least some of) the dark matter. This is unlike our framework, in which we introduced more new states ($\xi$ and $\phi_B$) to prevent overaccumulation of $\chi$. Therefore, the limits they derive from BBN and CMB do not apply here, though we agree that this would be an interesting and important avenue to explore. 
The neutron star cooling constraint derived there makes very rough assumptions about how heat from decays is deposited into the neutron star, with the implicit assumption that increases in the temperature of the core of the neutron star lead to commensurate increases in the observed effective temperature. A more sophisticated analysis appears in Ref.~\cite{McKeen:2021jbh}, though in the particular case of $n-n'$ mixing~\cite{Berezhiani:2018udo,Goldman:2022brt,Goldman:2022rth}. We note that the response of the star is sensitive to the precise connections between visible and dark sectors, as shown explicitly in Ref.~\cite{Goldman:2022rth}; here we focus on $n\to\chi \gamma$ with subsequent $\chi$ decay, removing energy from the star, though 
Fermi heating from the neutron decay is also present. The thermal transport and cooling in neutron stars demands careful investigation; for instance, BNV decays lead to $\beta$-disequilibrium, which leads to neutrino cooling via (direct and modified) Urca processes, which impact how the energy released in the decays is deposited back into the SM fluid. We cannot say at this time whether a constraint from thermal heating is more stringent or not.
While we agree that old, cold neutron stars should constrain this model, the details are intricate and dark process dependent and we decline to include such constraints here.
\section{Implications for Models of Baryogenesis and Dark Matter} 
\label{sec:dark_decay:baryogen_implications}
The prospect of explaining the origins of both the dark matter abundance and the cosmic baryon asymmetry within a single dynamical framework is a beguiling one. Different possibilities have existed for some time, and many share a common feature: there is a dark-sector baryon that carries baryon number and into which SM baryons can decay. A particularly intriguing variant is that of 
$B$-mesogenesis~\cite{Elor:2018twp,Alonso-Alvarez:2021qfd,Elor:2022jxy}. It proceeds in the early universe from late time, out of equilibrium production of $B$ mesons (with equal fractions of $b$ and ${\bar b}$ quarks) that evolve under SM CP-violating processes before decaying to a SM baryon and a dark fermion carrying the opposite sign of baryon number. Thus: no new sources of CP violation of the SM are required; the baryon number of the universe is conserved --- it is just sequestered into visible and dark sectors with opposite baryon number; and it occurs late in the history of the universe in that occurs after the QCD phase transition, making it possible to realize hadronic states, and before the epoch of big-bang nucleosynthesis. Finally, it is an example of a testable mechanism of baryogenesis~\cite{Barrow:2022gsu}, in that its essential features are subject to direct experimental investigation. Particularly, its reliance on the SM mechanism of CP violation (albeit new CPV sources could enter) implies that the branching ratios of $B$ mesons in SM baryons and the dark fermion (antibaryon) cannot be too small, with the expectation that the branching fractions can roughly be no less than ${\rm Br}(B_{s,d}^0 \to {\bar \chi} {\cal B}) \gtrsim 10^{-5}$ or ${\rm Br}(B^+ \to {\bar \chi} {\cal B}^{(+)}) \gtrsim 10^{-6}$~\cite{Elor:2022jxy}, which is compatible with the expectation ${\rm Br}(B_{s,d}^0 \to {\bar\chi} {\cal B} {\cal M})\gtrsim 10^{-4}$ for meson ${\cal M}$ due to the latter's larger phase space~\cite{Alonso-Alvarez:2021qfd}. The expected theoretical window in $\chi$ mass is $0.94 \,\rm GeV < m_{\chi} < 4.34 \,\rm GeV$~\cite{Alonso-Alvarez:2021qfd}, which is larger than what we provide in Eq.~(\ref{eq:chimasswindow}), because in their dark-sector model 
$\xi$ and $\phi_B$ have a limited mass difference to avoid washout of the produced baryon asymmetry. Studies from Belle~\cite{Belle:2021gmc} and BaBar~\cite{BaBar:2023rer} limit the available parameter space for ${\bar\chi} ubs$ couplings in the mass region of $1-4.4\,\rm GeV$, and it is anticipated that the remaining parameter space can be probed at Belle-II~\cite{BaBar:2023rer}. This model is particularly close to the model we study.
In this paper we have established severe limits on the $\varepsilon_{n\chi}$ and $\varepsilon_{\Lambda\chi}$ mixing parameters for $\chi$ masses satisfying $m_{\chi} \lesssim 1400\,\rm MeV$, as shown in Fig.~\ref{fig:DS(CMF)_eps_results}. In this mass region and for the regions of hidden-sector parameter space we have chosen, our limits constrain the flavor structure of models of $B$-mesogenesis, and we now turn to those and their implications.

Different UV completions of $B$-mesogenesis models fare differently in light of our constraints. Here we consider versions in which only one extra particle is needed. For example, in Ref.~\cite{Elor:2018twp}, a color-triplet, SU(2)$_{\rm L}$ singlet scalar with the SM quantum numbers $(3,1,-1/3)$ is used, though a scalar of form $(3,1,+2/3)$~\cite{Alonso-Alvarez:2021qfd} or a vector of form $(3,2,-1/6)$~\cite{Alonso-Alvarez:2021oaj} are noted alternatives. We do not consider this list exhaustive. The two scalars are just the leptoquarks we have noted in Sec.~\ref{sec:dark_decay:models}: $S_1^\ast$ and $\bar{S}_1^*$~\cite{Fornal2018PhRvL.120s1801F,Fajfer:2020tqf}. The phenomenology of these specific models has been studied, and in order to explain the baryon asymmetry, the dark matter abundance, and all empirical constraints, including those on $|\Delta F|=2$ meson mixing, a rich flavor pattern of couplings to quarks is needed~\cite{Alonso-Alvarez:2021qfd}.

To determine the implications of our constraints, we first note the structure of the Lagrangian for each UV completion, following Ref.~\cite{Alonso-Alvarez:2021oaj}, though we write our 2-spinors as in Ref.~\cite{Gardner:2018azu} and employ the conventions given there. Denoting the new scalars as $Y_{\rm Y}$ and the new vector as $X^\mu$, we have 
\begin{eqnarray}
&& {\cal L}_{Y_{\frac{2}{3}}} \supset - y_{d_a d_b} \epsilon_{\alpha\beta\gamma} Y^{\alpha}_{\frac{2}{3}} d^{\beta}_a d^{\gamma}_b 
- y_{\chi u_c} Y^{\alpha\,*}_{\frac{2}{3}} \chi^c
u^{\alpha}_c + \rm h.c. \,, 
\label{model23}\\
&& {\cal L}_{Y_{\!{-\frac{1}{3}}}} \supset - y_{u_a d_b} \epsilon_{\alpha\beta\gamma} Y^{\alpha}_{-\frac{1}{3}} u^{\beta}_a d^{\gamma}_b 
- y_{\chi d_c} Y^{\alpha\,*}_{-\frac{1}{3}} \chi^c d^{\alpha}_c 
- y_{Q_a Q_b} \epsilon_{\alpha\beta\gamma} Y^{\alpha}_{-\frac{1}{3}} \left( Q^{\beta}_a \varepsilon Q^{\gamma}_b  \right)
+ \rm h.c. \,, 
\label{model13} \\
&&{\cal L}_{X} \supset - y_{Q_a d_b} \epsilon_{\alpha\beta\gamma} \left( X^{\alpha}_{\mu} \varepsilon Q^{\beta}_a \right) \sigma^{\mu} d^{\gamma}_b 
- y_{\chi Q_c} \left( X^{\dagger\, \alpha}_{\mu} Q^{\alpha}_c \right) \sigma^{\mu} \chi^c 
+ \rm h.c. \,,
\label{modelX}
\end{eqnarray}
where $\varepsilon$ is an antisymmetric tensor in the two-spinor indices and $\chi^c$, noting Eq.~(\ref{eq:quarkchi_nochiral}) and its accompanying footnote ($\#4$), is a right-handed field. With the ${\cal B}$ assignments of $-2/3$ for the scalars $Y_{\frac{2}{3}}$ and $Y_{-\frac{1}{3}}$ and ${\cal B}=1$ for $\chi$, the noted interactions conserve baryon number. In Refs.~\cite{Alonso-Alvarez:2021qfd,Fajfer:2020tqf} $y_{Q_a Q_b}$ (for each $a,b$) is taken to be zero. The color structure of the first term of Eq.~(\ref{model23}) requires that the product of $d$-like quarks be antisymmetric in the generation indices $a,b$, which follows because we have assumed the scalar is a color triplet. As for the last case, the vector $X^\mu$ can be written in two-spinor form as~\cite{Alonso-Alvarez:2021oaj}
\begin{equation}
    X^\mu = \left( \stackrel{{Y^\mu_{\frac{2}{3}}}}{{}_{Y^\mu_{\!{-\frac{1}{3}}}}} \right)
\end{equation}
and thus through Eq.~(\ref{modelX}) we see that both scalars couple to left-handed quarks. We have defined our scalar-fermion couplings in the flavor basis, rather than the mass basis, but in the case of couplings to right-handed quarks no distinction needs be made. However, in the case of couplings to left-handed quarks we need to rotate the fields to the mass basis, to parallel the treatment of the charged weak current in the SM. As a result, a flavor diagonal coupling to a left-handed quark of a single flavor can engender a contribution to a flavor-changing neutral current (FCNC). In the example of $Z'$ models, satisfying FCNC constraints with a large $Z'$ coupling requires nearly flavor-universal couplings~\cite{Dobrescu:2014fca}, where we note that in the flavor universal limit the unitary structure of the Cabibbo-Kobayashi-Maskawa (CKM) matrix makes the FCNC couplings vanish. We will see that this effect does not appear here because our scalars do not ever couple to two left-handed quarks of the same flavor. Replacing a left-handed flavor state $d^i$ with a combination of mass states via $V_{ij} d_{j}$, with $V$ the CKM matrix, we see that the $X^\mu$ completion does lead to a FCNC of form~\cite{Alonso-Alvarez:2021oaj}
\begin{equation}
    {\cal L}_{X\,;\,\rm FCNC} \supset - y_{Q_a d_b} Y_{\frac{2}{3}\,\mu} V_{a a'}\bar{d}^c_{a'}
    \gamma^\mu P_R d_b \,,
\end{equation}
where we have employed 4-component notation. This interaction engenders not only $|\Delta F|=2$ meson-mixing but also structures 
such as $B_{(s)}\to {\bar K}$ or $B_{(s)} \to \pi^0$ at tree level, which can be probed through $B$ decay studies. We also see explicitly that the structure of the vertex does not require a flavor universal coupling to control the size of the effect. Thus there are no particular flavor conspiracies in satisfying the $|\Delta F|=2$ constraints, and to determine the impact of the constraints we have found on the mixing parameters $\varepsilon_{n\chi}$ and $\varepsilon_{\Lambda \gamma}$ on these models, it suffices to consider the contributions to these quantities from the scalar-fermion couplings within a particular UV complete model.

\begin{figure}[htb]
    \centering
    \subfigure[]{\includegraphics[width=0.496\linewidth]{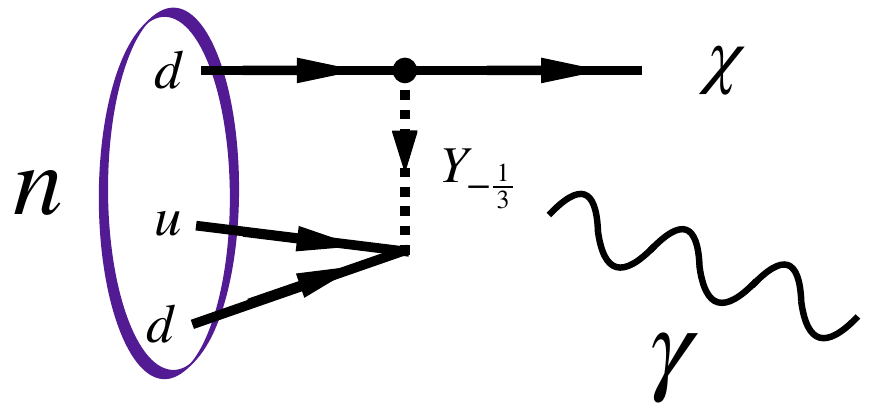}}
 \subfigure[]{\includegraphics[width=0.496\linewidth]{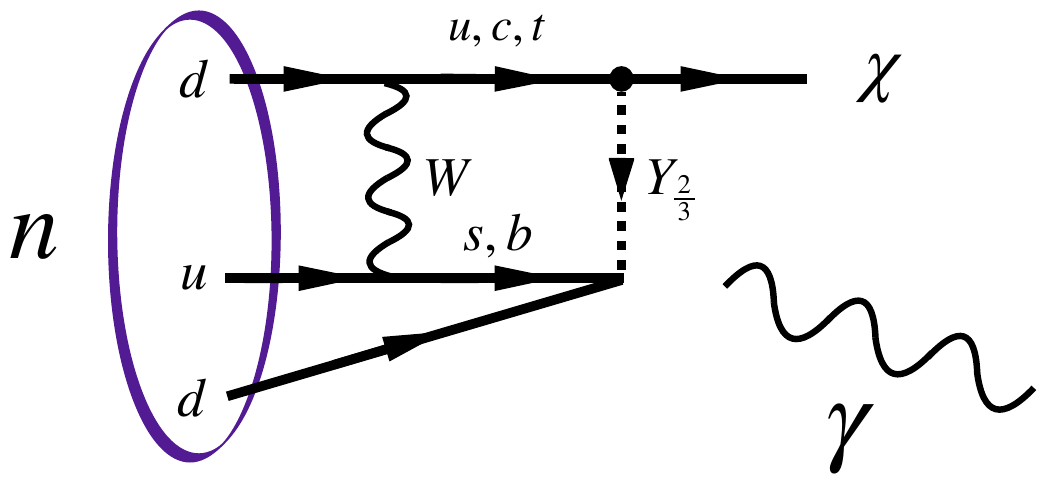}}
     \subfigure[]{\includegraphics[width=0.496\linewidth]{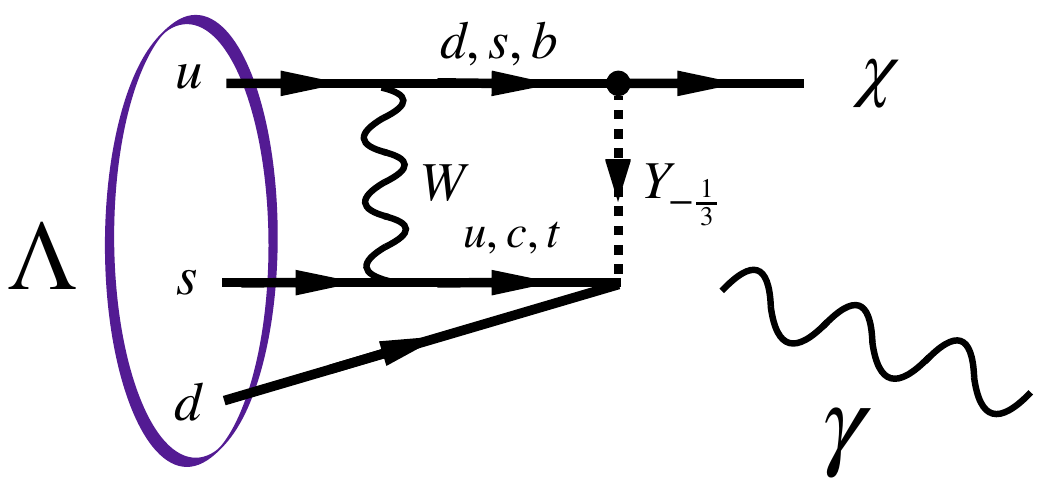}}
 \subfigure[]{\includegraphics[width=0.496\linewidth]{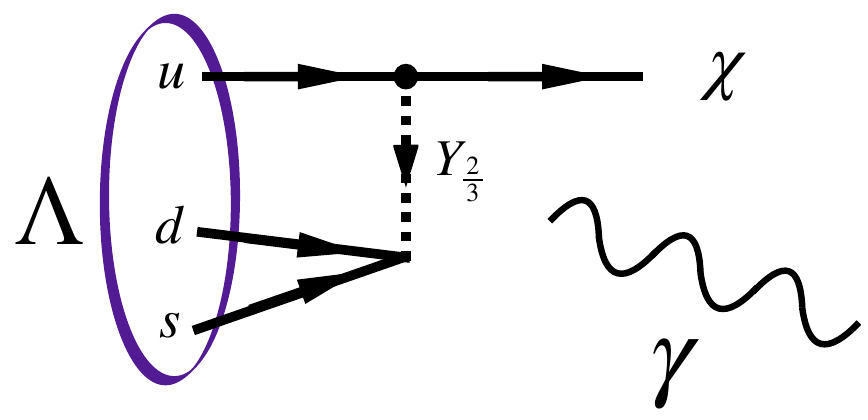}}
    \caption{ (Color Online) Feynman diagrams for $n\to \chi\gamma$ and $\Lambda\to \chi\gamma$ decays as mediated by the baryon-number-carrying scalars $Y_{\frac{2}{3}}$ and $Y_{-\frac{1}{3}}$ as discussed in text, after Ref.~\cite{Fajfer:2020tqf}.}
    \label{fig:nLchigammaY}
\end{figure}

Considering, then, the flavor structure of the couplings in Eq.~(\ref{model23}) we see that $n\to \chi\gamma$ cannot occur at tree level (via valence quarks), and a loop graph with $W$ and $Y_{\rm Y}$ exchange is needed to generate the process~\cite{Fajfer:2020tqf}. The opposite situation is true for $\Lambda\to \chi\gamma$, with Eq.~(\ref{model23}) and Eq.~(\ref{model13}) yielding that process at tree level and one-loop level, respectively. Pertinent Feynman diagrams are illustrated in Fig.~\ref{fig:nLchigammaY}, replacing the illustration of Fig.(\ref{fig:nchigamma}). Noting Eqs.~(\ref{eq:nchimix}) and (\ref{eq:dark_decay:vacuum:vacdec}), it is apparent that the mixing parameters $\varepsilon_{n\chi}$ and $\varepsilon_{\Lambda\chi}$ depend very differently on the underlying scalar-fermion couplings in the two cases --- we refer to Ref.~\cite{Fajfer:2020tqf} for explicit expressions. In particular, the one-loop diagrams bring in a coupling to the $b$ quark as well, in the combinations 
\begin{equation}
    y_{db} y_{\chi u} \,; \quad 
     y_{db} y_{\chi c} \,;\quad 
 y_{db} y_{\chi t} \,,
\end{equation} 
each of which could saturate the bound we have found for $\varepsilon_{n\chi}$. 
Turning to $\Lambda \to \chi\gamma$ decay, as shown in Fig.~\ref{fig:nLchigammaY}(d), we note, in  
contrast, that this process yields a constraint on $y_{sd}y_{\chi u}$. However, 
additional couplings enter at one-loop level, to
which two graphs contribute. Their form follows  from replacing one or the other of the $d$ quarks in the graph of Fig.~\ref{fig:nLchigammaY}(b) with an $s$ quark. The one-loop graphs also probe $b$-quark combinations, 
specifically 
%
\begin{eqnarray}
     y_{sb} y_{\chi u} \,;\quad 
     y_{sb} y_{\chi c} \,;\quad 
     y_{sb} y_{\chi t} 
    \label{eq:andflavor23} \,.
\end{eqnarray}

Turning to the model in Eq.~(\ref{model13}), 
we see, in contrast, that $n\to \chi\gamma$ does occur at tree level, and a loop graph with $W$ and $Y_{\rm Y}$ exchange is needed to generate $\Lambda\to \chi\gamma$ decay, with an example 
illustrated in Fig.~\ref{fig:nLchigammaY}. If 
we replace the $s$ quark in the initial state with a $d$
quark, we have a contribution to $n\to\chi\gamma$ decay as well. Consequently, our limit on $\varepsilon_{n\chi}$ 
constrains 
the $b$-quark 
combinations 
\begin{eqnarray}
     y_{td} y_{\chi b} \,;\quad 
     y_{cd} y_{\chi b} \,;\quad 
     y_{ud} y_{\chi b}     \,.
\end{eqnarray}
appearing from the loop graph. In contrast, 
our limit on $\varepsilon_{\Lambda\chi}$ constrains the $b$-quark couplings
\begin{eqnarray}
     y_{ts} y_{\chi b} \,;\quad 
     y_{cs} y_{\chi b} \,;\quad 
     y_{us} y_{\chi b}     \,.
\end{eqnarray}

In regards to the mechanism of $B$-mesogenesis, operators with the flavor combinations $\chi b u d$, $\chi b us$, $\chi b cd$, and $\chi bcs$ are pertinent, and they take one of three forms~\cite{Alonso-Alvarez:2021qfd}
\begin{equation}
{\cal \theta}^{(1)}_{ij} = (\chi b) (u_i d_j) \,, \quad
{\cal \theta}^{(2)}_{ij} = (\chi d_j) (u_i b) \,, \quad 
{\cal \theta}^{(3)}_{ij} = (\chi u_i) (d_j b) \,,
\end{equation}
where $i \in u,c$, $j \in d,s$, and the colors have been contracted to form a color singlet in each case. We see that the $Y_{2/3}$ scalar is uniquely 
associated with the ${\cal \theta}^{(3)}_{ij}$ operators, and the 
$Y_{1/3}$ scalar is associated with the other two operators. 
A successful description of the baryon asymmetry that also satisfies 
the various,
existing experimental constraints 
implies non-trivial flavor structure 
in the $y^2$ couplings~\cite{Alonso-Alvarez:2021oaj,Alonso-Alvarez:2021qfd}. 
Noting the 
suggested flavor-structure 
solutions of Fig.~12 in Ref.~\cite{Alonso-Alvarez:2021qfd} with $m_Y = 1.5\, \rm TeV$, 
we see that their $Y_{\frac{2}{3}}$ solution 
would require $y_{sb}y_{\chi u}$ 
to be of ${\cal O}(1)$ (and $y_{db}y_{\chi u}$ to be of ${\cal O}(10^{-3})$), 
but their $Y_{-\frac{1}{3}}$ solution would 
require $y_{cb}y_{\chi s}$ or possibly 
$y_{tb}y_{\chi s}$ to be of ${\cal O}(1)$. Since we cannot 
constrain either of these latter two 
combinations, we see that our studies can constrain the
$Y_{\frac{2}{3}}$ scalar scenario more severely, 
though we defer 
a detailed numerical analysis of our constraints on the flavor structure
of all of the $y^2$ couplings in both scalar scenarios to a subsequent paper. 
\section{Summary}
\label{sec:summary}
BNV has not yet been observed in terrestrial experiments, and its deep ties to explanations of the observationally well-established cosmic baryon asymmetry~\cite{Sakharov:1967dj} argue persuasively for its investigation on broader fronts. Previously, we have considered how it might eventually be discovered through precision measurements of neutron star observables, particularly those of changes in the binary-pulsar period, familiar from tests of general relativity~\cite{Berryman:2022zic}. Thus far we have found limits, and they are macroscopic ones, in that they emerge from the consideration of a neutron star as a whole. Such constraints miss a concrete connection to particle physics, and it is badly needed: regardless of whether we continue to constrain or, finally, discern the existence of BNV (in contradistinction to a failure of general relativity) from these studies, further theoretical progress on the problem of BNV requires constraints on the particle physics models of BNV themselves. In this paper, we have developed just such a connection, using a concrete description of the neutron star interior based on a relativistic mean-field theory in hadronic degrees of freedom~\cite{Walecka:1974qa,Serot:1984ey,Serot:1997xg} that successfully confronts existing macroscopic properties of neutron stars~\cite{Dexheimer:2008ax}. Within this context, we have developed how to assess the rates for BNV particle processes in dense matter, and we present explicit rates for benchmark processes, particularly ${\cal B} \to \chi\gamma$, considering its rate both at local points within a neutron star as well as its volume rate after integration over the structure of the entire star, up to its crust. Although our in-medium formalism is germane to the evaluation of any particle process in the dense medium of a neutron star, the focus of this paper --- noting current sensitivities --- is that of apparent BNV through baryon decays to hidden-sector particles. Finally, with this in place, we match the computed rate to our inferred limits on anomalous binary-period lengthening, i.e., how the binary itself spins down, to set one-sided limits at 2$\sigma$ on the mixing parameters $\varepsilon_{{\cal B}\chi}$, for individual binary-pulsar systems, as well as a combined limit for all of the studied systems. 

As a result of these studies, we discover that neutron stars open new windows on the study of BNV, probing $m_{\chi}$ parameter space not accessible to terrestrial nucleon decay experiments, due to experimental limitations in the detection of a final-state photon. More than this, the dense nuclear medium admits the study of regions for which $m_{\chi}$ exceeds the vacuum mass of the nucleon, as well as the possibility of probing strange baryon decays. Our final limits are reported in Figs.~\ref{fig:BR_limits_wide} and \ref{fig:zoomed_in_limits}. We observe that in the regions of parameter space to which proton decay (nuclear stability) experiments are sensitive~\cite{KamLAND:2005pen,Super-Kamiokande:2015pys}, they exceed the limits we set by nearly twenty orders of magnitude. In contrast, however, our neutron star limits exceed the sensitivity of those from terrestrial $\Lambda$ and neutron $\beta$-decay experiments by a comparably large amount. Let us emphasize that our limits are likely upper bounds, and hence are conservative, in that they are determined by the electromagnetic decay ${\cal B}\to {\chi}\gamma$ alone, although the particle physics models we study do admit the possibility of ${\cal B}\to \chi \,+\, \rm meson(s) $ decays as well. This latter set of decays has no reason to be negligible compared to the electromagnetic decays in rate --- and we note Ref.~\cite{Alonso-Alvarez:2021oaj} for specific examples computed within (in-vacuum) chiral EFT~\cite{Claudson:1981gh}. As a result, we would expect larger ${\cal B}$ decay rates for fixed $\varepsilon_{{\cal B\chi}}$, but the challenges in realizing a suitable theoretical assessment of the hadronic channels prompt the conservative approach we have espoused in this paper. 

We now turn to an assessment of the limitations in our approach. One key question concerns the largest value of $\varepsilon_{{\cal B}\chi}$, $\varepsilon_{{\cal B}\chi}^{\rm max}$, we can possibly limit with our formalism, in which the SM drives the dynamical response of the neutron star to BNV. (In our work, dark-sector interactions drive the removal of $\chi$, so that the neutron star survival constraints on the mass of $m_{\chi}$ noted in Refs.~\cite{McKeen:2018xwc,Motta:2018rxp,Baym2018PhRvL.121f1801B} do not operate.) We believe a realistic assessment of $\varepsilon_{{\cal B}\chi}^{\rm max}$ requires a study of neutron star heating from relatively fast rates of BNV, the complexities of which lie beyond the scope of this paper. We note, however, the outcomes of terrestrial neutron $\beta$-decay searches~\cite{Tan:2019mrj}, shown in Fig.~\ref{fig:zoomed_in_limits}, as well as limits arising from constraints due to the charged-current structure of the SM~\cite{Czarnecki:2019mwq}, noted in Eq.~(\ref{eq:nexoticCC}). Since $n \to \chi \gamma$ does not derive from a SM weak process in any way, a ${\rm Br}(n\to \chi\gamma)$ limit of ${\cal O}(10^{-3})$ implies a limit on $\varepsilon_{n\chi}$ of ${\cal O}(10^{-9})$! Thus we think these limits are severe enough that determining $\varepsilon_{{\cal B}\chi}^{\rm max}$ precisely is not an immediate concern, but, rather, an important topic for future investigation. 

Another potential limitation may be our use of a relativistic mean-field theory framework~\cite{Walecka:1974qa,Serot:1984ey,Serot:1997xg} in which to describe the nuclear medium within a neutron star. This approach is computationally tractable and readily allows for the treatment of more sophisticated models of the nucleon-nucleon interaction than those in which it was first devised. We have employed the chiral SU(3) hadronic model of Refs.~\cite{Dexheimer:2008ax,dexheimer_2017,Dexheimer:2020rlp} in this paper. This is admittedly a model that is not QCD, and our ability to assess the errors predicated by this choice is rather limited. We have, however, studied how our results change within a family of EoSs, namely DS(CMF)~1-8 EoSs~\cite{compose_CMF1,compose_CMF8}, to which it can be connected. Moreover, frankly, there is no other alternative for the treatment of dense nuclear matter, though this may ultimately change~\cite{Drischler:2019xuo}. We note that the use of chiral effective theory has been championed in this regard~\cite{Drischler:2021kxf}, but its applicability does not stretch much beyond that of nuclear saturation density. In the future, it may be advantageous to consider EoSs that blend the chiral effective field theory and relativistic mean-field theory approaches~\cite{Alford:2022bpp}. Nevertheless, given our interest in order-of-magnitude estimates, we believe that our choice is also reasonably realistic. 

Different paths beckon as opportunities for future work. We believe that studies of neutron star heating from BNV is important not only to discerning the limits of our existing formalism, but also, crucially, to interpreting what a significant observation of anomalous binary spin down might mean. It strikes us that theoretical heating studies and concomitant observational studies of neutron star cooling may be the only tangible way to tell a failure of general relativity, in some undetermined way, from BNV. 
These can be complex. 
Generally speaking, the star will respond to perturbations placed on it in ways that need not be adiabatic. There can be, for example, structural changes in the neutron star and work done by gravity and pressure on the fluid as it readjusts itself. There are also other reactions that occur in response to the removal of a neutron which contribute to extra heating and cooling. For example, although thermal energy of $O(E_F)$ is dumped in the star once a neutron is removed, through a dark decay or oscillation process, this does not characterize a net gain of thermal energy by the system. For that we need to include the cooling from Urca processes, which by assumption are faster than our neutron disappearance (apparent BNV) rates, because 
the thermal evolution of neutron star is determined by the total rate of heating or cooling.

As for other possibilities, we could consider how our results could change if the neutron star were a hybrid star, containing a quark core~\cite{Dexheimer:2020rlp}, or how viable models with a significant $\chi$ admixture in the neutron star (albeit constrained by Eq.~(\ref{eq:nexoticCC})~\cite{Czarnecki:2019mwq}), such as that of Ref.~\cite{Strumia:2021ybk}, could be addressed through modifications of our formalism. As for future terrestrial experiments that could complement the studies of this paper, it strikes us that empirical studies of the lifetime of SM-stable composites, such as atomic hydrogen, or of the $^9$Be nucleus, could yield fruitful results. 

\begin{acknowledgments}
We thank the Network for Neutrinos, Nuclear Astrophysics, and Symmetries (\href{https://n3as.berkeley.edu}{N3AS}) for an inspiring environment. J.M.B.~acknowledges support from the National Science Foundation, Grant PHY-1630782, the Heising-Simons Foundation, Grant 2017-228 and the U.S. Department of Energy Office of Science under award number DE-SC00018327. J.M.B.~also thanks the Institute for Nuclear Theory (INT) at the University of Washington for its kind hospitality and stimulating research environment. This research was supported in part by the INT’s U.S.~Department of Energy grant No. DE-FG02-00ER41132. S.G. and M.Z. acknowledge partial support from the U.S. Department of Energy Office of Science under contract DE-FG02-96ER40989. S.G. thanks N3AS and the Theoretical Division at Fermilab for gracious hospitality and lively environments 
and acknowledges support from N3AS and the Universities Research Association during sabbatical visits while completing this paper. 
\end{acknowledgments}

\appendix
\section{Dark Baryon Removal in a Neutron Star}
\label{app:dark_removal}

In this section we consider two scenarios for the removal of $\chi$
from the neutron star, so that the condition $n_{\chi}(r)\ll n(r)$ can be satisfied: one in which the annihilation rate is much faster than the self-interactions which help establish a thermal equilibrium, and another in which self-interactions of $\chi$ are much faster than its annihilation rate.

We first consider the scenario in which dark particles have a non-thermal distribution at the time of their annihilation. If we ignore the effects due to radial redistribution of $\chi$ after their production and prior to their annihilation, their number density ($n_{\chi}$) would approximately satisfy
\begin{equation}
    \dot{n}_{\chi}(t) = n_i(t) \times \Gamma_{\rm BNV} - n_{\chi}^2(t) \langle \sigma v \rangle, \label{eq:macro_bnv:density:fast_ann}
\end{equation}
in which $n_i(t)$ is the decaying baryon number density which we take to be constant on short timescales, and $\langle \sigma v \rangle$ is the annihilation cross section averaged over the distribution of $\chi$. The asymptotic value for $\chi$ number density  (at times $t \gg 1/\sqrt{n_i\, \Gamma_{\rm BNV}\, \langle \sigma v\rangle}$) is then equal to $n_{\chi}^{\infty} = \sqrt{n_i \, \Gamma_{\rm BNV} / \langle \sigma v \rangle}$, which relative to the local baryon number density $n(r)$, is given by
\begin{equation}
    \frac{n_{\chi}^{\infty}(r)}{n(r)} = 1.4 \times 10^{-15} \sqrt{f_i(r)}
    \left(\frac{n_{\rm sat}}{n(r)}\right)^{1/2} \left(\frac{\Gamma_{\rm BNV}}{10^{-10}\, {\rm yr}^{-1}}\right)^{1/2} \left(\frac{ 10^{-26}\, {\rm cm}^3\, {\rm s}^{-1} }{\langle \sigma v \rangle}\right)^{1/2},
    \label{eq:macro_bnv:assum:non_therm_chi}
\end{equation}
in which $f_i(r) \equiv n_i(r) / n(r) < 1$ is the fraction of baryon $i$ relative to the total baryon number density, $n_{\rm sat} = 0.15\, {\rm fm}^{-3}$ is the nuclear saturation density, and we used the scale of the canonical weak-scale cross section ($10^{-26} \, {\rm cm}^3\, {\rm s}^{-1}$) for comparison. We can see that this ratio is negligible for the reference values in this equation if $ \langle \sigma v\rangle \gg 10^{-56} \, {\rm cm}^3\, {\rm s}^{-1}$. 

We can generalize Eq.~\eqref{eq:macro_bnv:density:fast_ann} to scenarios in which the redistribution of $\chi$'s , after their production and prior to their annihilation, is not negligible, by noting that the total $\chi$-population satisfies
\begin{equation}
    \dot{N}_{\chi}(t) = B_i(t) \times \Gamma_{\rm BNV} - C_{\rm ann} N_{\chi}^2(t), \label{eq:macro_bnv:assum:N_chi_eq}
\end{equation}
in which $B_i(t)$ is the number of decaying baryons of type $i$, and $C_{\rm ann}$ is the annihilation rate per particle, such that the total annihilation rate is identified as $\Gamma_{\rm ann} \equiv C_{\rm ann} N_{\chi}^2/2$. We are interested in short timescales during which $B_i(t)$ can be taken as a constant ($t \ll \Gamma_{\rm BNV}^{-1}$). In this case, the solutions to Eq.~\eqref{eq:macro_bnv:assum:N_chi_eq}, assuming $N_{\chi}(0)=0$, are given by
\begin{equation}
    N_{\chi}(t) = \sqrt{\frac{B_i\, \Gamma_{\rm BNV} }{C_{\rm ann}}} \tanh \left( \sqrt{B_i\, \Gamma_{\rm BNV}\,C_{\rm ann} } \, t \right), \qquad\qquad t \ll \Gamma_{\rm BNV}^{-1}\label{eq:macro_bnv:assum:N_chi_sol}
\end{equation}
in which the timescale for achieving an equilibrium between the production and annihilation of $\chi$ ($\dot{N}_{\chi}(\tau_{\infty}) \approx 0$) can be identified as $\tau_{\infty} = 1 / \sqrt{ B_i \, \Gamma_{\rm BNV} \, C_{\rm ann}}$, which can be achieved for $\tau_{\infty} < t \ll \Gamma_{\rm BNV}^{-1}$, if 
\begin{equation}
    \frac{\Gamma_{\rm BNV}}{B_i } \ll C_{\rm ann}. \label{eq:macro_bnv:assum:timescale}
\end{equation}
The total number of $\chi$s can then be approximated by its equilibrium value given by $N_{\chi}^{\infty} = \sqrt{B_i\, \Gamma_{\rm BNV}  / C_{\rm ann}}$. We can see that if the condition in Eq.~\eqref{eq:macro_bnv:assum:timescale} holds, then $N_{\chi}^{\infty} \ll B_i$. 

We now calculate $C_{\rm ann}$ in the scenario in which the annihilation rate of $\chi$ is slower than its self-interaction rate, and the $\chi$'s are distributed spherically with an average radius of $R_{\chi}$, according to Boltzmann distribution. Using the virial theorem and assuming a radially uniform distribution of background neutron star matter (over $R_{\chi}$) with an average energy density $\bar{\cal E}$, we can write
\begin{equation}
    n_{\chi}(r) = n_{\chi}(0)\, e^{-r^2 / R_{\chi}^2}, \quad \quad R_{\chi} = \sqrt{\frac{3\, k_B\, T_{\chi}}{2\pi \, G\, \bar{\cal E}\, m_{\chi}}},
\end{equation}
in which $k_B$ is the Boltzmann constant, and $T_{\chi}$ is the dark sector temperature. The total annihilation rate ($\Gamma_{\rm ann}$) and $N_{\chi}$ can then be evaluated as
\begin{align}
    \Gamma_{\rm ann} =& \frac{1}{2} \int_0^{R_{\chi}} 4\pi r^2 \left[n_{\chi}(r) \right]^2\,\langle \sigma v\rangle\, dr =
    0.24\, \left(\frac{k_B\, T_{\chi}}{G\, \bar{\cal E}\, m_{\chi}}\right)^{3/2}\, \langle \sigma v\rangle \, [n_{\chi}(0) ]^2,\\
    N_{\chi} =&  \int_0^{R_{\chi}} 4\pi r^2 n_{\chi}(r)\, dr = 0.78\, \left(\frac{k_B\, T_{\chi}}{G\, \bar{\cal E}\, m_{\chi}}\right)^{3/2}\, n_{\chi}(0),
    \label{eq:macro_bnv:assum:N_chi_eq_int}
\end{align}
in which $\langle\sigma v\rangle$ is the thermally averaged annihilation cross section. Using the definition of $C_{\rm ann}$ we have
\begin{equation}
    C_{\rm ann} \equiv \frac{2\, \Gamma_{\rm ann}}{N_{\chi}^2} = 0.78\, \left(\frac{G\, \bar{\cal E}\, m_{\chi}}{k_B\, T_{\chi}}\right)^{3/2}\, \langle \sigma v\rangle,
\end{equation}
and an equilibrium between the production and annihilation can be achieved on timescales $t \ll \Gamma_{\rm BNV}^{-1}$, if (see Eq.~\eqref{eq:macro_bnv:assum:timescale})
\begin{equation}
    \langle \sigma v \rangle \gg 2 \times 10^{-55} \left(\frac{10^{57}}{B_i}\right) \, \left(\frac{ T_{\chi}}{ m_{\chi} }\right)^{3/2}  \left(\frac{10^{15} \, {\rm g} / {\rm cm}^3}{ \bar{\cal E} }\right)^{3/2} \left(\frac{\Gamma_{\rm BNV}}{10^{-10}\, {\rm yr}^{-1}}\right)\, {\rm cm}^3\, {\rm s}^{-1}, \label{eq:macro_bnv:assum:timescale:xsec}
\end{equation}
in which $T_{\chi}$ and $m_{\chi}$ have the same units. 
We can also find the equilibrium value for $\chi$ number density at the core by combining the definition of equilibrium number, $N_{\chi}^{\infty} = \sqrt{B_i\, \Gamma_{\rm BNV}  / C_{\rm ann}}$, with Eq.~\eqref{eq:macro_bnv:assum:N_chi_eq_int} to arrive at
\begin{equation}
    \frac{n_{\chi}^{\infty}(0)}{n_{\rm sat}} = 8 \times 10^{-16} \left( \frac{B_{57}\,  \Gamma_{\rm BNV}}{10^{-10}\, {\rm yr}^{-1}}\right)^{1/2}\left( \frac{10^{-26}\, {\rm cm}^3\, {\rm s}^{-1}}{\langle \sigma v \rangle}\right)^{1/2} \left( \frac{ \bar{\cal E}}{ 10^{15}\, {\rm g} / {\rm cm}^3}\right)^{3/4} \left( \frac{ m_{\chi}}{ T_{\chi}}\right)^{3/4},
\end{equation}
in which we defined $B_{57} \equiv B_i / 10^{57} \sim \mathcal{O}(1)$. Assuming the reference values in this equation, we can see that this ratio is about $10^{-13}\,\left({\rm MeV}/T_{\chi}\right)^{3/4}$ for $m_{\chi} \sim \mathcal{O}({\rm GeV})$. Therefore, we have shown that the self-annihilation of $\chi$ can be very effective at keeping its concentration negligible. This concludes the analysis of the necessary conditions on $\chi$ self-annihilation cross section imposed by Eq.~\eqref{eq:macro_bnv:assum:local:noacc}. The explicit forms of decay and annihilation rates for $\chi$ in terms of our model parameters are presented in Sec.~\ref{sec:app:JMB:annihilation}.

\section{SU(3) CMF models for the EoS}
\label{app:CMF}

The Lagrangian density of the class of CMF models we employ is given by~\cite{Dexheimer:2008ax}
\begin{equation}
    \mathcal{L} = \mathcal{L}_{\rm Kin} + \mathcal{L}_{\rm Int}  + \mathcal{L}_{\rm Self}  + \mathcal{L}_{\rm SB}, \label{eq:medium:MFT:CMF:Lag}
\end{equation}
in which $\mathcal{L}_{\rm Kin}$ contains the usual kinetic terms for baryons and leptons, $\mathcal{L}_{\rm Int}$ is due to the baryon-meson interactions which are given by 
\begin{equation}
    \mathcal{L}_{\rm Int} = - \sum_i \overline{\psi}_i \left( g_{i\omega} \gamma_0 \langle\omega^0\rangle + g_{i\phi} \gamma_0 \langle\phi^0\rangle + 2 g_{i\rho} \gamma_0 I_{3i} \langle\rho_3^0\rangle + m_i^* \right) \psi_i\,.
\end{equation}
We note $\psi_i$ denotes a baryon of species $i$ with an effective mass $m_i^*$ and an isospin 3-component $I_{3i}$, and the expectation value is evaluated in the ground state. The last two terms in Eq.~\eqref{eq:medium:MFT:CMF:Lag}, i.e., $\mathcal{L}_{\rm Self}$ and $\mathcal{L}_{\rm SB}$, contain the self-interactions of scalar and vector mesons  and explicit chiral symmetry breaking terms respectively. The explicit expressions are given in Eqs.~(3), (4), and (5) of Ref.~\cite{Dexheimer:2008ax}. The baryon effective masses are generated by the scalar meson VEVs, except for a small explicit mass term $\delta m_i \sim 150$ MeV, and are given by
\begin{equation}
    m_i^* = g_{i\sigma} \langle \sigma\rangle + 2 g_{i\delta} \langle \delta_3\rangle I_{3i} +  g_{i\zeta} \langle \zeta\rangle + \delta m_i,
\end{equation}
in which $\delta_3$ is the isospin 3-component of ${\bm \delta}$. The time-component of baryon self-energy is given by
\begin{equation}
    \Sigma^0_i =  g_{i\omega} \langle \omega^0 \rangle + g_{i\rho} \langle \rho^0_3 \rangle I_{3i}  + g_{i\phi} \langle \phi^0 \rangle.
\end{equation}
We refer to Fig.~\ref{fig:medium:MFT:meff_sigma} for a plot of 
$m^*$ and $\Sigma^0$ with density. 
The coupling constants are chosen~\cite{dexheimer_2017, compose_CMF1,compose_CMF8} to reproduce the hadron vacuum masses, the nuclear saturation properties (density $n_{\rm sat} = 0.15\, {\rm fm}^{-3}$, binding energy per nucleon $B/A= -16.00\, {\rm MeV}$, compressibility $K = 300\, {\rm MeV}$), the symmetry energy ($E_{\rm sym} = 30\, {\rm MeV}$), and hyperon potentials. Furthermore, the pion and kaon decay constants constrain the scalar meson VEVs.


Our chosen class of EoSs contain
variations 
that depend on the included  degrees of freedom, and 
they are 
given in Table~\ref{tab:medium:MFT:CMF:choices} for convenience. Thus in order to explore how our results vary with the EoS, we employ the choices given there. 
\begin{table}[t!]
    \centering
    \begin{tabular}{|c|c|c|c|c|c|c|c|c|}
    \hline 
    EoS & DoF  & Add. Int. & $L$ (MeV) & $U_{\Lambda} \, ({\rm MeV})$ & $U_{\Sigma} \, ({\rm MeV})$ & $U_{\Xi} \, ({\rm MeV})$ & $U_{\Delta} \, ({\rm MeV})$ & $M_{\rm max}\, (M_{\astrosun})$\\
        \hline\hline
        $1$ & N$+$Y & -- & $88$ & $-28$ & $5$ & $-18$ & -- & $2.07$\\\hline
        $2$ & N & -- & $88$ & $-28$ & $5$ & $-18$ & -- & $2.13$\\\hline
        $3$ & N$+$Y & $\omega \rho$ & $75$ & $-28$ & $5$ & $-18$ & -- & $2.00$\\\hline
        $4$ & N & $ \omega \rho$ & $75$ & $-28$ & $5$ & $-18$ & -- & $2.05$\\\hline
        $5$ & N$+$Y & $ \omega \rho + \omega^4$ & $75$ & $-27$ & $6$ & $-17$ & -- & $2.07$\\\hline
        $6$ & N & $\omega \rho + \omega^4$ & $75$ & $-27$ & $6$ & $-17$ & -- & $2.11$\\\hline
        $7$ & N$+$Y$+\Delta$ & $ \omega \rho + \omega^4$ & $75$ & $-27$ & $6$ & $-17$ & $-64$ & $2.07$\\\hline
        $8$ & N$+\Delta$ & $ \omega \rho + \omega^4$ & $75$ & $-27$ & $6$ & $-17$ & $-64$ & $2.09$\\\hline
    \end{tabular}
    \caption{The set of CMF EoS variants taken from Ref.~\cite{compose_CMF1,compose_CMF8}; we refer to them as DS(CMF)-1 through DS(CMF)-8, respectively, in later use. The second and third columns describe the degrees of freedom (DoF); nucleons (N), hyperons (Y), and delta resonances ($\Delta$); and the additional vector interactions (``Add. Int.") beyond the standard terms ($\mathcal{L}_{\rm Self}$)  that are included for each EoS respectively. The fourth column represents the assumed value for symmetry energy ($E_{\rm sym}$) slope ($L$). The fifth to eighth columns are the single-particle hyperon potentials, and the last column is the maximum neutron star mass ($M_{\rm max}$), or the TOV mass, that can be generated. }
    \label{tab:medium:MFT:CMF:choices}
\end{table}
The set of EoSs that we utilize has also been extended to include crusts based on a zero-temperature unified EoS~\cite{Gulminelli:2015csa} at $\beta$--equilibrium with similar values of the symmetry energy slope ($L$), in which the effective interactions are Skyrme forces Rs~\cite{PhysRevC.33.335} (EoS 1-2) and SkMP~\cite{PhysRevC.40.2834} (EoS 3-8) with cluster energy functionals taken from Ref.~\cite{Danielewicz:2008cm}.
\section{In-Medium Electromagnetic Form Factors}
\label{app:form_fac}
In this appendix we derive the general form for electromagnetic interactions of baryons in the context of hadronic RMF models, and explicitly show how the electric charge and magnetic moment are to be identified from the scattering amplitudes of baryons off of electric and magnetic potentials respectively. We start from the Dirac equation~\eqref{eq:medium:MFT:dirac}, $\slashed{p} u(p^*) = (m^* + \slashed{\Sigma}) u(p^*)$, in which we suppressed the spin index $\lambda$, to write
\begin{align}
    \overline{u}({p'}^*) \left( \frac{i\sigma^{\mu\nu} \left(p'_{\nu} - p_{\nu}\right) }{2m^*}\right) u(p^*) 
    &= \overline{u}({p'}^*) \left( \gamma^{\mu} - \frac{ {p'}^{\mu} + p^{\mu}}{2m^*} + \frac{\Sigma^{\mu}}{m^*} \right) u(p^*),
\end{align}
in which $\sigma^{\mu\nu} \equiv (i/2) [\gamma^{\mu}, \gamma^{\nu}]$ and $\gamma^{\mu}$ are the usual Dirac matrices. The in-medium Gordon decomposition is then given by
\begin{align}
    \overline{u}({p'}^*) \gamma^{\mu} u(p^*) = \overline{u}({p'}^*) \left( \frac{ {p'}^{\mu} + p^{\mu}}{2m^*} + \frac{i\sigma^{\mu\nu} q_{\nu} }{2m^*} - \frac{\Sigma^{\mu}}{m^*} \right) u(p^*)
\end{align}
in which we defined $q_{\nu} \equiv p'_{\nu} - p_{\nu}$. The general form of a vector interaction vertex, $\Gamma^{\mu}$, can be written as
\begin{align}
    \Gamma^{\mu} = \gamma^{\mu} A + \left({p'}^{\mu} + p^{\mu}\right) B + q^{\mu} C + D \Sigma^{\mu},
\end{align}
in which $A, B, C, D$ are functions of scalar quantities (e.g., $q^2$). Applying the Ward identity, $q_{\mu} \Gamma^{\mu} = 0$, plus ${p'}^{*2} = p^{*2} = m^{*2}$ and ${p'}^2 - p^2 = 2 q\cdot \Sigma$,  yields $C = 0$ and $2B = D$. The electromagnetic vertex factor can then be written as
\begin{align}
    \Gamma^{\mu} = \gamma^{\mu} F_1^*(q^2) + \frac{i\sigma^{\mu\nu} q_{\nu} }{2m^*} F_2^*(q^2), 
\end{align}
in which $F_{1,2}^*$ are in principle distinct from their vacuum counterparts $F_{1,2}$. 

We now show how the electric charge can be identified in the scattering amplitude of a baryon from a Coulomb potential $A_{\mu} = (\Phi(x), \vec{0})$. Employing equations $\overline{u}(k^*, \lambda) u(k^*, \lambda) = 2 m^*$ and $\overline{u}(k^*, \lambda) \gamma^0 u(k^*, \lambda) = 2 E^*(k^*)$, this amplitude can be written as
\begin{equation}
   i\mathcal{M}  = -i e F_1^*(0) \tilde{\Phi}(q)  \left(\frac{E^*}{m^*}\right) 2m^* \chi^{\dagger} \chi,
\end{equation}
in which $E^* = \sqrt{{m^*}^2 + (\vec{p}^{\, *})^2}$, and $\chi$ is the Pauli spinor. The electric charge ($q$) can then be identified, by considering this scattering in the c.v. frame ($\vec{p}^{\, *} = 0$), as $q = F^*_1(0) (E^{*,({\rm c.v.})} / m^*) = F^*_1(0)$. This can also be understood from the time component of spin-independent conserved EM current $J^{0} = \overline{\psi} \gamma^{0} \psi= 2 E^*$, with the Lorentz invariant electric charge defined in the c.v. frame ($E^{*,({\rm c.v.})} = m^*$).

Similarly, we can identify the magnetic moment from the scattering amplitude of a baryon from a static magnetic field potential $A_{\mu} = (0, \vec{A})$ at small momentum transfers ($q^2 \approx 0$), which is given by
\begin{equation}
   i\mathcal{M}  = +i e \overline{u}({p'}^*) \left[  \gamma^i F_1^*(0) + \frac{i \sigma^{i\nu} q_{\nu}}{2m^*} F_2(0)\right] u(p^*) \tilde{A}^i(0). \label{eq:app:form_fac:mag_amp}
\end{equation}
The first term can be written as
\begin{equation}
    \begin{split}
   \overline{u}({p'}^*)\gamma^i u(p^*) =& \left(E^* + m^*\right) \left[ \chi^{\dagger}, \chi^{\dagger} \frac{\vec{\sigma}\cdot \vec{p}^{\,'*}}{E^* + m^*} \right]     \begin{pmatrix}
    0 & \sigma^i\\
    \sigma^i & 0
   \end{pmatrix}  
    \begin{pmatrix}
    \eta \\
    \frac{\vec{\sigma}\cdot \vec{p}^*}{E^* + m^*} \eta
   \end{pmatrix} \\
   =& \chi^{\dagger} \left[ \sigma^i \vec{\sigma} \cdot \vec{p}^{\,*} +  \vec{\sigma} \cdot \vec{p}\,'^{*} \sigma^i\right] \eta,
   \end{split}
\end{equation}
in which $\sigma^i$ are the Pauli matrices, and $\chi$, $\eta$ represent the spin states. This expression can be further simplified using $\sigma^i \sigma^j = \delta^{ij} + i \epsilon^{ijk} 
\sigma^k$,
such that
\begin{equation}
    \overline{u}({p'}^*)\gamma^i u(p^*) = \chi^{\dagger} \left[ \left(p^* + p'^*\right)^i - i \epsilon^{ijk} \left(p'^* - p^*\right)^j \sigma^k\right] \eta.
\end{equation}
The $F_2$ term in the scattering amplitude ( Eq.~\eqref{eq:app:form_fac:mag_amp}) already contains a factor of $q$, and so we can evaluate it using the leading order expansion of the spinors in the non-relativistic limit ($\vec{p}^{\,*} \ll m^*$), which is given by $u(\vec{p}^{\,*}=0) = \sqrt{2 m^*} (\chi, 0)^T$. We also note that
\begin{align}
    \frac{i}{2m^*} \sigma^{ij} q_j &= \frac{i \epsilon^{ijk}}{2m^*} \sigma^k q_j, \label{eq:app:form_fac:sigma_j}\\
    \frac{i}{2m^*} \sigma^{i0} q_0 &= \frac{q_0}{2m^*} 
    \begin{pmatrix}
    0 & \sigma^i\\
    \sigma^i & 0
   \end{pmatrix}, \label{eq:app:form_fac:sigma_0}
\end{align}
such that the spin-dependent contribution from Eq.~\eqref{eq:app:form_fac:sigma_0}, i.e., $\overline{u}({p'}^*) (\sigma^{i0} q_0)u(p^*)$ is proportional to $q_0 q^j$, which is subdominant to other terms.
The term from Eq.~\eqref{eq:app:form_fac:sigma_j}, i.e., $\overline{u}({p'}^*) (\sigma^{ij} q_j)u(p^*)$ is given by
\begin{equation}
   \overline{u}({p'}^*)\left(  \frac{i}{2m^*} \sigma^{ij} q_j  \right) u(p^*) = i \epsilon^{ijk} q_j \left(\chi^{\dagger} , 0\right)  
   \begin{pmatrix}
    \sigma^k \eta \\
    0
   \end{pmatrix} = i \epsilon^{ijk} q_j \chi^{\dagger} \sigma^k \eta.
\end{equation}
The amplitude in Eq.~\eqref{eq:app:form_fac:mag_amp} can then be written as (note $q_j = - q^j$)
\begin{equation}
\begin{split}
   i\mathcal{M}  =& - e \chi^{\dagger} \left\{  \epsilon^{ijk} q^i \tilde{A}^j(0) \sigma^k \left[F_1^*(0) + F_2^*(0)\right] \right\} \eta \\
   =& - 2 i e \left[F_1^*(0) + F_2^*(0)\right] S^k \tilde{B}^k,
\end{split}
\end{equation}
in which we defined the magnetic field by $\tilde{B}^k \equiv - i \epsilon^{ijk} q^i \tilde{A}^j$, spin by $\vec{S} \equiv (1/2) \chi^{\dagger} \vec{\sigma}\eta$, and the baryon g-factor can be identified as $g^* = 2\left[F_1^*(0) + F_2^*(0)\right]$.

\section{Nonrelativistic Limit of In-Medium Scattering}
\label{app:non-rel_scatt}
In this appendix we study the non-relativistic (NR) limit of the RMF model, and derive the elastic scattering formalism in the Born approximation. Since the medium effects in the RMF formalism resemble an electromagnetic interaction with a constant electromagnetic background field given by $eA^{\mu} \to \Sigma^{\mu}$, it is instructive to consider the NR limit of baryon EM interactions in medium. We explicitly show that the NR limit of the modified Dirac (Eq.~\eqref{eq:medium:MFT:dirac}) solutions under the influence of electromagnetism, reduces to the two-component Pauli spin theory, with replacements $m \to m^*$, $e\Phi \to e\Phi + \Sigma^0$, $e\vec{A} \to e\vec{A} + \vec{\Sigma}$, in which $\Sigma^0$ and $\vec{\Sigma}$ are the self-energies due to the medium effects, $e$ is the baryon electric charge, with $\Phi$ and $\vec{A}$ as the scalar and vector EM potentials respectively. We start from the Schrodinger equation, which can be written by denoting the Dirac wave-function ($\psi$) in two-component notation~\cite{bjorken1964relativistic}, $\psi = \left(\tilde{\varphi}, \tilde{\chi}\right)^T$, such that we have
\begin{equation}
    i \frac{\partial}{\partial t}
    \begin{pmatrix}
    \tilde{\varphi} \\
    \tilde{\chi}
   \end{pmatrix}  =
   \vec{\sigma} \cdot \vec{\pi} 
   \begin{pmatrix}
    \tilde{\chi} \\
    \tilde{\varphi} 
    \end{pmatrix} 
    +
    \left(e \Phi + \Sigma^0 \right)\begin{pmatrix}
    \tilde{\varphi} \\
    \tilde{\chi}
    \end{pmatrix} 
    +
    m^{*}\begin{pmatrix}
    \tilde{\varphi} \\
    -\tilde{\chi}
   \end{pmatrix}, \label{eq:app:non-rel_scatt:schrodinger}
\end{equation}
in Pauli-Dirac representation, with $\vec{\pi} \equiv \vec{p} - \vec{\Sigma} - e \vec{A}$. Using the definition $\left(\tilde{\varphi}, \tilde{\chi}\right) = \exp (-i m^* t) \left(\varphi,  \chi\right)$, we can rewrite Eq.~\eqref{eq:app:non-rel_scatt:schrodinger} as
\begin{equation}
    i \frac{\partial}{\partial t}
    \begin{pmatrix}
    {\varphi} \\
    {\chi}
   \end{pmatrix}  =
   \vec{\sigma} \cdot \vec{\pi} 
   \begin{pmatrix}
    {\chi} \\
    {\varphi} 
    \end{pmatrix} 
    +
    \left(e \Phi + \Sigma^0\right)\begin{pmatrix}
    {\varphi} \\
    {\chi}
    \end{pmatrix} 
    -2m^*
    \begin{pmatrix}
    0 \\
    {\chi}
   \end{pmatrix}.
\end{equation}
We note that in the NR limit, in which kinetic and interaction energies are much smaller than $m^*$, the second component $\chi$ is subdominant to the first component $\varphi$ and is approximately given by 
\begin{equation}
    \chi \approx \frac{\vec{\sigma} \cdot \vec{\pi} }{2m^{*}} \varphi.
\end{equation}
We also arrive at the Pauli equation governing the first component ($\varphi$):
\begin{equation}
\label{eq:app:non-rel_scatt:pauli}
    i \frac{\partial}{\partial t} \varphi = \left( \frac{\left(\vec{p} - \vec{\Sigma} - e \vec{A}\right)^2}{2 m^*} - \frac{e}{2m^*} \vec{\sigma} \cdot \vec{B} + e\Phi + \Sigma^0 \right) \varphi,
\end{equation}
in which $\vec{B} = \nabla \times A$. This expression can be further simplified for a weak uniform magnetic field ($\vec{A} = \vec{B} \times \vec{r}/2$) as
\begin{equation}
\label{eq:app:non-rel_scatt:pauli:weak}
    i \frac{\partial}{\partial t} \varphi = \left( \frac{|\vec{p}^{\,*}|^{2}}{2 m^*}  - \frac{e}{2m^*} \left( \vec{L}^{\, *} + 2\vec{S} \right) \cdot \vec{B} + e\Phi + \Sigma^0\right) \varphi,
\end{equation}
in which $\vec{p}^{\,*} = \vec{p} - \vec{\Sigma}$ is the kinetic three momentum, and $\vec{L}^* = \vec{r} \times \vec{p}^{\, *}$ and $\vec{S} = \vec{\sigma} / 2$ are the baryon's kinetic orbital angular momentum and spin respectively. Note that in the n.m. frame ($\vec{\Sigma}=0$) the canonical and kinetic three momenta are equal $\vec{p} = \vec{p}^{\, *}$. 

We now construct the elastic scattering formalism off of an arbitrary potential ($V$) in this NR limit, by turning off the electromagnetic fields, i.e., $\vec{A} = \Phi = 0$, for the rest of this discussion. From Eq.~\eqref{eq:app:non-rel_scatt:pauli}, we deduce the energy eigenvalues $E = |\vec{p}^{\, *}|^2/2m^* + \Sigma^0$, which agree with the NR expansion of Dirac energy eigenvalues given in Eq.~\eqref{eq:medium:MFT:dirac:energy}. The energy eigenfunctions in position space satisfy
\begin{equation}
    -\nabla^2 \varphi  + 2i \vec{\Sigma}\cdot \vec{\nabla} \varphi  + \left[ |\vec{\Sigma}|^2 - 2m^*\left(E - \Sigma^0\right)\right] \varphi = 0,
\end{equation}
with solutions of the form
\begin{equation}
    \varphi = e^{-i E t} \left[ A_1 e^{i \vec{p}\cdot\vec{x} } + A_2 e^{-i \left( \vec{p} - 2 \vec{\Sigma}\right)\cdot \vec{x}}\right],
\end{equation}
which can also be written in a more symmetric way in terms of $\vec{p}^{\,*}$. If we orient our coordinates such that $\vec{\Sigma}\cdot\vec{x} > 0$, then for a positive $\vec{p}$ ($\vec{p}\cdot\vec{x} > 0$) the first term is a plane wave moving to the right and the second term is a wave moving to the left. Therefore, we pick the first term for incident waves in the elastic scattering problem. Let $H_0$ be the Hamiltonian used in Eq.~\eqref{eq:app:non-rel_scatt:pauli} (with $\Phi = \vec{A} = 0$), and $| k^{(+)}\rangle$ be the state that satisfies the following Schrodinger equation in the presence of a potential $V$
\begin{equation}
    \left(E - H_0\right) | k^{(+)}\rangle = V | k^{(+)}\rangle,
\end{equation}
then, $| k^{(+)}\rangle$ can be found from the Lippmann-Schwinger equation:
\begin{equation}
    | k^{(+)}\rangle = | k\rangle + \frac{1}{E- H_0 + i\varepsilon} V | k^{(+)}\rangle.
\end{equation}
The momentum representation of operator $G \equiv (E- H_0 + i\varepsilon)^{-1}$ is given by
\begin{equation}
    \langle \vec{q}\,| G |\, \vec{q}{\,'}\rangle = \delta(\vec{q}-\vec{q}') \frac{2m^*}{\left(\vec{k} - \vec{\Sigma}\right)^2 - \left(\vec{q} - \vec{\Sigma}\right)^2 + i\varepsilon},
\end{equation}
and the position space representation is given by
\begin{equation}
    \langle \vec{r}\,| G |\, \vec{r}{\,'}\rangle = \int \frac{d^3 q}{\left(2\pi\right)^3} \frac{2m^*}{\left(\vec{k} - \vec{\Sigma}\right)^2 - \left(\vec{q} - \vec{\Sigma}\right)^2 + i\varepsilon} e^{i \vec{q} \cdot \left(\vec{r} - \vec{r}{\,'}\right)}.
\end{equation}
We define $\vec{R}\equiv \vec{r} - \vec{r}{\,'}$, $\vec{\xi}\equiv \vec{k} - \vec{\Sigma}$ and change the variable from $\vec{q}$ to $\vec{Q} \equiv \vec{q} - \vec{\Sigma}$ such that
\begin{equation}
\begin{split}
    \langle \vec{r}\,| G |\, \vec{r}{\,'}\rangle =& \int \frac{d^3 Q}{\left(2\pi\right)^3} \frac{2m^*}{\vec{\xi}^2 - \vec{Q}^2 + i\varepsilon} e^{i \left(\vec{Q} + \vec{\Sigma} \right) \cdot \vec{R}}\\
    =& \frac{m^* e^{i\vec{\Sigma}\cdot \vec{R}}}{4\pi^2 (i R)} \int_{-\infty}^{\infty} \frac{Q dQ}{\vec{\xi}^2 - \vec{Q}^2 + i\varepsilon} \left[ e^{i Q R} - e^{-i Q R} \right]\\
    =& \frac{-m^*}{2\pi R} e^{i \vec{\Sigma} \cdot \vec{R}} e^{i \xi R},
\end{split}
\end{equation}
in which we performed the angular integration in the second line, and the complex contour integration in the third line. To characterize the scattering problem at $r \to \infty$ we approximate the above expression for $(r'/r) \to 0$ using $R = |\vec{r} - \vec{r}{\,'}| \approx r - \hat{r} \cdot \vec{r}{\,'}$, such that
\begin{equation}
\begin{split}
    \langle \vec{r}\,| G | \vec{r}{\,'}\rangle =& \left(\frac{-m^*}{2\pi r}\right)  e^{i \left[ | \vec{k} - \vec{\Sigma}\, | r + \vec{\Sigma} \cdot \vec{r}\, \right]} \,  e^{-i \left[ | \vec{k} - \vec{\Sigma}\, | \hat{r} + \vec{\Sigma} \right] \cdot \vec{r}{\,'} }.
\end{split}
\end{equation}
We now write the asymptotic form of the Lippmann-Schwinger equation in position space as
\begin{equation}
    \psi_k\left(\vec{r}\,\right) \sim \varphi_k\left( \vec{r}\, \right) - \frac{m^*}{2\pi r} e^{i \left[ | \vec{k} - \vec{\Sigma}\, | r + \vec{\Sigma} \cdot \vec{r}\, \right]} \int d^3 r'\,  e^{-i \left[ | \vec{k} - \vec{\Sigma}\, | \hat{r} + \vec{\Sigma} \right] \cdot \vec{r}{\,'} }\, V(r') \psi_k (\vec{r}{\, '}),
\end{equation}
in which $\psi_k\left( \vec{r}\, \right) \equiv \langle \vec{r}\, | \vec{k}^{(+)}\rangle$ and $\varphi_k\left( \vec{r}\, \right)\equiv \left(2\pi\right)^{-3/2} \exp \left(i \vec{k} \cdot \vec{r}\, \right)$. The exponential outside of the integral in the second term is an ellipsoidal wave (stretched along $\vec{\Sigma}$) which becomes spherical in the n.m. frame ($\vec{\Sigma} = 0$). The exponent inside the integral is a vector pointing in the direction of $| \vec{k} -\vec{\Sigma}\, | \hat{r} + \vec{\Sigma}$, which reduces to the familiar $k \hat{r}$ term in the n.m. frame. We can see that the gradient of the ellipsoidal surface is equal to the vector in the exponent inside the integral since
\begin{equation}
    \nabla \left[ | \vec{k} - \vec{\Sigma}\, | r + \vec{\Sigma} \cdot \vec{r}\, \right] = | \vec{k} - \vec{\Sigma}\, | \hat{r} + \vec{\Sigma},
\end{equation}
which suggests that the exponent $\vec{k}{\,'} \equiv | \vec{k} - \vec{\Sigma}\, | \hat{r} + \vec{\Sigma}$ is the momentum of scattered particle in the direction of an observer at $r$. Note that the kinetic energy of the scattered particle is given by
\begin{equation}
    T(k') = \frac{\left(\vec{k}{\,'} - \vec{\Sigma} \right)^2}{2m^*} = \frac{\left(\vec{k} - \vec{\Sigma} \right)^2}{2m^*} = T(k),
\end{equation}
and the scattering is indeed elastic. We can therefore deduce the scattering amplitude by writing 
\begin{equation}
    \psi_k\left(\vec{r}\,\right) \sim \left(2\pi\right)^{-3/2} \left\{ e^{i \vec{k} \cdot\vec{r}} + \frac{e^{i \left[ | \vec{k} - \vec{\Sigma}\, | r + \vec{\Sigma} \cdot \vec{r}\, \right]}}{r}    f(k' | k)    \right\},
\end{equation}
in which 
\begin{equation}
    f(k' | k)    = - 4\pi^{2} m^* \int d^3 r\,  \varphi_{k'}^*(\vec{r}\,)\, V(r) \psi_k (\vec{r}),
\end{equation}
which is the Fourier transform of the potential in the Born approximation.

\section{In-Medium Compton Scattering}
\label{app:compton}
\begin{figure}[htb]
    \centering
    \includegraphics[width=0.7\textwidth]{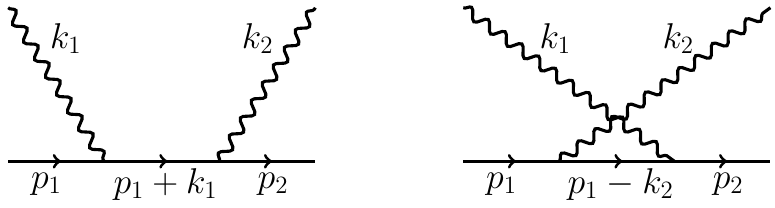}
    \caption{Feynman diagrams for the baryon Compton scattering ${\cal B}(p_1) + \gamma(k_1) \to {\cal B}(p_2) + \gamma(k_2)$.}
    \label{fig:compt:feyn}
\end{figure}

In this section we evaluate the Compton scattering cross section of baryons, ${\cal B}(p_1) + \gamma(k_1) \to {\cal B}(p_2) + \gamma(k_2)$ (see Fig.~\ref{fig:compt:feyn}) in neutron star medium, denoting the photon and baryon energies by $\omega_{1,2}$ and $E_{1,2}$ respectively. 
We first note that the second term in the baryon propagator defined in Eq.~\eqref{eq:medium:modific:baryon-propagator} vanishes since
\begin{equation}
    (p_1^* + k_1)^2 - (m_{\cal B}^*)^2 
    = 2 \left(E_{1}^* |\vec{k}_1| -  \vec{k}_1\cdot \vec{p}^{\,*}_1\right) = 2 |\vec{k}_1| \left( \sqrt{(\vec{p}^{\,*}_1)^2 + (m_{\cal B}^*)^2} - \hat{k}_1 \cdot \vec{p}^{\,*}_1\right) > 0,
\end{equation}
and similarly, it can be shown that $(p_1^* - k_2)^2 - (m_{\cal B}^*)^2$ is strictly negative. The amplitude for the diagrams shown in Fig.~\ref{fig:compt:feyn} can then be written as
\begin{equation}
    i \mathcal{M} = i \mathcal{M}_L + i \mathcal{M}_R,
\end{equation}
in which
\begin{equation}
\begin{split}
    i \mathcal{M}_L = -i \overline{u}(p_2) &\left( \gamma^{\mu} F_1^* + \frac{i \sigma^{\mu\nu} k_{2,\nu}}{2m_{\cal B}^*} F_2^*\right) \epsilon^*_{\mu}(k_2) \left( \frac{\slashed{p}^*_1 + \slashed{k}_1 + m}{(p_1^* + k_1)^2 - (m_{\cal B}^*)^2} \right) \\
    \times &\left( \gamma^{\nu} F_1^* + \frac{i \sigma^{\nu\alpha} k_{2,\alpha}}{2m_{\cal B}^*} F_2^*\right) \epsilon_{\nu}(k_1) u(p_1),
\end{split}
\end{equation}
and 
\begin{equation}
\begin{split}
    i \mathcal{M}_R = -i \overline{u}(p_2) &\left( \gamma^{\nu} F_1^* + \frac{i \sigma^{\nu\alpha} k_{1,\alpha}}{2m_{\cal B}^*} F_2^*\right) \epsilon_{\nu}(k_1) \left( \frac{\slashed{p}^*_1 - \slashed{k}_2 + m}{(p_1^* - k_2)^2 - (m_{\cal B}^*)^2} \right) \\
    \times &\left( \gamma^{\mu} F_1^* + \frac{i \sigma^{\mu\nu} k_{2,\nu}}{2m_{\cal B}^*} F_2^*\right) \epsilon^*_{\mu}(k_2) u(p_1),
\end{split}
\end{equation}
in which $F_{1,2}^*$ are the in-medium form factors. The interaction term in the amplitude can be simplified using 
\begin{equation}
    \left( \gamma^{\mu} F_1^* + \frac{i \sigma^{\mu\nu} k_{2,\nu}}{2m_{\cal B}^*} F_2^*\right) \epsilon_{\mu}^*(k_2) = \slashed{\epsilon}^*(k_2) F_1^* - \frac{F_2^*}{2m_{\cal B}^*} \slashed{\epsilon}^*(k_2) \slashed{k}_2,
\end{equation}
which follows from $\epsilon_{\mu}(k_1) k_1^{\mu} = \epsilon_{\mu}(k_2) k_2^{\mu} = 0$. The spin-averaged squared amplitudes simplify to
\begin{equation}
\begin{split}
    \overline{\left|\mathcal{M}_L\right|^2} = 
    \frac{1}{4 \left[(p_1^* + k_1)^2 - (m_{\cal B}^*)^2\right]^2}
    \, {\rm Tr}\, &\left[ \left( \slashed{p}^*_2 + m_{\cal B}^*\right)  \left( F_1^* + \frac{F_2^*}{2 m_{\cal B}^*} \slashed{k}_2\right) \gamma^{\mu} \left( \slashed{p}_1^* + \slashed{k}_1 + m_{\cal B}^*\right)   \gamma^{\nu} \right. \\
    & \times  \left( F_1^* - \frac{F_2^*}{2m_{\cal B}^*} \slashed{k}_1 \right) \left(\slashed{p}^*_1 + m_{\cal B}^*\right)  \left( F_1^* - \frac{F_2^*}{2m_{\cal B}^*} \slashed{k}_1 \right)  \\
    & \times  \left. \gamma_{\nu} \left( \slashed{p}_1^* + \slashed{k}_1 + m_{\cal B}^*\right) \gamma_{\mu} \left( F_1^* + \frac{F_2^*}{2m_{\cal B}^*} \slashed{k}_2\right)  
    \right],
\end{split}
\end{equation}
and 
\begin{equation}
\begin{split}
    \overline{\left|\mathcal{M}_R\right|^2} = 
    \frac{1}{4 \left[(p_1^* - k_2)^2 - (m_{\cal B}^*)^2\right]^2}
    \, {\rm Tr}\, &\left[ \left( \slashed{p}^*_2 + m_{\cal B}^*\right)  \left( F_1^* + \frac{F_2^*}{2 m_{\cal B}^*} \slashed{k}_1\right) \gamma^{\nu} \left( \slashed{p}_1^* - \slashed{k}_2 + m_{\cal B}^*\right)   \gamma^{\mu} \right. \\
    & \times \left( F_1^* - \frac{F_2^*}{2m_{\cal B}^*} \slashed{k}_2 \right) \left(\slashed{p}^*_1 + m_{\cal B}^*\right)  \left( F_1^* - \frac{F_2^*}{2m_{\cal B}^*} \slashed{k}_2 \right) \\
    & \times \left. \gamma_{\mu} \left( \slashed{p}_1^* - \slashed{k}_2 + m_{\cal B}^*\right) \gamma_{\nu} \left( F_1^* + \frac{F_2^*}{2m_{\cal B}^*} \slashed{k}_1\right)  
    \right],
\end{split}
\end{equation}
with the cross-term given by
\begin{equation}
\begin{split}
    \overline{\mathcal{M}_L\mathcal{M}_R^{\dagger}} = \frac{{\rm Tr}\, \left[ T_{\rm LR} \right]}{4 \left[(p_1^* - k_2)^2 - (m_{\cal B}^*)^2\right]\left[(p_1^* + k_1)^2 - (m_{\cal B}^*)^2\right]},
\end{split}
\end{equation}
in which 
\begin{equation}
\begin{split}
    T_{\rm LR} =& \left( \slashed{p}^*_2 + m_{\cal B}^*\right)  \left( F_1^* + \frac{F_2^*}{2 m_{\cal B}^*} \slashed{k}_2\right) \gamma^{\mu} \left( \slashed{p}_1^* + \slashed{k}_1 + m_{\cal B}^*\right)   \gamma^{\nu} \left( F_1^* - \frac{F_2^*}{2m_{\cal B}^*} \slashed{k}_1 \right) \left(\slashed{p}^*_1 + m_{\cal B}^*\right) \\
    &  \times \left( F_1^* - \frac{F_2^*}{2m_{\cal B}^*} \slashed{k}_2 \right) \gamma_{\mu} \left( \slashed{p}_1^* - \slashed{k}_2 + m_{\cal B}^*\right) \gamma_{\nu} \left( F_1^* + \frac{F_2^*}{2m_{\cal B}^*} \slashed{k}_1\right),  
\end{split}
\end{equation}
with $\overline{\mathcal{M}_L\mathcal{M}_R^{\dagger}}= \overline{\mathcal{M}_R\mathcal{M}_L^{\dagger}}$. We now define the following Mandelstam variables
\begin{align}
    s^* \equiv& \left(p_1^* + k_1\right)^2 = (m_{\cal B}^*)^{2} + 2 p_1^* \cdot k_1 = (m_{\cal B}^*)^{2} + 2 p_2^* \cdot k_2 \\ 
    t^* \equiv& \left(p_2^* - p_1^*\right)^2 = 2(m_{\cal B}^*)^{2} - 2 p_1^* \cdot p^*_2 = - 2 k_1 \cdot k_2 \\ 
    u^* \equiv& \left(k_2 - p_1^*\right)^2 = (m_{\cal B}^*)^{2} - 2 p_1^* \cdot k_2 = (m_{\cal B}^*)^{2} - 2 p_2^* \cdot k_1, 
\end{align}
such that $s^* + t^* + u^* = 2(m_{\cal B}^*)^2$. We suppress the superscripts (``*") of $m_{\cal B}^*$, $F_{1,2}^*$ in some of the following equations for convenience. The averaged amplitude-squared can be written as
\begin{equation}
    \overline{\left| \mathcal{M}\right|^2} = \frac{1}{16} \left( \frac{\rm I}{(p^*_1\cdot k_1)^2} + \frac{\rm II}{(p^*_1\cdot k_1)(p^*_1\cdot k_2)} + \frac{\rm III}{(p^*_1\cdot k_1)(p^*_1\cdot k_2)} + \frac{\rm IV}{(p^*_1\cdot k_2)^2}\right), \label{eq:app:compton:av_amp_sqr}
\end{equation}
in which 
\begin{align}
{\rm I}  =& 8 F_1^4 \left(m_{\cal B}^4 + m_{\cal B}^2 (3 s+u) - s u\right) + 4 F_2^2 F_1^2 \left(1-\frac{s}{m_{\cal B}^2}\right)
   \left[2 m_{\cal B}^4-m_{\cal B}^2 (3 s+u)+s (3 s-u)\right]\nonumber \\
   &\, - \frac{F_2^4 m_{\cal B}^2}{2}\left(1-\frac{s}{m_{\cal B}^2}\right)^3 \left(m_{\cal B}^2-u\right), \\
{\rm II} =& {\rm III} = F_2^4 \left(2- \frac{s+u}{m_{\cal B}^2}\right) \left(s u - m_{\cal B}^4\right) - 8 F_1^4 m_{\cal B}^2 \left(2 m_{\cal B}^2+s+u\right) \nonumber \\
   &\quad\quad - 2 F_2^2 F_1^2 \left(3 m_{\cal B}^2 (s+u) - 2 \left(s^2+s u+u^2\right) + \frac{s u (s+u)}{m_{\cal B}^2} -2 m_{\cal B}^4 \right), \\
{\rm IV} =&  8 F_1^4 \left(m_{\cal B}^4+m_{\cal B}^2 (s+3 u)- s u\right) + 4 F_2^2 F_1^2 \left(1-\frac{u}{m_{\cal B}^2}\right)
   \left[2 m_{\cal B}^4-m_{\cal B}^2 (s+3 u)+u (3 u-s)\right] \nonumber \\ 
   &\, - \frac{F_2^4 m_{\cal B}^2}{2} \left(m_{\cal B}^2-s\right) \left(1 - \frac{u}{m_{\cal B}^2}\right)^3,
\end{align}
in which we note that $\rm I$ and $\rm IV$ are related via $(s \leftrightarrow u)$ replacement. Equation~\eqref{eq:app:compton:av_amp_sqr} can then be written as
\begin{equation}
\begin{split}
      \overline{\left| \mathcal{M}\right|^2} =& 
      \frac{2F_1^4 \left[6 m_{\cal B}^8-m_{\cal B}^4 \left(3 s^2+14 s u+3 u^2\right) + m_{\cal B}^2 (s+u) \left(s^2+6 s u+u^2\right) - s u \left(s^2+u^2\right)\right]}{\left(m_{\cal B}^2-s\right)^2 \left(m_{\cal B}^2-u\right)^2}  \\
   &\, + \frac{F_2^4 \left[3 m_{\cal B}^8-m_{\cal B}^4 \left(s^2+8 s u+u^2\right)+4 m_{\cal B}^2 s u (s+u)-s^2 u^2\right] }{4 m_{\cal B}^4 \left(m_{\cal B}^2-s\right) \left(m_{\cal B}^2-u\right)} \\
   &\, + \frac{F_2^2 F_1^2 \left[2 m_{\cal B}^6 - 3 m_{\cal B}^4
   (s+u) + 2 m_{\cal B}^2 \left(s^2+s u+u^2\right) - s u (s+u)\right]}{m_{\cal B}^2\left(m_{\cal B}^2-s\right) \left(m_{\cal B}^2-u\right)}.  
 \end{split}
\end{equation}
\begin{figure}[htb]
    \centering
    \includegraphics[width=0.85\textwidth]{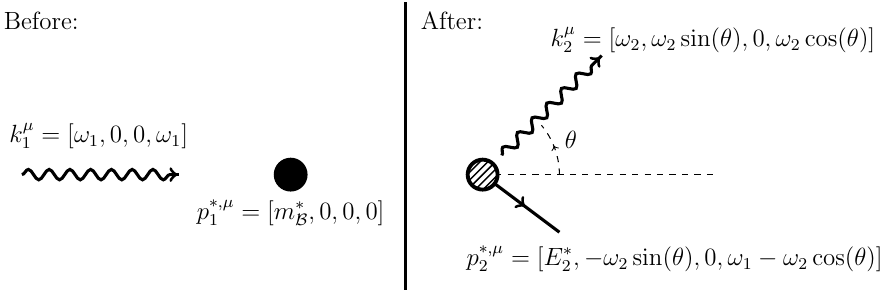}
    \caption{The Compton scattering in the rest (c.v.) frame of ${\cal B}(p_1)$. Note that even though there is a specific direction to the canonical momentum $\vec{p}_1^{\,\rm (c.v.)}\neq 0$, the amplitude depends only on $p^*_1$ for which $\vec{p}_1^{\,*, \rm (c.v.)} = 0$. Therefore, we have the freedom to choose the z-axis in the direction of the incoming photon. Evaluating the integrated cross section in Eq.~\eqref{eq:app:compton:diff_xsec} will require specifying the Fermi ellipsoid in the c.v. frame, which depends on $\vec{p}_1^{\,\rm (n.m.)}$ (see Eq.~\eqref{eq:app:compton:boost}).}
    \label{fig:app:compt:cvdiag}
\end{figure}
We now consider the Compton scattering in the rest (c.v.) frame of ${\cal B}(p_1)$ (see Fig.~\ref{fig:app:compt:cvdiag}), in which $\vec{p}_1^{\, *} = 0$. We first note that the relationship $k_1 \cdot k_2 = p^{*}_1 \cdot \left( k_1 - k_2\right)$, written in the c.v. frame, yields $\omega_1 \omega_2 \left( 1 - \cos \theta\right)= m_{\cal B}^* \left( \omega_1 - \omega_2 \right)$. We then arrive at the following kinematics in the c.v. frame
\begin{align}
    \omega_2 =& \frac{ \omega_1 }{ 1 + \frac{\omega_1}{m_{\cal B}^*} \left( 1 - \cos \theta\right)  },\\
    \vec{p}^{\, *}_2 =& \left[ -\omega_2 \sin\theta, 0, \omega_1 - \omega_2 \cos\theta \right],\\
    E^*_{\rm 2}=& \sqrt{ (m_{\cal B}^*)^2 + \omega_1^2 + \omega_2^2 - 2 \omega_1 \omega_2 \cos \theta}. 
\end{align}
which resembles the familiar Compton's formula. We use these kinematical relationships to write Eq.~\eqref{eq:app:compton:av_amp_sqr} in terms of $\omega_1$ and the scattering angle ($\theta$) in the c.v. frame as
\begin{equation}
    \begin{split}
        \overline{\left| \mathcal{M}\right|^2} =&  16 F_1^4 m_{\cal B} \bigg\{\frac{m_{\cal B}^3}{4} \left( \cos (2 \theta ) + 3 \right) + \omega_1 \sin^2\left(\theta/2\right)  m_{\cal B}^2 (\cos (2 \theta ) + 3) \\
        &\qquad \qquad + \omega_1^2 \left[m_{\cal B} (\cos (2 \theta ) + 5 ) - 2 \omega _1 (\cos (\theta)-1)\right] \sin^4\left(\theta/2\right)  \bigg\}\\ 
   & -8 F_1^2 F_2^2 m_{\cal B} \bigg\{
  \frac{\omega_1^3}{2} \left[\cos (2 \theta) -8 \cos (\theta ) + 7\right] \sin^2\left(\theta/2\right) -  m_{\cal B} \omega_1^2 \left[5\cos (\theta)-7\right]  \sin^2\left(\theta/2\right) \\
  &\qquad \qquad \qquad + 8 \omega_1 m_{\cal B}^2 \sin^2\left(\theta/2\right) + 2 m_{\cal B}^3 \bigg\} \\
   & + F_2^4 m_{\cal B} \omega _1^2 \left[ 5-\cos (2 \theta ) \right] \left[ m_{\cal B} -\omega_1 (\cos(\theta)-1) \right].
    \end{split}
\end{equation}

The phase space integrals over the final states (see Eq.~\eqref{eq:medium:modific:phase_space_int}) can be written as 
\begin{equation}
    \begin{split}
        \int d\Pi_2 &= \int \frac{d^3 k_2}{\left(2\pi\right)^3} \frac{1}{2 \omega_2} \frac{d^3 p^*_2}{\left(2\pi\right)^3} \frac{1}{2 E^*_2} \left(2\pi\right)^4 \delta^4 \left( k_2 + p^*_2 - k_1 - p^*_1\right) \, [1- f_{\cal B}(\vec{p}_2) ]\\
        &= \int \frac{\omega_2 d\omega_2 \, d\Omega_{2}}{16 \pi^2 E^*_2} \delta\left(\omega_2 + \sqrt{(m_{\cal B}^*)^2 + \omega^2_1 + \omega^2_2 - 2 \omega_1 \omega_2 \cos \theta} - \omega_1 - m_{\cal B}^*\right) [1-f_{\cal B}(\vec{p}_2)]\\
        &= \int \frac{d\Omega_{2}}{16 \pi^2 } \frac{\omega_2}{E^*_2 + \omega_2 - \omega_1 \cos \theta} [1-f_{\cal B}(\vec{p}_2)] = \int \frac{d\Omega_{2}}{16 \pi^2 } \frac{\omega_2^2}{m_{\cal B}^*\, \omega_1} [1-f_{\cal B}(\vec{p}_2)],  
    \end{split}
\end{equation}
in which $d\Omega_{2} = d\cos (\theta)\, d\phi$ is the differential solid angle of $\vec{k}_2$ in the c.v. frame, and $f_{\cal B}(\vec{p}_2)$ is the Pauli blocking factor for the outgoing baryon. The shape of the Fermi surface in a general frame (such as c.v.) changes from being spherical to an ellipsoid.
The general form of $f_{\cal B}(\vec{p}_2)$ in an arbitrary frame is given by $\theta(E_F^* - p_2^{*\mu} B_{\mu}/n_{\cal B})$~\cite{HOROWITZ1987613}, in which $E_{F, {\cal B}}^* \equiv \sqrt{p_{F, {\cal B}}^{2} + (m_{\cal B}^*)^{2}}$, $p_{F, {\cal B}}$ is the Fermi momentum defined in the n.m. frame, $B^{\mu}$ is the baryon current density defined below Eq.~\eqref{eq:medium:modific:baryon-propagator}, and $n_{\cal B}$ is the baryon number density. Evaluating the invariant argument of the step-function in the n.m. frame yields $\theta(E_F^* - E_{2}^{*, {\rm (n.m.)}})$, in which 
\begin{equation}
    E_{2}^{*, ({\rm n.m.})} = \left(\frac{E_{1}^{*, ({\rm n.m.})}\, E_2^{*, ({\rm c.v.})} }{m_{\cal B}^*}\right) + \left(\frac{\vec{p}_1^{\, \rm (n.m.)}  \cdot \vec{p}_2^{\, *, ({\rm c.v.})}}{m_{\cal B}^*}\right). \label{eq:app:compton:boost}
\end{equation}
We see that even though the amplitude in the c.v. frame depends only on $\theta$, integrating over the azimuthal angle ($\phi$) requires the explicit coordinates of the initial baryon ${\cal B}(p_1)$ momentum in the n.m. frame, $\vec{p}_1^{\, ({\rm n.m.})}$, in our chosen coordinate in Fig.~\ref{fig:app:compt:cvdiag}. Using Eq.~\eqref{eq:medium:modific:incident_beam_velocity}, and noting $v_{\mathcal{B}}^* = 0$, $v_{\mathcal{A}}^* = 1$ in our chosen frame (c.v.), the in-medium Compton scattering differential cross section can be written as 
\begin{equation}
    \frac{d \sigma}{d\Omega_2} = \frac{\omega_2^2\,  \overline{\left| \mathcal{M}\right|^2}}{64 \pi^2 \omega_1^2\, (m_{\cal B}^*)^2 } [1-f_{\cal B}(\vec{p}_2)], \label{eq:app:compton:diff_xsec}
\end{equation} 
in which we recover the Klein-Nishina~\cite{klein-nishina,peskin2018introduction} formula if we set $f_{\cal B}(\vec{p}_2) = 0$, $F_1 = e$, $F_2 = 0$ and replace $m_{\cal B}^*$ by $m_e$. 

\section{Fermion Mixing in Dense Matter}
\label{app:mixing}
In this appendix we evaluate the eigenvalues of a system consisting of a neutral baryon (${\cal B}$) and a dark fermion ($\chi$) with a mixing term between them, in the context of the RMF framework. We suppress the superscript (``*") in the baryon's effective mass ($m_{\cal B}^*$) for convenience. The Lagrangian for this system is presented in Eq.~\eqref{eq:dark_decay:medium:method:toy_mix_L}, with conjugate momenta given by
 \begin{equation}
     \Pi_{{\cal B}, \chi} = \frac{\partial \mathcal{L}}{\partial \dot{\psi}_{{\cal B}, \chi}} = i \psi^{\dagger}_{{\cal B}, \chi},
 \end{equation}
 and a coupled set of equation of motion 
 \begin{align}
     \left( i \slashed{\partial} - \slashed{\Sigma}_{\cal B} - m_{\cal B}\right) \psi_{\cal B} =&  \varepsilon \psi_{\chi}, \label{eq:app:mixing:coupled:Dirac:n}\\
    \left( i \slashed{\partial} - m_{\chi}\right) \psi_{\chi} =&  \varepsilon \psi_{{\cal B}}.
    \label{eq:app:mixing:coupled:Dirac:chi}
 \end{align}
 Note that the baryon current $J^{\mu}_{\cal B} \equiv \overline{\psi}_{\cal B} \gamma^{\mu} \psi_{\cal B}$ satisfies
  \begin{equation}
     \begin{split}
         \partial_{\mu} J^{\mu}_{\cal B} =& \left(\partial_{\mu} \overline{\psi}_{\cal B}\right) \gamma^{\mu} \psi_{\cal B} + \overline{\psi}_{\cal B} \slashed{\partial} \psi_{\cal B} = i \varepsilon \left( \overline{\psi}_{\chi} \psi_{\cal B} - \overline{\psi}_{{\cal B}} \psi_{\chi} \right),
     \end{split}
 \end{equation}
 and as expected is not conserved: rather, the combined current $J^{\mu} \equiv J^{\mu}_{\cal B} + J^{\mu}_{\chi}$ is conserved. The conserved energy-momentum prescribed by the Noether's theorem~\cite{Noether1918, peskin2018introduction} is given by
\begin{align}
    H =& \int d^3 x \left[ \overline{\psi}_{\cal B}\left( i \vec{\gamma} \cdot \vec{\nabla} + \slashed{\Sigma}_{\cal B} + m_{\cal B} \right) \psi_{\cal B} + \overline{\psi}_{\chi}\left( i \vec{\gamma} \cdot \vec{\nabla} + m_{\chi} \right) \psi_{\chi} +  \varepsilon \left(\overline{\psi}_{{\cal B}} \psi_{\chi} + \overline{\psi}_{\chi} \psi_{{\cal B}}\right)\right],  \label{eq:app:mixing:hamilton} \\
    \vec{P} =& \int d^3 x\, \left[ \psi_{\cal B}^{\dagger} \left( -i\vec{\nabla} \right)\psi_{\cal B} + \psi_{\chi}^{\dagger} \left( -i\vec{\nabla} \right)\psi_{\chi} \right]. 
\end{align}
We expand each of the fields in terms of four modes $\omega_{1,2}^{\pm}(\vec{k})$ as 
\begin{align}
    \psi(x, t) = \sum_{s} \int \frac{d^3k }{\sqrt{2(2\pi)^3}} \Bigg[ &\alpha \left( a_1(k, s) u_1(k, s) e^{- i \omega_1^{(+)} t + i k \cdot x} 
    +
    b_1^{\dagger}(k, s) v_1(k, s) e^{-i \omega_1^{(-)} t - i k \cdot x}  \right) \nonumber\\
    +& \beta \left(a_2(k, s)  u_2(k, s) e^{- i \omega_2^{(+)} t + i k\cdot x} + b_2^{\dagger}(k, s)  v_2(k, s) e^{-i \omega_2^{(-)} t - i k\cdot x}  \right)
    \Bigg], \label{eq:app:mixing:fourier}
\end{align}
in which $\psi(x, t)$ stands for $\psi_{{\cal B},\chi}$, $a_{1,2}$ and $b_{1,2}$ are the annihilation operators for particles and anti-particles, $\omega$ stands for $\omega(\vec{k})$, we note the inequality $\omega(\vec{k}) \neq \omega(-\vec{k})$ if $\vec{\Sigma}_{\cal B} \neq 0$ (see Eq.~\eqref{eq:medium:MFT:dirac:energy}), and the fact that in the presence of medium ($\Sigma^0_{\cal B} \neq 0$) the particle and anti-particle energies are not equal anymore (e.g., see Eq.~(2.40) of~\cite{HOROWITZ1987613}). The coefficients $\alpha$ and $\beta$ can be found by requiring that the Hamiltonian in Eq.~\eqref{eq:app:mixing:hamilton} is diagonal. The spinors $u(p,s)$ and $v(p,s)$ satisfy (see Eq.~(2.33) of Ref.~\cite{Serot:1984ey})
\begin{align}
    u^{\dagger}(p,s) u(p,s') =& v^{\dagger}(p,s) v(p,s') = \delta_{ss'},\\
    \sum_{s} u(p,s) \overline{u}(p,s) =& \frac{\slashed{p} + m}{2 E(p)}, \\
    \sum_{s} v(p,s) \overline{v}(p,s) =& \frac{\slashed{p} - m}{2 E(p)},
\end{align}
in which $m$ stands for $m_{\cal B}, m_{\chi}$, and we make the replacements $p \to p^*_{\cal B}$, $E \to E^*_{\cal B}$ for the baryon. We can combine Eqs.~\eqref{eq:app:mixing:coupled:Dirac:n} with~\eqref{eq:app:mixing:coupled:Dirac:chi} and arrive at the following equation after multiplying the left-hand-side by $\left(\slashed{P} + m_{\chi}\right) \left(\slashed{P}^* + m_{{\cal B}}\right)$:
\begin{equation}
\begin{split}
\left(P^2 - m_{\chi}^2\right) \left[ (P - \Sigma_{\cal B})^2 - m_{\cal B}^2\right] \psi_{\cal B} 
    =& \varepsilon^2 \left( 2 P_{\mu} \left( P^{\mu} - \Sigma^{\mu}_{\cal B}\right) + 2 m_{\cal B} m_{\chi} - \varepsilon^2 \right) \psi_{\cal B}, \label{eq:app:mixing:Dirac:comb}
\end{split}
\end{equation}
in which we note the definition $P_{\mu} = i \partial_{\mu} = ( H, -\vec{P})$. We now plug in the field expansion from Eq.~\eqref{eq:app:mixing:fourier} into Eq.~\eqref{eq:app:mixing:Dirac:comb} to arrive at the equation governing the spectrum of $\omega_{1,2}^{\pm}$ modes:
\begin{align}
    \left(\omega^{2} - k^2 - m_{\chi}^2\right) \left[ (\omega - \Sigma^0_{\cal B})^2 - (\vec{k} - \vec{\Sigma}_{\cal B})^2 - m_{\cal B}^2\right] 
    \label{eq:app:mixing:fmix:omega}
    = 2\varepsilon^2 &\left( \omega^2 - k^2 - \omega \Sigma^0_{\cal B} + \vec{k} \cdot \vec{\Sigma}_{\cal B} + m_{\cal B} m_{\chi} \right. \nonumber \\
    & \quad \left. - \frac{\varepsilon^2}{2} \right),
\end{align}
We denote the solutions in the absence of mixing ($\varepsilon=0$) by $\omega_0$, and solve Eq.~\eqref{eq:app:mixing:fmix:omega} using a perturbation series in powers of $\delta \equiv \varepsilon / \omega_0$: $\omega = \sum_{i=0}^{\infty} \omega_i \delta^i$, with the zeroth order equation yielding
\begin{equation}
   \mathcal{O}(\delta^0):\quad  \left[\left(\omega_0 - \Sigma^0_{\cal B}\right)^2 -(\vec{k} - \vec{\Sigma}_{\cal B})^2-m_{\cal B}^2 \right]  \left(\omega_0^2 - k^2-m_{\chi}^2 \right) = 0, \label{eq:app:mixing:fmix:zeroth}
\end{equation}
such that we get the usual spectrum for ${\cal B}$ and $\chi$:
\begin{align}
    \label{eq:app:mixing:fmix:zeroth_chi}
    \omega_{0}^{(\pm)}(\chi) =& \pm \sqrt{k^2 + m_{\chi}^2},\\
    \omega_{0}^{(\pm)}({\cal B}) =& \Sigma^0_{\cal B} \pm \sqrt{\left(\vec{k} - \vec{\Sigma}_{\cal B}\right)^2 + m_{{\cal B}}^2}.
    \label{eq:app:mixing:fmix:zeroth_n}
\end{align}
For the rest of this discussion we consider the energy spectrum in the n.m. frame ($\vec{\Sigma}_{\cal B}^{\, ({\rm n.m.})}=0$). Denoting $\Sigma^{0, ({\rm n.m.})}_{\cal B}$ by $\Sigma_0$ for convenience, we can rewrite  Eq.~\eqref{eq:app:mixing:fmix:omega} as
\begin{equation}
    \begin{split}
    &\omega^4 - 2 \Sigma_0 \omega^3 - \left[ 2 k^2 + m_{\cal B}^2 + m_{\chi}^2 - \Sigma_0^2 + 2 \varepsilon ^2 \right] \omega^2 
    + 2\Sigma_0 \left[ \left(k^2 + m_{\chi}^2\right) + \varepsilon^2 \right] \omega \\
        &+ \left(k^2 + m_{\chi }^2\right) \left(k^2 + m_{\cal B}^2 - \Sigma_0^2\right) + \varepsilon ^2 \left(2 k^2 - 2
   m_{\cal B} m_{\chi}\right)  + \varepsilon^4 = 0. 
\end{split}
\end{equation}
We note that both of the parenthesis in Eq.~\eqref{eq:app:mixing:fmix:zeroth} can be simultaneously equal to zero, if $k$ satisfies the following condition at zeroth order: 
\begin{equation}
    |\vec{k}| = \frac{ \sqrt{\left[ \left( \Sigma^0 - m_{\cal B}\right)^2 - m_{\chi}^2 \right]\left[ \left( \Sigma^0 + m_{\cal B}\right)^2 - m_{\chi}^2 \right]} } { 2\Sigma^0 }, \label{eq:app:mixing:fmix:transition:mom}
\end{equation}
We divide our solutions into two sets: the normal solutions for which the zeroth order condition in Eq.~\eqref{eq:app:mixing:fmix:transition:mom} is not satisfied, and those for which Eq.~\eqref{eq:app:mixing:fmix:transition:mom} is satisfied, which we denote by a ``*" superscript. First we write down the general equations to third order in the perturbation. The first order equation is given by
\begin{equation}
    \mathcal{O}(\delta^1):\quad   2 \omega_0 \omega_1 \left[\left(\omega_0 - \Sigma_0\right)^2 -k^2-m_{\cal B}^2 \right] + 2 \omega_1 \left(\omega_0 - \Sigma_0\right) \left(\omega_0^2 -k^2-m_{\chi}^2\right) = 0.
\end{equation}
Given the $\mathcal{O}(\delta^0)$ equation in Eq.~\eqref{eq:app:mixing:fmix:zeroth}, we conclude that either $\omega_1 = 0$ or the condition in Eq.~\eqref{eq:app:mixing:fmix:transition:mom} is satisfied. The second order equation yields:
\begin{equation}
    \begin{split}
 \mathcal{O}(\delta^2):\quad  &\left(\omega_1^2 + 2 \omega_0 \omega_2\right)  \left[\left(\omega_0 - \Sigma_0\right)^2 -k^2-m_{\cal B}^2 \right] - 2 \omega_0^2\left(m_{\cal B} m_{\chi} - \Sigma_0 \omega_0 + \omega_0^2 - k^2 \right) \\
   &\, + \left(2 \omega_2 \left(\omega_0 - \Sigma_0\right) + \omega_1^2\right)
   \left(\omega_0^2 - k^2 - m_{\chi}^2\right) + 4 \omega_0
   \omega_1^2 \left(\omega_0 - \Sigma_0\right) = 0.
\end{split}
\end{equation}
Finally, the third order equation is given by
\begin{equation}
    \begin{split}
    \mathcal{O}(\delta^3):\quad   &\left(2 \omega_1 \omega_2+2 \omega_0 \omega_3\right)
   \left[\left(\omega_0 - \Sigma_0\right)^2 -k^2-m_{\cal B}^2 \right] + 2\omega_1 \left(\omega _1^2+2 \omega_0 \omega_2\right) \left(\omega_0 - \Sigma_0\right)\\
   &+ \left(2 \omega _3 \left(\omega_0 - \Sigma_0\right) + 2 \omega_1 \omega_2\right) \left(\omega_0^2-k^2-m_{\chi}^2 \right) \\ 
   & - 2 \omega_0^2 \omega_1\left(2 \omega_0 - \Sigma_0\right)+2 \omega _1 \omega_0 \left(2 \omega _2
   \left(\omega_0 - \Sigma_0\right)+\omega _1^2\right) = 0.
   \end{split}
\end{equation}
First, let us assume that the condition in Eq.~\eqref{eq:app:mixing:fmix:transition:mom} doesn't hold, in which case $\mathcal{O}(\delta^1)$ condition yields $\omega_1 = 0$. We then solve for $\omega_2$ in $\mathcal{O}(\delta^2)$:
\begin{align}
    \omega_2^{(+)} (\chi) =& \frac{\sqrt{k^2 + m_{\chi}^2}
   \left(m_{\chi} \left(m_{\cal B}  + m_{\chi} \right) - \Sigma_0 \sqrt{k^2+m_{\chi}^2} \right)}{\left(\sqrt{k^2+m_{\chi}^2} - \Sigma_0\right)^2 - k^2 - m_{\cal B}^2},\\
    \omega_2^{(-)} (\chi) =& -\frac{\sqrt{k^2 + m_{\chi}^2}
   \left(m_{\chi} \left(m_{\cal B} + m_{\chi} \right) + \Sigma_0 \sqrt{k^2+m_{\chi}^2} \right)}{\left(\sqrt{k^2 + m_{\chi}^2} + \Sigma_0\right)^2- k^2 - m_{\cal B}^2},\\
  \omega_2^{(+)} ({\cal B}) =& 
   \frac{\left(\sqrt{k^2+m_{\cal B}^2} + \Sigma_0\right)^2
   \left(m_{\cal B} \left(m_{\cal B} + m_{\chi}\right) + \Sigma_0 \sqrt{k^2+m_{\cal B}^2}\right)}{\sqrt{k^2 + m_{\cal B}^2}\left[\left(\sqrt{k^2 + m_{\cal B}^2} + \Sigma_0\right)^2 - k^2 - m_{\chi}^2\right]},\\
    \omega_2^{(-)} ({\cal B}) =& -\frac{\left(\sqrt{k^2+m_{\cal B}^2} - \Sigma_0\right)^2
   \left(m_{\cal B} \left(m_{\cal B} + m_{\chi}\right) - \Sigma_0 \sqrt{k^2+m_{\cal B}^2}\right]}{\sqrt{k^2 + m_{\cal B}^2}\left(\Sigma_0 \left(\Sigma_0 - 2 \sqrt{k^2 + m_{\cal B}^2} \right) + m_{\cal B}^2 - m_{\chi}^2\right)},
\end{align}
which after plugging into the $\mathcal{O}(\delta^3)$ equation yields $\omega_3 = 0$, and so the energies of $\chi$ and ${\cal B}$ to third order are given by
\begin{align}
    \label{eq:app:mixing:fmix:normal:chi_pos}
    \omega^{(+)}(\chi) =& \sqrt{k^2+m_{\chi}^2} + \frac{\varepsilon^2 \left(m_{\chi} \left(m_{\cal B} + m_{\chi}\right) - \Sigma_0 \sqrt{k^2+m_{\chi}^2} \right)}{\sqrt{k^2+m_{\chi}^2}
   \left[\left(\sqrt{k^2 + m_{\chi}^2} - \Sigma_0\right)^2-k^2-m_{\cal B}^2\right]} + \mathcal{O}(\delta^4),\\
   \label{eq:app:mixing:fmix:normal:chi_neg}
    \omega^{(-)}(\chi) =& -\sqrt{k^2+m_{\chi}^2} - \frac{\varepsilon^2 \left(m_{\chi} \left(m_{\cal B} + m_{\chi}\right) + \Sigma_0 \sqrt{k^2+m_{\chi}^2} \right)}{\sqrt{k^2+m_{\chi}^2}
   \left[\left(\sqrt{k^2 + m_{\chi}^2} + \Sigma_0\right)^2-k^2-m_{\cal B}^2\right]} + \mathcal{O}(\delta^4),\\
    \label{eq:app:mixing:fmix:normal:n_pos}
    \omega^{(+)}({\cal B})=& \Sigma_0 + \sqrt{k^2+m_{\cal B}^2}  
     + \frac{\varepsilon ^2 \left(m_{\cal B} \left(m_{\cal B} + m_{\chi}\right) + \Sigma_0 \sqrt{k^2+m_{\cal B}^2}\right)}{\sqrt{k^2 + m_{\cal B}^2}
   \left[\left(\sqrt{k^2+m_{\cal B}^2} + \Sigma_0\right)^2- k^2 - m_{\chi}^2\right]} + \mathcal{O}(\delta^4),\\
    \label{eq:app:mixing:fmix:normal:n_neg}
    \omega^{(-)}({\cal B})=& \Sigma_0 -\sqrt{k^2+m_{\cal B}^2}  
     - \frac{\varepsilon ^2 \left(m_{\cal B} \left(m_{\cal B} + m_{\chi}\right) - \Sigma_0 \sqrt{k^2+m_{\cal B}^2}\right)}{\sqrt{k^2 + m_{\cal B}^2}
   \left[m_{\cal B}^2 - m_{\chi}^2 - \Sigma_0 \left(2 \sqrt{k^2+m_{\cal B}^2} - \Sigma_0\right) \right]} + \mathcal{O}(\delta^4),
\end{align}
in which the negative energy solutions would be interpreted as antiparticles. We now consider the second set of solutions assuming that Eq.~\eqref{eq:app:mixing:fmix:transition:mom} holds. The zeroth order equation yields
\begin{align}
    \omega_{0}^*(\chi) = \omega_{0}^*({\cal B}) =&  \frac{m_{\chi}^2 - m_{\cal B}^2 + \Sigma_0^2}{2 \Sigma^0}.\label{eq:app:mixing:fmix:transition:e}
\end{align}
The first order equation $\mathcal{O}(\delta^1)$ is trivial, and the second order equation yields:
\begin{equation}
\omega_1^* = \pm \sqrt{ \frac{\left(m_{\cal B}+m_{\chi}\right) \left[ 2 \Sigma_0^2 m_{\cal B} + \left(m_{\chi}-m_{\cal B}\right) \left(m_{\cal B}+m_{\chi}\right)^2\right] - \Sigma_0^4}{4 \left(m_{\chi}^2-m_{\cal B}^2-\Sigma_0^2\right)} },  \label{eq:app:mixing:fmix:transition:first_order}
\end{equation}
which we plug into the third order equation to arrive at
\begin{equation}
    \omega_2^*= \frac{ \left(m_{\chi }^2 - m_{\cal B}^2\right)^3 - \Sigma_0^2
   \left(m_{\chi} - m_{\cal B}\right) \left(m_{\cal B}+m_{\chi}\right)^3}{4 \Sigma_0 \left(m_{\cal B}^2-m_{\chi}^2+\Sigma_0^2\right)^2},
\end{equation}
such that the full energy spectrum is given by
\begin{equation}
\begin{split}
    \omega^{*}_{(\pm)} =& \frac{\Sigma_0^2 - m_{\cal B}^2 + m_{\chi}^2}{2 \Sigma_0} 
     \pm \varepsilon \Sigma_0 \sqrt{\frac{\left(m_{\cal B}+m_{\chi}\right)^2-\Sigma_0^2}{\left(m_{\cal B}^2 - m_{\chi}^2\right)^2-\Sigma_0^4}}\\
    &+ \frac{\varepsilon^2 \Sigma_0 \left(m_{\chi} - m_{\cal B}\right) \left(m_{\cal B}+m_{\chi }\right)^3
   \left[\left(m_{\cal B}-m_{\chi }\right)^2-\Sigma_0^2\right]}{\left[\left(m_{\cal B}^2-m_{\chi}^2\right)^2-\Sigma_0^4\right]^2}  + \mathcal{O}(\delta^3). \label{eq:app:mixing:fmix:transition:all_orders}
\end{split}
\end{equation}
We can see that the first order term breaks the degeneracy by splitting the energies. 

\section{Baryon Decays to $\chi + \gamma$}
\label{sec:app:JMB:decay}
\setcounter{equation}{0}

Here we present the full calculation of the decay of a baryon to $\chi + \gamma$. 

\subsection{Matrix Element}

The matrix element for this process is
\begin{equation}
	i \mathcal{M} = \frac{i\varepsilon_{{\cal B}\chi} g_{\cal B} e}{4 m_{\cal B}^*} \overline{u}_\chi(k_\chi) \frac{1}{\slashed{k}_\chi - \slashed{\Sigma}_{\cal B} - m_{\cal B}^*} \slashed{\epsilon}^* \slashed{k}_\gamma u(p^*_{\cal B}) \,, 
\end{equation}
where we note 
\begin{equation}
	(\slashed{k}_\chi - \slashed{\Sigma}_{\cal B} - m_{\cal B}^*)^{-1} = \frac{\slashed{k}_\chi - \slashed{\Sigma}_{\cal B} + m_{\cal B}^*}{(k_\chi - \Sigma_{\cal B})^2 - (m_{\cal B}^*)^2} \equiv \frac{\slashed{k}^* + m_{\cal B}^*}{(k^*_{\chi})^2 - (m_{\cal B}^*)^2}
\end{equation}
and we define the quantity $k^*_{\chi} \equiv k_\chi - \Sigma_{\cal B} = p_{\cal B}^* - k_\gamma$. Note in the neutron star medium that energy-momentum conservation of the total canonical 
momentum still holds: $p_{\cal B}^{\mu} = k_{\gamma}^{\mu} + k_{\chi}^{\mu}$. Consideration of the kinematics show that we need only consider the first term in the full baryon propagator given in Eq.~\eqref{eq:medium:modific:baryon-propagator}. 

We then find the spin-summed matrix element to be
\begin{align}
    |\mathcal{M}|^2 & = \frac{\varepsilon_{{\cal B}\chi}^2 g_{\cal B}^2 e^2 (p_{\cal B}^* \cdot k_\gamma)}{(m_{\cal B}^*)^2 [(m_{\cal B}^*)^2 - (k^*_{\chi})^2]^2} \big\{ [(m_{\cal B}^*)^2 - (k^*_{\chi})^2] (k_\chi \cdot k_\gamma) \nonumber\\
    & \qquad \qquad\qquad\qquad\qquad\qquad + 2 (k^*_{\chi} \cdot k_\gamma) (k^*_{\chi} \cdot k_\chi + m_{\cal B}^* m_{\chi}) \big\} \\
    & = \frac{\varepsilon_{{\cal B}\chi}^2 g_{\cal B}^2 e^2 (p_{\cal B}^* \cdot k_\gamma)}{2(m_{\cal B}^*)^2 (p_{\cal B}^* \cdot k_\gamma)^2} \left[(p_{\cal B}^* \cdot k_\gamma) (k_\chi \cdot k_\gamma) + (k^*_{\chi} \cdot k_\gamma) (k^*_{\chi} \cdot k_\chi + m_{\cal B}^* m_{\chi}) \right] \\
    & = \frac{\varepsilon_{{\cal B}\chi}^2 g_{\cal B}^2 e^2}{2(m_{\cal B}^*)^2} \left[ (p_{\cal B}^* \cdot k_\chi) + m_{\cal B}^* m_{\chi} \right],
\end{align}
where we note the useful relations $(k^*_{\chi})^2 = (m_{\cal B}^*)^2 - 2 (p_{\cal B}^* \cdot k_\gamma)$ and $(k^*_{\chi} \cdot k_\gamma) = (p_{\cal B}^* \cdot k_\gamma)$.

\subsection{Integrated Rates}

We now address full integral over phase space,
\begin{equation}
	\frac{dn_{\cal B}}{d\tau} = -\int \frac{d^3 \vec{p}_{\cal B}}{(2\pi)^3 (2E^*_{\cal B})} \frac{d^3 \vec{k}_\chi}{(2\pi)^3 (2E_\chi)} \frac{d^3 \vec{k}_\gamma}{(2\pi)^3 (2E_\gamma)} f_{\cal B}(\vec{p}_{\cal B}) \times |\mathcal{M}|^2 \times (2\pi)^4 \delta^{(4)}\left(p_{\cal B} - k_\chi - k_\gamma \right). \label{eq:master_decay_rate1}
\end{equation}

In the main text, we presented the rate as an integral over the baryon Fermi sphere of the dilated widths of individual baryons. Here, we will contrast this approach with a more straightforward evaluation of this integral and demonstrate that these yield consistent results, as expected.

Our first step in the evaluation of the rate is to separate the integrals over the $\chi$ and $\gamma$ phase spaces and evaluate these first:
\begin{align}
    \dfrac{dn_{\cal B}}{d\tau} & = -\int_0^{p_{F,n}} \dfrac{p^2 \, dp}{4\pi^2 E^*_{\cal B}}\, 
    {\cal G}(p); 
    \\
    {\cal G}(p) & = \int \dfrac{d^3 \vec{k}_\chi}{(2\pi)^3 (2E_\chi)} \dfrac{d^3 \vec{k}_\gamma}{(2\pi)^3 (2E_\gamma)} |\mathcal{M}|^2 \times (2\pi)^4 \delta^{(4)}\left(p_{\cal B} - k_\chi - k_\gamma \right).
\end{align}
We have simplified the first integral by noting that it only depends on the magnitude of the three-momentum $|\vec{p}_{\cal B}| \equiv p$, and that we only integrate within the neutron Fermi sphere. We tackle this second integral by computing it in the c.m. frame of the decaying neutron. We note, however, that the matrix element depends on $p_{\cal B}^*$, which has a nonvanishing spatial component, though we will find that this is not relevant for the ultimate evaluation of the integral.

We begin by articulating the boost between the n.m.~frame and the c.m.~frame. We denote the 4-momentum of a baryon in the n.m.~frame by $p_{\cal B}^{\rm (n.m.)} = \left(E_{\cal B}, \vec{p}_{\cal B}\right)$, and the vector self-energy as  $\Sigma_{\cal B}^{\rm (n.m.)} = \left(\Sigma_{\cal B}^0, \vec{0}\right)$, and express the c.m.~frame 4-momentum $p_{\cal B}^{\rm (c.m.)}$ only in terms of the quantities in the n.m.~frame. The boost from the n.m.~to the c.m.~frame is parameterized by
\begin{equation}
    \gamma = \dfrac{E_{\cal B}^{*} + \Sigma^0_{\cal B}}{\sqrt{s}}, \quad \gamma \beta = \dfrac{|\vec{p}_{\cal B}|}{\sqrt{s}}, \quad s \equiv (m_{\cal B}^*)^2 + 2 E_{\cal B}^* \Sigma^0_{\cal B} + (\Sigma^0_{\cal B})^2, \label{eq:app:decay:boost_factors}
\end{equation}
such that 
\begin{equation}
    p_{\cal B}^{\rm (c.m.)} = \Lambda \cdot p_{\cal B}^{\rm (n.m.)} = \left( \begin{array}{cc} \gamma & -\gamma \beta \\ -\gamma \beta & \gamma \end{array} \right) \left( \begin{array}{c} E_{\cal B}^* + \Sigma^0_{\cal B} \\ |\vec{p}_{\cal B}| \hat{z} \end{array} \right) = \left( \begin{array}{c} \sqrt{s} \\ \vec{0} \end{array} \right).
\end{equation}
Therefore, we write $p_{\cal B}^*$ in the c.m.~frame as
\begin{equation}
	p_{\cal B}^{*, \rm{(c.m.)}} = \frac{1}{\sqrt{s}} \left( \begin{array}{c} (m_{\cal B}^*)^2 + E_{\cal B}^* \Sigma^0_{\cal B} \\ |\vec{p}_{\cal B}| \Sigma^0_{\cal B} \hat{z} \end{array} \right).
\end{equation}
With 
\begin{equation}
	E_\chi^{\rm{(c.m.)}} = \dfrac{s + m_{\chi}^2}{2\sqrt{s}};  \qquad |\vec{k}_\chi^{\rm{(c.m.)}}| = \dfrac{s - m_{\chi}^2}{2\sqrt{s}}.
\end{equation}
we may write
\begin{equation}
	p_{\cal B}^{*, \rm{(c.m.)}} \cdot k_\chi^{\rm{(c.m.)}} = \frac{[(m_{\cal B}^*)^2 + E_{\cal B}^* \Sigma^0_{\cal B}](s + m_{\chi}^2)}{2s} - \frac{(|\vec{p}_{\cal B}| \Sigma^0_{\cal B})(s - m_{\chi}^2)}{2s} \cos\theta^*,
\end{equation}
in which $\cos\theta^*$ is the angle between $\vec{k}_\chi^{\rm{(c.m.)}}$ and $(\vec{p}_{\cal B})^{*, \rm{(c.m.)}}$, and we note that the second term vanishes once an integration over the direction of $\vec{k}_\chi^{\rm{(c.m.)}}$ is performed. We consider next the form of the energy delta function when transformed to momentum; using $k\equiv |\vec{k}_\gamma^{\rm{(c.m.)}}| = |k_\chi^{\rm{(c.m.)}}|$, we have
\begin{equation}
	\delta\left(\sqrt{s} - k - \sqrt{k^2 + m_{\chi}^2}\right) = \delta\left(k - \dfrac{s - m_{\chi}^2}{2\sqrt{s}}\right) \frac{E_\chi}{\sqrt{s}}.
\end{equation}
Putting these pieces together, we arrive at
\begin{align}
	{\cal G}(p) & = \int \frac{k \, dk}{4\pi E_\chi} \delta\left(k - \dfrac{s - m_{\chi}^2}{2\sqrt{s}}\right) \frac{E_\chi}{\sqrt{s}} \times \frac{\varepsilon_{n\chi}^2 g_{\cal B}^2 e^2}{2(m_{\cal B}^*)^2} \frac{[(m_{\cal B}^*)^2 + E_{\cal B}^* \Sigma^0_{\cal B}](s + m_{\chi}^2) + 2sm_{\cal B}^* m_{\chi}}{2s} \\
	& = \frac{\varepsilon_{{\cal B}\chi}^2 g_{\cal B}^2 e^2}{32\pi(m_{\cal B}^*)^2} \left(\frac{s-m_{\chi}^2}{s^2}\right) \left\{ s[(m_{\cal B}^*)^2 + E_{\cal B}^* \Sigma^0_{\cal B} + 2m_{\cal B}^* m_{\chi}] + m_{\chi}^2[(m_{\cal B}^*)^2 + E_{\cal B}^* \Sigma^0_{\cal B}] \right\}. \label{eq:app:JMB:decay:gamma}
\end{align}
We can therefore write Eq.~\eqref{eq:master_decay_rate1} as
\begin{equation}
    \dfrac{dn_{\cal B}}{d\tau} = -\int_0^{p_{F,{\cal B}}} \frac{p^2 \, dp}{4\pi^2 E^*_{\cal B}} {\cal G}(p) = -\int_{m_{\cal B}^*}^{E_{F,{\cal B}}} \frac{\sqrt{(E_{\cal B}^*)^2 - (m_{\cal B}^*)^2} dE_{\cal B}^*}{4\pi^2} \, {\cal G}(p),
\end{equation}
which after using the definitions in Eq.~\eqref{eq:dark_decay:medium:dim_less_def} turns into the expression given in Eq.~\eqref{eq:app:JMB:decay:tot_rate}. Using Eq.~\eqref{eq:medium:decay:rate_boosts}, we can also write the individual baryon decay rate in the c.m.~frame $\Gamma_{\rm c.m.} (p_{\cal B})$ as

\begin{equation}
    \begin{split}
        \Gamma_{\rm c.m.} (p_{\cal B}) = \left(\frac{g_{\cal B}^2 e^2 \varepsilon_{{\cal B} \chi}^2}{128\pi\, m_{\cal B}^*}\right) \frac{1 + \sigma^2 + 2 x \sigma - \mu^2}{\left( 1 + \sigma^2 + 2 x\sigma \right)^{3/2} \left( 1 + x \sigma \right)} &\left[ \left(1 + \sigma^2 + 2 \sigma x\right) \left( 1 + \sigma x + 2 \mu \right) \right.\\
        &\,\, \left. + \mu^2 (1 + \sigma x)\right] \,.
        \label{eq:app:decay:cm_rate}
    \end{split}
\end{equation}
We note that if the self-energy were to vanish ($\sigma=0$) we would recover the vacuum decay rate reported in Eq.~(\ref{eq:dark_decay:vacuum:vacdec}). 

\bibliography{macro4micro}

\end{document}